\documentclass[12pt]{iopart}
\usepackage{iopams}

\usepackage{bm}
\usepackage{amsthm,amsfonts,epsfig,setspace}
\usepackage{colortbl,color}
\usepackage[margin=1in]{geometry}
\usepackage{epstopdf}
\usepackage{amssymb}
\usepackage{mathrsfs}
\usepackage{latexsym,bm}
\usepackage{amsthm}
\usepackage{float}
\usepackage{graphicx}
\usepackage{CJK}
\usepackage{graphics}
\usepackage{color}
\usepackage{subfigure}
\usepackage[english]{babel}

\usepackage{algorithm}
\usepackage{algpseudocode}

\usepackage{booktabs}
\usepackage{diagbox,multirow,arydshln}
\usepackage{array}
\usepackage{hhline}
\usepackage{makecell}
\usepackage{float}
\usepackage{framed}
\usepackage{varwidth}
\usepackage[export]{adjustbox}
\usepackage{colortbl}
\usepackage{sidecap}
\usepackage[section]{placeins}
\usepackage{framed}
\usepackage{caption}
\usepackage{placeins}
\usepackage{needspace}

\newcommand{\correct}{\color{black}}

\newcommand {\pic}{\setlength{\unitlength}{1cm}
\begin{picture}(4,0)
\thicklines \put(-0.5,0){\line(1,0){16}}
\end{picture}
}
\newtheorem{thm}{Theorem}[section]

\begin{document}
\title[Multi-physics joint inversion using full-waveform and gravity data]
{A level-set structural approach for multi-physics joint inversion using full-waveform and gravity data}

\author{Xingyu Deng\textsuperscript{1}, Jianfeng Zhao\textsuperscript{2}, Wenbin Li\textsuperscript{1}, Jianwei Ma\textsuperscript{2,3}}
\address{\textsuperscript{1} School of Science, Center of Geophysics, Harbin Institute of Technology, Shenzhen, Shenzhen, 518055, China}
\address{\textsuperscript{2} School of Mathematics, Center of Geophysics, Harbin Institute of Technology, Harbin, 150001, China}
\address{\textsuperscript{3} School of Earth and Space Sciences, Peking University, Beijing, 100871, China}
\ead{liwenbin@hit.edu.cn, jma@hit.edu.cn}

\begin{abstract}
This paper presents a level-set based structural approach for the joint inversion of full-waveform and gravity data. The joint inversion aims to integrate the strengths of full-waveform inversion for high resolution imaging and gravity inversion for detecting density contrasts over extensive regions. Although common studies typically only observe full-waveform inversion assisting gravity inversion, we propose three key points that enable gravity data to complement full-waveform data in the joint inversion. (i) Based on the well-posedness theorem, we consider a volume mass distribution where the density-contrast value is imposed as a priori information, ensuring that the gravity data provide meaningful information. (ii) We utilize a level-set formulation to characterize the shared interface of wave velocity and density functions, connecting multi-physics datasets via the structural similarity of their inversion parameters. (iii) We develop a balanced and decaying weight to regulate the influence of multi-physics datasets during joint inversion. This weight comprises a balanced part that accounts for the differing scales of full-waveform and gravity data, and a decaying part designed to effectively utilize the features and advantages of each dataset.
\end{abstract}
\maketitle

\section{Introduction}
The inverse problem of joint inversion stems from the challenge of accurately imaging subsurface structures in geophysical exploration, where the joint inversion of multi-physics datasets has emerged as an effective strategy to mitigate the inherent limitations of individual methods \cite{delberchi16}. Full-waveform inversion (FWI) is a powerful tool for constructing high-resolution images due to its ability to exploit the entire content of seismic recordings \cite{virope09}. However, FWI often suffers from local minima caused by the inaccuracy of initial model and the lack of low-frequency components in seismic data \cite{yonliahuazhe18, konrovetc23, metbroetc16}. It struggles to correctly resolve deep or extensive structures, especially in regions with sharp interfaces where complex wave phenomena such as scattering and reflections arise. In contrast, gravity inversion is well suited to imaging large-scale features \cite{liold98}, and it is sensitive to density contrasts associated with interfaces \cite{liluqia16}.

{\correct The joint inverse problem involves multiple systems of equations. A key challenge is to establish mathematical relationships between different inversion parameters, i.e., wave velocity $c$ and density $\rho$ in the joint inversion of full-waveform and gravity data.  There are typically two types of joint inversion approaches for connecting multi-physics parameters. A direct approach is to specify an explicit formula between their values \cite{bir61, garetc74, savrodmas82, niejac00, afnkoknak02, siletc20}. For example, in \cite{niejac00}, the authors consider a linear formula $c=b\rho+a$; in \cite{afnkoknak02}, the authors derive a log-linear formula $\rho=a\mathrm{ln}c+b$; in \cite{siletc20}, the authors employ a power function $\rho=\rho_0 c^{k_0}$. This type of approach is termed the compositional approach \cite{leletc12}, as it relies on an explicit formula derived from geologic or petrophysical composition to link multi-physics parameters. A key limitation of this approach, however, is the lack of a universally applicable formula; the velocity-density relationships can vary with factors such as medium type, temperature, and pressure \cite{niejac00}.}

{\correct A more general approach is to consider structural similarity between inversion parameters \cite{heretc95, habold97, galetc03, galmej04, mooetc11, abugaohabliu12,crestagha18}. Although velocity $c$ and density $\rho$ represent distinct physical properties, they correspond to the same subsurface region and thus exhibit comparable structural features. For example, if there is an interface, e.g., the boundary of a salt dome, such structural discontinuity should appear in both $c$ and $\rho$. This idea forms the basis of the structural approach, which uses geometric correlations to establish connections between multi-physics parameters. A prominent structural approach is the cross-gradient method \cite{galmej04, abugaohabliu12, crestagha18, jia20}. This method assumes that if two physical properties, such as $c(\mathbf{r})$ and $\rho(\mathbf{r})$, share structural features, their gradients $\nabla c$ and $\nabla\rho$ should align with each other. Consequently, their cross-product, $\nabla c\times\nabla\rho$, should be close to $\mathbf{0}$. The cross-gradient method imposes a penalization on the cross-product of gradients to achieve structural similarity between inversion parameters. Another useful approach is based on curvatures of model parameters \cite{habold97}, using curvature to indicate the structural similarity, and penalizing the difference of curvatures of multi-physics parameters. 
Both the cross-gradient and curvature-based methods rely on penalty terms (cross-product of gradients, or difference of curvatures) to enforce structural similarity, which is often insufficient. Excessive penalization can dominate the total energy and suppress data misfit terms, whereas weak penalization fails to ensure meaningful structural correspondence. Ultimately, penalty-based approaches provide only an indirect and often inadequate mechanism for enforcing structural similarity. In this work, we develop a joint inversion algorithm based on level-set interface inversion, employing the level-set method \cite{oshset88} to describe the interface structures of wave-velocity and density functions, characterizing their structural similarity directly through shared interfaces.}

The level-set method can naturally handle topological changes of interfaces, and is widely used for solving inverse problems involving interface structures. For instance, it is employed in inverse scattering to determine the geometry of obstacles \cite{litlessan98,dorles06}, in seismic tomography to recover reflectors and velocity discontinuities \cite{lileuqia14}, and in inverse gravimetry to delineate the domain of density contrasts \cite{isaleuqia11,liqia21}. 
{\correct Moreover, the concept of the level-set method has led to novel algorithms for binary tomography \cite{kadvan19}.}
In this work, we utilize the level-set function to describe the shared interface between wave-velocity and density functions. By explicitly enhancing the characterization of shared interface, we can establish a strong structural coupling between the wave-velocity and density functions. Moreover, in the level-set formulation it is straightforward to impose a priori information on model parameters.

To achieve effective integration of gravity and full-waveform data, it is essential to derive unambiguous information from the gravity data. The inverse problem of gravimetry is notoriously ill-posed and lacks inherent resolution \cite{isa90}. Consequently, the well-posedness theorem of inverse gravimetry should be studied, ensuring that the gravity data can yield reliable information for joint inversion. In addition, since the joint inversion employs multi-physics datasets, determining a weighting strategy to regulate the influence of each dataset is a critical challenge. We will address these problems in this work.

The rest of the paper is organized as follows. In section \ref{sec2}, we explain mathematical models of the joint inverse problem, and discuss the way of using gravity data to extract unambiguous information. In section \ref{sec3}, we propose the level-set structural approach for joint inversion, including the level-set formulation, regularization techniques, and optimization strategies. In section \ref{sec4}, we develop a weighting term to regulate the influence of full-waveform and gravity data, which is crucial to the success of multi-physics joint inversion. Section \ref{sec5} provides computational results, where the units of physical quantities involved in the joint inversion are specified. Finally, in section \ref{sec6}, we draw our conclusions.

\section{Joint inversion of full-waveform and gravity data} \label{sec2}
\subsection{Modeling of the joint inversion}
We consider an acoustic wave equation for modeling of the waveform data,
\begin{equation} \label{eqn1}
\nabla^2p(\mathbf{r},t)-\frac{1}{c^2(\mathbf{r})}\frac{\partial^2p(\mathbf{r},t)}{\partial t^2}=s(\mathbf{r},t)\,,
\end{equation}
where $p$ denotes the wave pressure field, $c$ is the wave velocity, and $s(\mathbf{r},t)$ denotes the source term; $\mathbf{r}$ and $t$ are spatial and temporal coordinates, respectively. A band-limited point source term is considered,
\begin{equation}\label{eqn2}
s(\mathbf{r},t)=\delta(\mathbf{r}-\mathbf{r}_s) \,{\correct s_0(t-t_0)}\,,
\end{equation}
where $\delta(\cdot)$ denotes the Dirac delta function, $\mathbf{r}_s$ is the point source location, and {\correct $s_0(t-t_0)$ is the Ricker wavelet: $s_0(t-t_0)= \left(1 - 2\pi^2 f_0^2 (t-t_0)^2 \right) \cdot e^{-\pi^2 f_0^2 (t-t_0)^2}$.} For each point source $\mathbf{r}_s$, the waveform data $p(\mathbf{r},t)$ are acquired along space and time: $\Gamma_p\times [0,T]$, where $\Gamma_p$ denotes the measurement surface.

The gravity potential can be modeled by the integral equation,
\begin{equation}\label{eqn3}
U(\tilde{\mathbf{r}})=\gamma \int_{\Omega} K(\tilde{\mathbf{r}},\mathbf{r})\,\rho(\mathbf{r})\mathrm{d}\mathbf{r}\,,
\end{equation}
where $\rho(\mathbf{r})$ denotes the anomalous density function, $\gamma$ is a constant related to the universal gravitational constant, and $U(\tilde{\mathbf{r}})$ is the generated gravity potential. The integral kernel $K(\tilde{\mathbf{r}},\mathbf{r})=K(|\tilde{\mathbf{r}}-\mathbf{r}|)$ is the fundamental solution of Laplace's equation,
\begin{equation}\label{eqn4}
K(\tilde{\mathbf{r}},\mathbf{r})=\left\{\begin{array}{ccc}
-\frac{1}{2\pi}\mathrm{ln}|\tilde{\mathbf{r}}-\mathbf{r}| &,&\tilde{\mathbf{r}},\mathbf{r} \in\mathbf{R}^2,\\
\frac{1}{4\pi|\tilde{\mathbf{r}}-\mathbf{r}|} &,&\tilde{\mathbf{r}},\mathbf{r} \in\mathbf{R}^3.
\end{array}\right.
\end{equation}
Here, we consider the vertical component of gravity acceleration, which is the commonly used gravity data in geophysical explorations,
\begin{equation}\label{eqn5}
g_z:=\frac{\partial U(\tilde{\mathbf{r}})}{\partial \tilde{z}}=\gamma \int_{\Omega} K_z(\tilde{\mathbf{r}},\mathbf{r})\,\rho(\mathbf{r})\mathrm{d}\mathbf{r}\,, 
\end{equation}
where $K_z(\tilde{\mathbf{r}},\mathbf{r}):=\frac{\partial K}{\partial\tilde{z}}=-\frac{1}{2^{d-1}\pi}\frac{\tilde{z}-z}{|\tilde{\mathbf{r}}-\mathbf{r}|^d},\ d=2,3$, and $\tilde{z}$ denotes the vertical component of the spatial coordinate $\tilde{\mathbf{r}}$. The gravity data $g_z$ are acquired along a measurement surface $\Gamma_g$.

Given the full-waveform data $p(\mathbf{r},t)$ along $\Gamma_p\times [0,T]$, and the gravity acceleration $g_z$ along $\Gamma_g$, the joint inversion aims to recover both the wave velocity $c(\mathbf{r})$ and the density function $\rho(\mathbf{r})$.

\subsection{\correct A crucial guide for the use of gravity data}
%{Let the gravity data provide information}
The gravity data have low resolution, and the inverse problem of gravity is severely ill-posed in the Hadamard sense, e.g.,  the same measurement data can correspond to quite different density distributions. To make the gravity and full-waveform data complement each other, it is crucial to extract unambiguous information from the gravity data; otherwise, the joint inversion may only reflect scenarios where FWI assists gravity.

The following uniqueness result provides a math insight into the way of using gravity data \cite{isa90, liqia21,cheli24}.
\begin{thm}\label{thm1}
Let $\Omega_0$ be a convex domain with analytic (regular) boundary, $\Gamma_g\subset\partial\Omega_0$  be a nonempty hyper-surface, and $\Omega\subset\Omega_0$ be a bounded domain with connected $\mathbf{R}^d\setminus\overline{\Omega}$. 
Consider a volume mass distribution with the density function $\rho(\mathbf{r})=f(\mathbf{r})\chi_D(\mathbf{r})$; $D\subset\Omega$ denotes the domain of mass anomaly, which admits piecewise smooth boundaries, and $\chi_D$ is a characteristic function : $\chi_D(\mathbf{r})=1$, $\mathbf{r}\in D$; $\chi_D(\mathbf{r})=0$, $\mathbf{r}\notin D$. Given the modulus of gravity acceleration, $|\nabla U|$, on $\Gamma_g$, and given $f\ge0$ in $\Omega$, the domain $D$ can be uniquely determined if one of the following constraints is satisfied:

(1) $D$ is star-shaped with respect to its center of gravity, and $f$ is constant;

(2) $D$ is convex in one direction, e.g. in $x_d$, where $x_d$ denotes a component of the spatial coordinate $\mathbf{r}=(x_1,\cdots,x_d)\in\mathbf{R}^d$, and $f$ is constant;

(3) $D$ is convex in $x_d$, $f$ does not depend on $x_d$, $f\in C(\Omega)$, and $\Omega\subset \mathrm{supp}\,f$;

(4) $D$ is convex, $f\in L_1(\Omega)$, and $0<f$ on $\Omega$.
\end{thm}

Theorem \ref{thm1} indicates that the gravity inversion requires the density-contrast value $f(\mathbf{r})$, and then the domain of mass can be uniquely determined under certain constraints. This result suggests that we should consider a volume mass distribution in the joint inversion, and impose the density-contrast value $f(\mathbf{r})$ as a priori information. Then the gravity data can provide helpful information for the joint inversion although we only employ the vertical component of $\nabla U$.

{\correct Theorem \ref{thm1} also rigorously requires the domain $D$ to be star-shaped, convex, or convex in one direction. Among these, `convex in one direction' is the most applicable, as the other two are too restrictive and exclude scenarios with disjoint objects. However, we will not enforce these geometric constraints on $D$ in the joint inversion algorithm. Rigorously imposing such constraints would make the inversion algorithm overly complicated and impractical for real-world applications. Instead, it should be understood that the closer the solution's conditions are to the requirements of Theorem \ref{thm1}, the more accurate the information provided by the gravity data. Thus, Theorem \ref{thm1} serves as a crucial guide for the use of gravity data in joint inversion.}

\section{The level-set structural approach} \label{sec3}

\subsection{Level-set formulation}
Theorem \ref{thm1} suggests a volume mass distribution with the anomalous density $\rho(\mathbf{r})=f(\mathbf{r})\chi_D(\mathbf{r})$, and so one can extract meaningful insights from the gravity data. In the joint inversion, we assume that the wave velocity $c(\mathbf{r})$ and the anomalous density $\rho(\mathbf{r})$ have the same interface structure, corresponding to internal discontinuities in the medium. Thus, we consider a similar formulation for the velocity function: $c(\mathbf{r})=c_1(\mathbf{r})\chi_D(\mathbf{r})+c_2(\mathbf{r}) (1-\chi_D(\mathbf{r}))$. The boundary of the domain $D$ depicts the shared interface of $c(\mathbf{r})$ and $\rho(\mathbf{r})$, and this structural similarity links the two parameters in the joint inversion. Note that $D$ is not necessarily connected.

We use a level-set method to describe the domain $D$ and its interface in $c(\mathbf{r})$ and $\rho(\mathbf{r})$,
\begin{equation}\label{eqn6}
\rho(\mathbf{r})=f(\mathbf{r})H(\phi(\mathbf{r})), \quad c(\mathbf{r})=c_1(\mathbf{r})H(\phi(\mathbf{r}))+c_2(\mathbf{r}) (1-H(\phi(\mathbf{r})))\,.
\end{equation}
Here, $\phi$ is the level-set function, which can be a signed distance to the interface $\partial D$,
\begin{equation}\label{eqn7}
\phi(\mathbf{r})=\left\{\begin{array}{ccc}
\mathrm{dist}(\mathbf{r},\partial D)&,&\mathbf{r} \in \bar{D}\\
-\mathrm{dist}(\mathbf{r},\partial D)&,&\mathbf{r} \in \bar{D}^c
\end{array}\right.;
\end{equation}
$H(\cdot)$ is the Heaviside function: $H(\phi)=1,\ \phi\ge0$, and $H(\phi)=0,\ \phi<0$.  The zero level set $\{\mathbf{r}\mid\phi(\mathbf{r})=0\}$ depicts the interface $\partial D$, and $H(\phi(\mathbf{r}))$ expresses the characteristic function $\chi_D(\mathbf{r})$. In the joint inversion the level-set function $\phi$ links the parameters $c(\mathbf{r})$ and $\rho(\mathbf{r})$.

The following derivatives are useful:
\begin{equation}\label{eqn8}
\hspace{-60pt}
\frac{\partial\rho}{\partial\phi}=f(\mathbf{r})\delta(\phi(\mathbf{r})), \  \frac{\partial c}{\partial\phi}=(c_1(\mathbf{r})-c_2(\mathbf{r}))\delta(\phi(\mathbf{r})),\ 
\frac{\partial c}{\partial c_1}=H(\phi(\mathbf{r})), \  \frac{\partial c}{\partial c_2}=1-H(\phi(\mathbf{r})),
\end{equation}
where $\delta(\cdot)$ denotes the Dirac delta function; one should use a numerical delta function in computations. Since the full-waveform inversion can cause instabilities near the sharp interface, we suggest the following smooth version of the delta function and its corresponding Heaviside,
\begin{equation}\label{eqn9}
H_\tau(\phi)=\frac{1}{2}\left(\tanh\frac{\phi}{\tau}+1 \right)\,,\quad \delta_\tau(\phi)=H_\tau'(\phi)=\frac{1}{2\tau}\frac{1}{\cosh^2\frac{\phi}{\tau}}\,,
\end{equation}
where the small parameter $\tau$ controls the thickness of interface, e.g., $\tau$ is taken as the mesh size of discretization. The choice of these two functions is important to the performance of the joint inversion.
%  and we will not impose the constraints in Theorem \ref{thm1}, which is the requirement of pure gravity inversion.

\subsection{Data fitting and regularization}
Now we propose the energy function for the joint inversion. Suppose that there are $N_s$ point-source wave excitations, and define the measured waveform data as $p_i^*(\mathbf{r},t)$, $i=1,\cdots, N_s$; let $g_z^*(\tilde{\mathbf{r}})$ denote the measured gravity data. We have two data-fitting terms in the joint inversion,
\begin{equation}\label{eqn10}
\sum_{i=1}^{N_s}\int_{\Gamma_p\times [0,T]} \left|p_i(\mathbf{r},t)-p_i^*(\mathbf{r},t) \right|^2\mathrm{d}\mathbf{r}\,\mathrm{d}t\,,\quad
\mathrm{and}\ \  \int_{\Gamma_g} \left|g_z(\tilde{\mathbf{r}})-g_z^*(\tilde{\mathbf{r}}) \right|^2 \mathrm{d}\tilde{\mathbf{r}}\,,
\end{equation}
where $p_i(\mathbf{r},t)$ and $g_z(\tilde{\mathbf{r}}) $ denote the predicted waveform data and gravity data, respectively. In practice, it is more natural to use a discretized formulation. Let $N_r$ denote the number of receivers along $\Gamma_p$, $N_t$ denote the number of sampling instances along $[0,T]$, and $N_m$ denote the number of measurements along $\Gamma_g$. We consider the following data-fitting terms:
\begin{equation}\label{eqn11}
E_p:=\frac{1}{N_s N_r N_t}\sum_{i=1}^{N_s}\sum_{j=1}^{N_r}\sum_{k=1}^{N_t} \left|p_{i,j,k}-p_{i,j,k}^* \right|^2\,;\quad
E_g:=\frac{1}{N_m} \sum_{j=1}^{N_m} \left|g_{z,j}-g_{z,j}^*\right|^2 \,.
\end{equation}
The subscripts $i,j,k$ in $p$ and $p^*$ indicate the $i$-th point source, the $j$-th receiver, and the $k$-th instance, respectively; the subscript $j$ in $g_z$ and $g_z^*$ indicates the $j$-th measurement.

The data misfit function in the joint inversion is as follows,
\begin{equation}\label{eqn12}
E_d=E_p+\omega\cdot E_g\,,
\end{equation}
where $\omega$ is a weighting term which manages the contributions of the two data-fitting terms. Since the waveform data and the gravity data have distinct units, and the scales of $E_p$ and $E_g$ can differ significantly, the choice of $\omega$ is crucial to the success of joint inversion. We will discuss it in detail in Section \ref{sec4}.

%%%%%
The following regularizations are useful in the level-set joint inversion.

\subsubsection{Level-set reinitialization.} \label{subsubrein}
Reinitialization  aims to maintain the level-set function $\phi$ as a signed-distance function, as shown in equation (\ref{eqn7}), so that $\phi$ is well behaved near the zero level set. A signed-distance function satisfies $|\nabla\phi|=1$, and a standard way to perform reinitialization is to solve the following PDE system \cite{sussmeosh94,oshfed06},
\begin{equation}\label{eqn13}
\left\{
\begin{array}{l}
\frac{\partial\Phi}{\partial\xi}+\mathrm{sign}(\phi)\left( |\nabla\Phi|-1\right)=0 \,,\\
\Phi\mid_{\xi=0}=\phi \,,
\end{array}
\right.
\end{equation}
where $\mathrm{sign}(\cdot)$ denotes the signum function. Ideally, equation (\ref{eqn13}) should be solved to steady state in the pseudo-time direction $\xi$. But since the reinitialization is applied to $\phi$ repeatedlly in its iteration, one only needs to solve equation (\ref{eqn13}) for several $\Delta\xi$ steps, e.g. 5 steps, in every reinitialization. The solution $\Phi$ is then used to replace the original $\phi$.

\subsubsection{Penalization on the measure of interface.} \label{sec3.2.2}
The inversion of full-waveform data tends to exhibit instability near the interface, and adding the following regularization term to $E_d$ helps to achieve better results,
\begin{equation}\label{eqn14}
E_\phi:=\frac{1}{2}\int_\Omega |\nabla\phi|^2\mathrm{d}\mathbf{r}\,.
\end{equation}
{\correct The primary goal is to penalize the measure (length or area) of the interface defined by the zero level set $\{\mathbf{r}\mid\phi(\mathbf{r})=0\}$. A standard way to measure this is with the functional: $\mathcal{L}=\int_{\Omega}\delta(\phi)|\nabla\phi|\mathrm{d}\mathbf{r}$. However, directly using $\mathcal{L}$ is problematic because its Fr\'echet derivative has a complicated, nonlinear expression: $\frac{\partial\mathcal{L}}{\partial\phi}\doteq-|\nabla\phi|\nabla\cdot\left(\frac{\nabla\phi}{|\nabla\phi|}\right)$; this expression can lead to numerical instability in iterations. 
Since the level-set reinitialization is applied to $\phi$ (see section \ref{subsubrein}), it maintains the property $|\nabla\phi|\doteq 1$. Under this condition, the complicated derivative simplifies to $\frac{\partial\mathcal{L}}{\partial\phi}\doteq -\Delta\phi$. This is identical to the Fr\'echet derivative of the regularization term $E_\phi$ as shown in equation (\ref{eqn14}). Therefore, we use $E_\phi$ to approximate the effect of $\mathcal{L}$. Penalizing $E_\phi$ tends to shrink the length or area of the interface, and thus prevents the formation of irregularities like burrs and sharp corners, leading to a more stable and regular shape evolution.}

\subsubsection{Regularization on the other parameters.}
Imposing regularization on the parameters $c_1(\mathbf{r})$ and $c_2(\mathbf{r})$ is necessary in their iterations. We suggest the total-variation (TV) regularization with 2-1 norm for $c_1$ and $c_2$,
\begin{equation} \label{eqn15}
E_{c_i}:=\int_\Omega |\nabla c_i| \mathrm{d}\mathbf{r}\,,\quad i=1,2\,.
\end{equation}
{\correct It regularizes $c_i$, $i=1,2$, without over-smoothing them. Compared} to the square of gradient, $\int_\Omega |\nabla c_i|^2 \mathrm{d}\mathbf{r}$, the 2-1 norm TV regularization allows for non-smoothness in $c_i$. Moreover, {\correct because the sharp interface is already represented by the level-set function,} we do not require the 1-1 norm TV regularization for $c_i$, $\int_\Omega \sum_{k=1}^d |\partial_k  c_i| \mathrm{d}\mathbf{r}$, which poses more challenges in optimization.

\subsection{An Adam approach for optimization}
The joint inversion is performed by solving the optimization problem,
\begin{eqnarray}
\mathrm{argmin}_{\phi,c_1,c_2} E_{total}=E_p+\omega\cdot E_g+\lambda_\phi E_\phi+\lambda_{c_1} E_{c_1}+\lambda_{c_2} E_{c_2}\,, \label{eqn16}\\
\mathrm{s.t.\ the\ model\ equations\ (\ref{eqn1}),\, (\ref{eqn3})\ and\ the\ level\ set\ formulation\ (\ref{eqn6})\,.} \nonumber
\end{eqnarray}
$E_{total}$ is a total-energy functional combining the data-fitting and regularization terms, where $\lambda_\phi$, $\lambda_{c_1}$ and $\lambda_{c_2}$ are parameters controlling the amount of regularization.

We propose an Adam algorithm \cite{kinba14} with adjustment coefficients for the optimization problem. Let $\Theta\in\{\phi,c_1,c_2\}$ denote the model parameter to be recovered. It is updated in the following way,
\begin{equation} \label{eqn17}
\Theta_{n+1}=\Theta_n-\alpha_\Theta\frac{\epsilon}{\sqrt{\hat{\mathbf{v}}_n}+\epsilon_0}\hat{\mathbf{m}}_n\,,
\end{equation}
where
\begin{eqnarray}
\hat{\mathbf{m}}_n=\frac{\mathbf{m}_n}{1-\beta_1^n} , \quad \mathbf{m}_n=\beta_1\mathbf{m}_{n-1}+(1-\beta_1)\mathbf{G}_n \,, \label{eqn18} \\
\hat{\mathbf{v}}_n=\frac{\mathbf{v}_n}{1-\beta_2^n} , \quad  \mathbf{v}_n= \beta_2\mathbf{v}_{n-1}+(1-\beta_2)\mathbf{G}_n\odot\mathbf{G}_n \,, \label{eqn19} \\
\mathbf{G}_n:= \frac{\partial E_{total}}{\partial_\Theta}\ \mathrm{at\ iteration}\ n\,. \label{eqn20}
\end{eqnarray}
$\beta_1,\beta_2\in[0,1)$ are decay rates for moment estimates, e.g., $\beta_1=0.9$, $\beta_2=0.999$; $\epsilon$ is a step size, e.g., $\epsilon=10^{-2}$; $\epsilon_0$ is a small constant for stabilization, e.g., $\epsilon_0=10^{-8}$. In equation (\ref{eqn19}), the symbol $\odot$ denotes element-wise multiplications. In the updating formula (\ref{eqn17}), we introduce an adjustment coefficient: $\alpha_\Theta>0$, which can take different values for different model parameters $\Theta$. Heuristically, the coefficient can be viewed as being absorbed into the gradient $\mathbf{G}_n$; since $\alpha_\Theta>0$, the negative gradient is still in the descent direction. In practice, we observe that the Adam approach has significantly better performance than the plain gradient descent approach.

In equations (\ref{eqn18}) and (\ref{eqn19}), the initial moment $\mathbf{m}_0$ and the initial second moment $\mathbf{v}_0$ are simply taken as $\mathbf{0}$, and the key information for Adam is the Fr\'echet derivative  $\mathbf{G}_n= \frac{\partial E_{total}}{\partial_\Theta}$. According to equation (\ref{eqn16}), it holds that,
\begin{eqnarray}
\frac{\partial E_{total}}{\partial\phi}=\frac{\partial E_p}{\partial\phi}+\omega\cdot\frac{\partial E_g}{\partial\phi}+\lambda_\phi\frac{\partial E_\phi}{\partial\phi}\,, 
\label{eqn21}\\
\frac{\partial E_{total}}{\partial c_i}=\frac{\partial E_p}{\partial c_i}+\lambda_{c_i}\frac{\partial E_{c_i}}{\partial c_i}\,,\quad i=1,2 \,. \label{eqn22}
\end{eqnarray}
Firstly, 
\begin{equation} \label{eqn23}
\frac{\partial E_p}{\partial\Theta}=\frac{\partial E_p}{\partial c}\cdot\frac{\partial c}{\partial\Theta}\,,\quad \Theta\in\{\phi,c_1,c_2\}\,,
\end{equation}
where $\frac{\partial c}{\partial\Theta}$ is computed according to equations (\ref{eqn8}) and (\ref{eqn9}), and the computation of $\frac{\partial E_p}{\partial c}$ is more complicated. Considering (\ref{eqn11}), we have
\begin{equation}
\frac{\partial E_p}{\partial c}=\frac{2}{N_sN_rN_t}\sum_{i=1}^{N_s}\sum_{j=1}^{N_r}\sum_{k=1}^{N_t} \left(p_{i,j,k}-p_{i,j,k}^* \right) \frac{\partial p_{i,j,k}}{\partial c}(\mathbf{r})\,.
\end{equation}
The Fr\'echet derivative $\frac{\partial p_{i,j,k}}{\partial c}(\mathbf{r})$ is evaluated under the constraint of the wave equation (\ref{eqn1}). And the perfectly matched layer (PML) boundary condition \cite{joh21} is imposed for (\ref{eqn1}); in particular, spatial derivatives $\frac{\partial}{\partial x_i}$ are replaced by $\frac{\partial}{\partial x_i}+\psi$, where $\psi$ is an operator defined at time step $t$ as $\psi^t=a\psi^{t-1}+b\left(\frac{\partial}{\partial x_i}\right)_t$, and $a$ and $b$ are values that are determined for each grid cell according to its location. Then an adjoint state method \cite{ple06} is employed to compute $ \frac{\partial p_{i,j,k}}{\partial c}(\mathbf{r})$ under the PDE constraint. All these techniques are integrated into Deepwave \cite{richardson_alan_2023}, which is an open-source Python library that implements forward modelling and backpropagation of wave equations in PyTorch. In this work, we utilize the code of Deepwave to compute $\frac{\partial p_{i,j,k}}{\partial c}(\mathbf{r})$ and $\frac{\partial E_p}{\partial c}$ numerically.

Next, we illustrate the calculation of $\frac{\partial E_g}{\partial\phi}$,
\begin{equation}  \label{eqn25}
\frac{\partial E_g}{\partial\phi}=\frac{\partial E_g}{\partial\rho}\cdot \frac{\partial \rho}{\partial\phi}\, ,
\end{equation}
where $\frac{\partial \rho}{\partial\phi}$ is computed according to equations (\ref{eqn8}) and (\ref{eqn9}). Considering equations (\ref{eqn11}) and (\ref{eqn5}), we have that
\begin{equation} \label{eqn24}
\frac{\partial E_g}{\partial\rho}=\frac{2}{N_m} \sum_{j=1}^{N_m} \left(g_{z,j}-g_{z,j}^*\right)\frac{\partial g_{z,j}}{\partial\rho} \nonumber = \frac{2 \gamma}{N_m} \sum_{j=1}^{N_m} \left(g_{z,j}-g_{z,j}^*\right)  K_z(\tilde{\mathbf{r}}_j,\mathbf{r})\, ,
\end{equation}
where $\tilde{\mathbf{r}}_j$ denotes spatial coordinate of the $j$-th measurement. Substituting (\ref{eqn24}) into (\ref{eqn25}), it holds that
\begin{equation}  \label{eqn27}
\frac{\partial E_g}{\partial\phi}=f(\mathbf{r})\cdot\delta_\tau(\phi(\mathbf{r}))\cdot\frac{2\gamma}{N_m} \sum_{j=1}^{N_m} \left(g_{z,j}-g_{z,j}^*\right)  K_z(\tilde{\mathbf{r}}_j,\mathbf{r})\, .
\end{equation}

The derivatives of regularization terms are straightforward. As discussed in \ref{sec3.2.2}, 
\begin{equation}
\frac{\partial E_\phi}{\partial\phi}=-\Delta\phi\,.
\end{equation}
And considering equation (\ref{eqn15}), we have that
\begin{equation}
\frac{\partial E_{c_i}}{\partial c_i}=-\nabla\cdot\left(\frac{\nabla c_i}{|\nabla c_i|} \right)\,;
\end{equation}
numerically, a small constant should be introduced into the denominator to prevent instability: $\frac{\partial E_{c_i}}{\partial c_i}\doteq-\nabla\cdot\left(\frac{\nabla c_i}{\sqrt{|\nabla c_i|^2+\epsilon_0}} \right)$.

\section{A strategy of balanced and decaying weight} \label{sec4}
It is crucial to determine the weighting of each dataset utilized in the multi-physics joint inversion. As shown in equation (\ref{eqn12}), we introduce a parameter $\omega$ to manage the contributions of two data-fitting terms: $E_d=E_p+\omega\cdot E_g$. The weighting parameter $\omega$ is designed in the following way,
\begin{equation} \label{eqn30}
\omega(n)=\omega_1(n)\cdot \omega_2(n)\,,
\end{equation}
where $n$ indicates the $n$-th iteration.

Firstly, the weighting parameter should balance the different scales of $E_p$ and $E_g$. And we introduce $\omega_1(n)$ as follows,
\begin{eqnarray} 
{\correct \omega_1(n)}&{\correct =}&{\correct \frac{\mathrm{average}\left|\left(\frac{\partial E_p}{\partial\phi}(\mathbf{r})\right)_n\right|}{\mathrm{average}\left|\left(\frac{\partial E_g}{\partial\phi}(\mathbf{r})\right)_n\right|}\,,\ \ \mathrm{or}} \label{eqn31} \\
\omega_1(n)&=&\frac{\mathrm{sup}\left|\left(\frac{\partial E_p}{\partial\phi}(\mathbf{r})\right)_n\right|}{\mathrm{sup}\left|\left(\frac{\partial E_g}{\partial\phi}(\mathbf{r})\right)_n\right|}\,.
\end{eqnarray}
{\correct Here, the average or supremum value is taken over $\mathbf{r}$ in the computational domain, and the supremum is a maximum in the discrete computation. The principle behind $\omega_1$ is to make $E_p$ and $E_g$ have a comparable effect on the update of $\phi$, such that $\frac{\partial E_p}{\partial\phi} \sim \omega_1\,\frac{\partial E_g}{\partial\phi}$. This is based on the fact that the level-set function $\phi$ represents the shared interface of $c(\mathbf{r})$ and $\rho(\mathbf{r})$, linking the full-waveform data and gravity data in the joint inversion. We introduce two formulas for $\omega_1$, because both of them are useful. The performance of one may slightly exceed the other in certain situations. In the numerical results of this paper, we uniformly use equation (\ref{eqn31}) for $\omega_1$.}

In addition, we propose a decaying parameter $\omega_2(n)$ in the following way,
\begin{equation}
\omega_2(n)=\omega_0\, e^{-\lambda n}\,,
\end{equation}
where $\omega_0$ and $\lambda$ are positive constants; $\omega_0$ gives the initial setup, and $\lambda$ defines the decaying rate. We observe that the inversions using full-waveform data and gravity data produce different effects. Full-waveform inversion tends to recover parameters along characteristics; for instance, as the sources and receivers are located in the upper side of the domain, the inversion using full-waveform data first resolves shallow-region structures, and then slowly resolves deep structures. On the other hand, gravity inversion quickly generates overall profile of the recovered structure. Therefore, we take $\omega_0>1$ in the parameter $\omega_2(n)$, so that gravity data are dominant in the initial iterations, and the inversion can rapidly recover overall structures in the whole domain. The other term $e^{-\lambda n}$ in $\omega_2(n)$ causes it to decay during iterations, allowing $E_p$ with the full-waveform data to dominate in later stages, given that the full-waveform data provide higher resolution in the inversion.
{\correct In short, $\omega_2(n)$ enables gravity data to dominate at early iterations and full-waveform data to dominate in later stages. The goal of $\omega_2(n)$ is different from that of relaxed penalty methods  \cite{golosh09, vanher13, essguavanetc18}, where the successively varying penalty parameter aims to enforce a constraint.}

We mention that the strategy of balanced and decaying weight is crucial for the success of  joint inversion. {\correct The complete joint inversion algorithm is summarized as follows.}

%%%%%%%%%%%%%%%%%%%
\pic\\
\noindent { \textsf{ Algorithm 1. Joint inversion of full-waveform and gravity data.
\begin{algorithmic}[1]
\State Initialize $\phi$, $c_1$ and $c_2$; freeze the density-contrast value $f$ based on prior information.
\State Construct $c(\mathbf{r})$ and $\rho(\mathbf{r})$ according to equation (\ref{eqn6}).
\State Compute $p(\mathbf{r},t)$ by solving equation (\ref{eqn1}) with the PML boundary condition; compute $g_z$ according to equation (\ref{eqn5}).
\State Evaluate $\frac{\partial E_p}{\partial\Theta}$ ($\Theta\in\{\phi,c_1,c_2\}$) according to (\ref{eqn23}); evaluate $\frac{\partial E_g}{\partial\phi}$ according to (\ref{eqn27}).
\State Determine the balanced and decaying weighting term $\omega(n)$ by equation (\ref{eqn30}).
\State Evaluate the Fr\'echet derivatives $\frac{\partial E_{total}}{\partial\Theta}$ ($\Theta\in\{\phi,c_1,c_2\}$) according to (\ref{eqn21}) and (\ref{eqn22}).
\State Update the model parameters $\Theta\in\{\phi,c_1,c_2\}$ according to equation (\ref{eqn17}).
\State Perform level-set reinitialization for $\phi$ by solving equation (\ref{eqn13}), and use $\Phi$ to replace $\phi$.
\State Go back to step 2 until the stopping criterion is achieved: $E_{total}<\epsilon_{stop}$, or the number of iterations exceeds the specified value  $n_{\mathrm{max}}$.
\end{algorithmic}
} } \pic
%%%%%%%%%%%%%%%%%%%

\section{Numerical computation} \label{sec5}
\subsection{\correct Computational setup and units}
%{Units of physical quantities and some other settings}
The multi-physics joint inversion involves different types of data and quantities. Table \ref{Tab1} lists the units of the physical quantities used in the computation.

\begin{table}[H]
		\centering
		\begin{tabular}{cccccc} 
			\toprule
			$\mathbf{r}$      &  $t$  &  $c(\mathbf{r})$   &  $\rho(\mathbf{r})$    &  $p(\mathbf{r},t)$  & $g_z$ \\
			\midrule
			kilometer (km)       & second (s) & km/s  & g/cm$^3$  &  pascal (Pa)  &  milliGal (mGal)                  \\
			\bottomrule
		\end{tabular}\caption{Units of physical quantities used in the computation.}
		\label{Tab1}
\end{table}
\vspace{-10pt}
 In equation (\ref{eqn1}), the source term $s(\mathbf{r},t)$ is then measured in Pa/km$^2$. And we multiply it by $10^6$ to simulate the real situation, i.e., the Ricker wavelet in $s(\mathbf{r},t)$ is in the form of 
\begin{equation} \label{eqn33}
\correct s_0(t-t_0)=10^6\cdot \left(1 - 2\pi^2 f_0^2 (t-t_0)^2 \right) \cdot e^{-\pi^2 f_0^2 (t-t_0)^2}.
\end{equation}
{\correct{In numerical computation, we choose the peak frequency $f_0=5$\,Hz, and $t_0=\frac{1}{f_0}=0.2$\,s. Since seismic data typically lack frequencies below $3$\,Hz, we apply a high-pass filter with a $3$\,Hz cutoff to $s_0(t-t_0)$ to simulate this reality. Figure \ref{Fig_RickerWavelet} plots the Ricker wavelet with and without the low-frequency cutoff, where the black solid line shows the filtered wavelet that we use in numerical computation.}}

In equation (\ref{eqn5}), the constant $\gamma$ is $\gamma=2^{d-1}\pi\cdot\gamma_0$, where $\gamma_0$ denotes the universal gravitational constant: $\gamma_0=6.67384\times 10^{-8}\mathrm{cm}^3\mathrm{g}^{-1}\mathrm{s}^{-2}$. When $\mathbf{r}$ is in km and $\rho(\mathbf{r})$ is in g/cm$^3$, the value of $\gamma_0$ can be simply taken as $6.67384$, and the obtained $g_z$ is naturally in mGal.
 
To demonstrate the effect of joint inversion, we will compare it with the inversion using full-waveform data only. As a benchmark, the full-waveform inversion is performed using the state-of-the-art Deepwave package \cite{richardson_alan_2023}, where we employ the 1-1 norm TV regularization for $c(\mathbf{r})$, $\int_\Omega \sum_{k=1}^d |\partial_k  c| \mathrm{d}\mathbf{r}$, to enhance its sharp interface without the level-set formulation.

\subsection{Results}
\subsubsection{Example 1.}
\begin{figure}[htbp!] % !ht 表示尽量放在当前位置或页顶
%    \captionsetup{font={color=red}} %\correct
    \centering
    \includegraphics[width=0.50\textwidth]{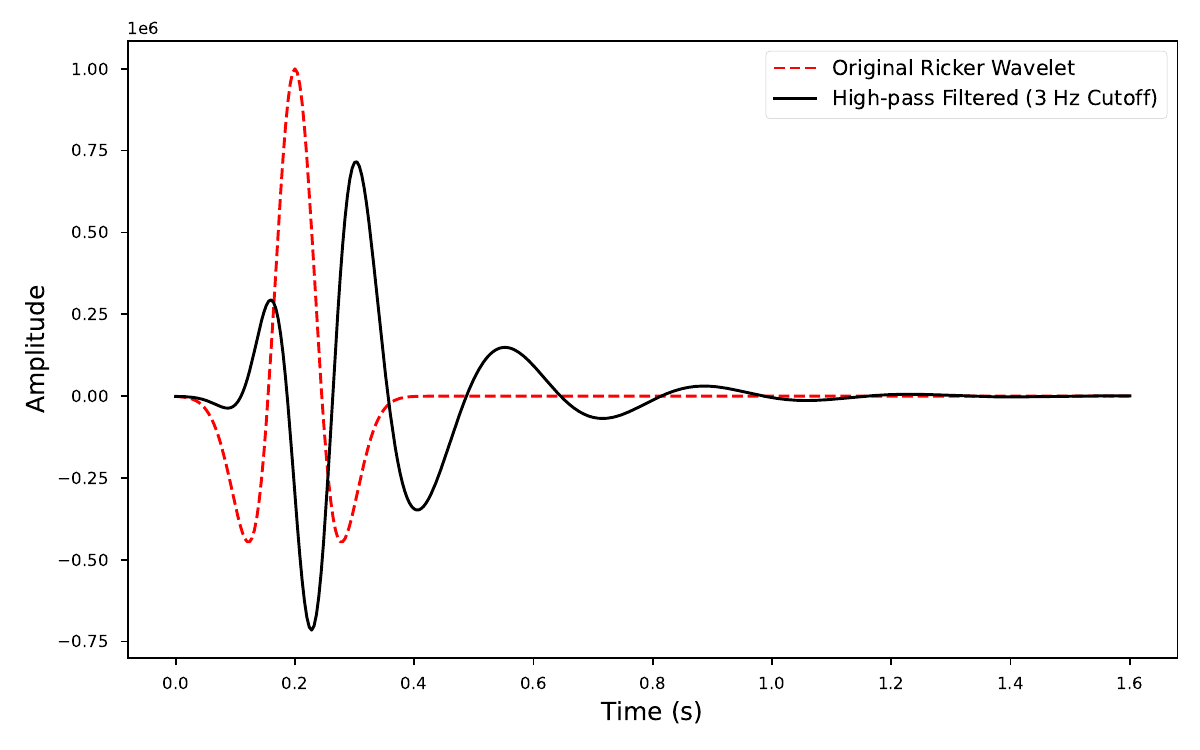} % 控制宽度为正文宽度的60%
    \caption{Ricker wavelet with and without the low-frequency cutoff. The red dashed line plots the original Ricker wavelet as shown in equation (\ref{eqn33}) with $f_0=5$\,Hz and $t_0=0.2$\,s; the black solid line plots the high-pass filtered wavelet with a $3$\,Hz cutoff, which we use in numerical computation.}
    \label{Fig_RickerWavelet}
\end{figure}

\begin{figure}[htbp!]
\centering
(a){{\includegraphics[scale=0.24,angle=0]{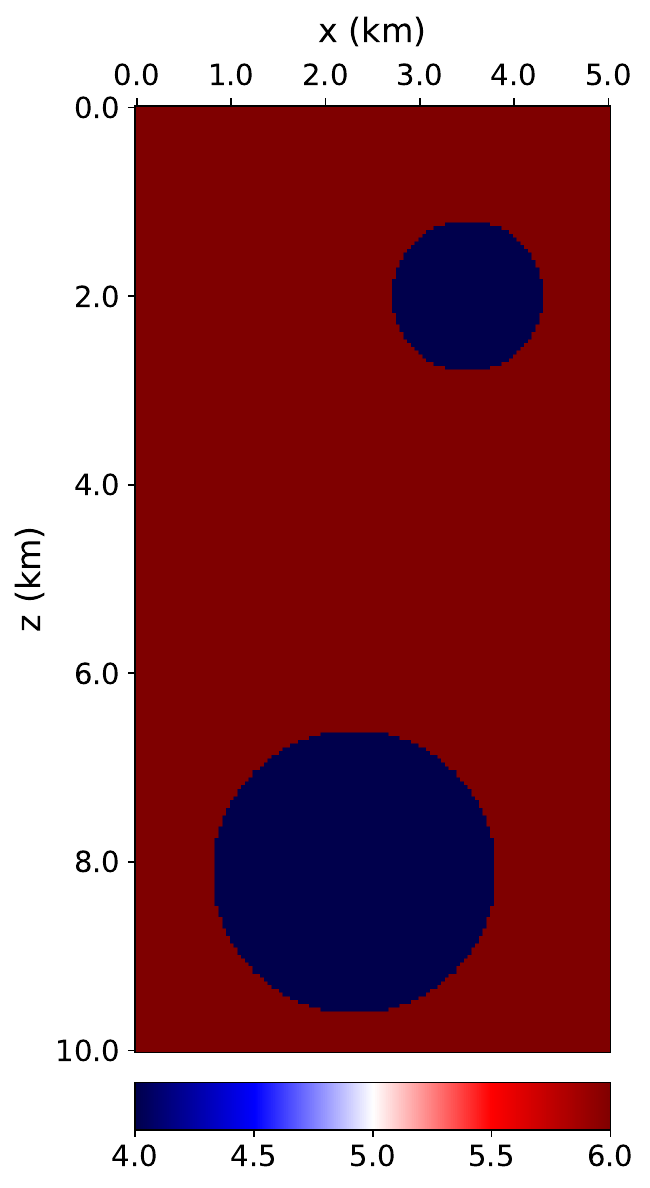}}}
(b){{\includegraphics[scale=0.24,angle=0]{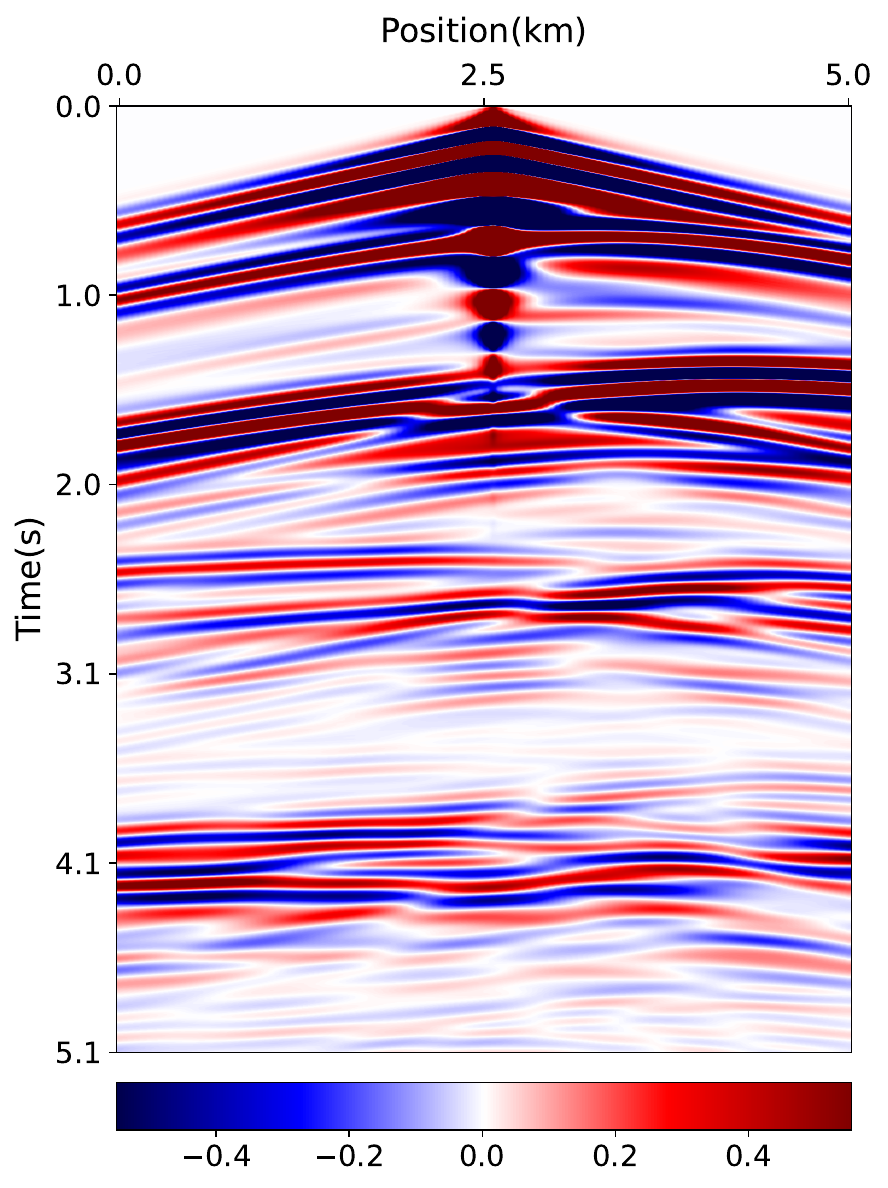}}}
(c){{\includegraphics[scale=0.24,angle=0]{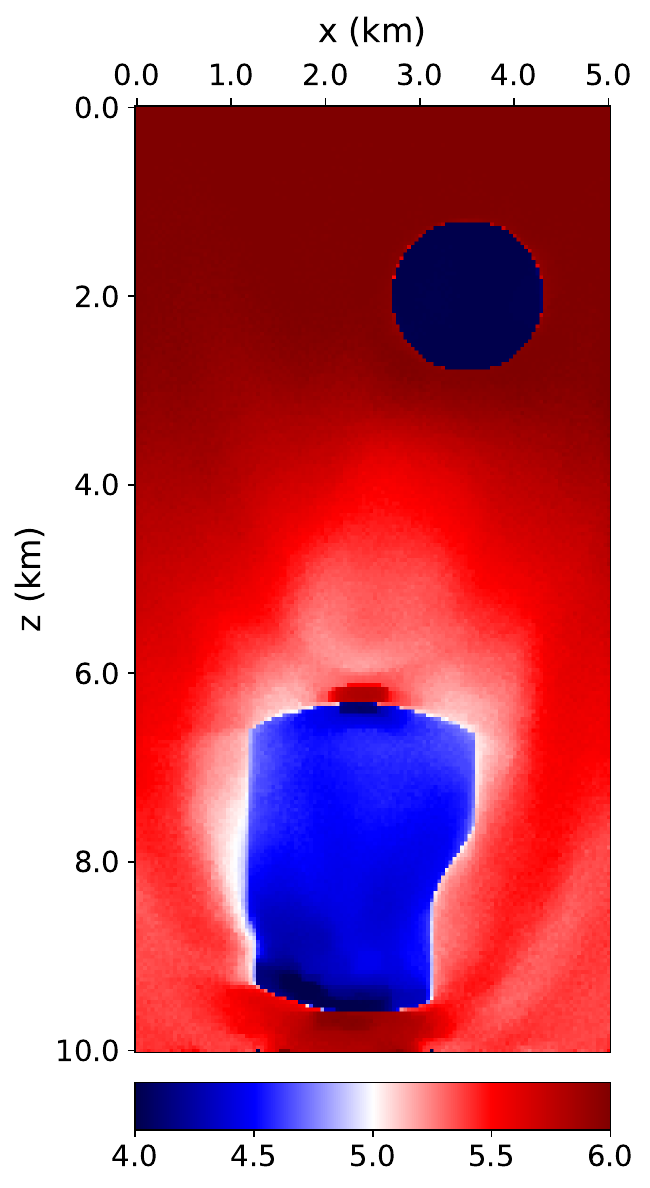}}}
{\correct (d){{\includegraphics[scale=0.4,angle=0]{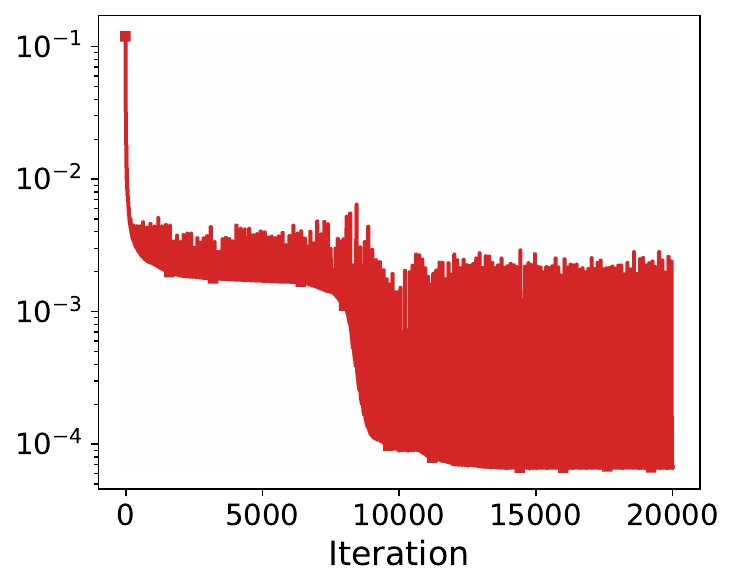}}}}
\caption{Example 1: velocity model and full-waveform inversion result. {\correct The initial velocity guess is shown in Figure \ref{Fig4}\,(a).} (a) True velocity; (b) waveform data for the 16\,th source of 30, where the color scale is clipped between its $5\%$ and $95\%$ quantiles for enhanced visualization; (c) recovered solution of FWI after 20,000 iterations; {\correct (d) data misfit in the FWI.}}
\label{Fig2}
\end{figure}

We illustrate the scenario that gravity data can complement full-waveform data in the joint inversion. 
{\correct Figure \ref{Fig2}\,(a) shows the synthetic velocity model. The computational domain is $\Omega=[0,5]\times[0,10]$\,km, where we denote the 2D spatial coordinate as $\mathbf{r}=(x,z)$.} The point sources and receivers are located along $z=0$\,km; there are 30 sources with $x$-coordinates $x=0:0.172:5$\,km, and 124 receivers with $x$-coordinates $x=0:0.04:5$\,km. {\correct The source wavelet is the high-pass Ricker wavelet, as shown in Figure \ref{Fig_RickerWavelet} by the black solid line.}
{\correct To simulate the waveform data, we solve equation (\ref{eqn1}) using a time step size of $\Delta t=0.003$\,s and a spatial mesh size of $h=0.04$\,km. This combination satisfies the Courant-Friedrichs-Lewy (CFL) condition for the 2D wave equation, which requires $\Delta t \le \frac{1}{c_{\mathrm{max}}}\frac{h}{\sqrt{2}}$.}
The total recording time is $5.1$\,s, with a sampling interval of $0.003$\,s. {\correct Figure \ref{Fig2}\,(b) illustrates the waveform data for the 16\,th source of 30;} since the waveform data can contain extreme outliers, the color scale of the waveform data is clipped between its $5\%$ and $95\%$ quantiles for enhanced visualization.

Figure \ref{Fig2}\,(c) plots the recovered solution of full-waveform inversion, {\correct using the initial velocity guess shown in Figure \ref{Fig4}\,(a)}.
{\correct Figure \ref{Fig2}\,(d) illustrates the convergence history of the full-waveform inversion. We perform an excessive number of iterations to ensure the solution reaches definitive convergence, thereby providing the best possible benchmark against which to measure the improvement of later joint inversion.
It shows that the FWI performs excellently in inverting the velocity model with shallow structure, but yields a poor solution for the deep structure. In fact, this synthetic model is designed with a lower velocity in the circular region, making the deep structure undetectable in surface data.}

% \begin{figure}[htbp!]
% \centering
% (a){{\includegraphics[scale=0.24,angle=0]{figure/EXP11_vtrue.pdf}}}
% (b){{\includegraphics[scale=0.24,angle=0]{figure/EX1_1b.pdf}}}
% (c){{\includegraphics[scale=0.24,angle=0]{figure/EX1_1c.pdf}}}
% \caption{\bf{Delete??}\correct{Example 1: velocity model with shallow structure. Full-waveform inversion results. (a) True model; (b) waveform data for the 16\,th source of 30, where the color scale is clipped between its $5\%$ and $95\%$ quantiles for enhanced visualization; (c) recovered solution of FWI after 20,000 iterations.}}
% \label{Fig1}
% \end{figure}

\begin{figure}[htbp!]
\centering
(a){{\includegraphics[scale=0.24,angle=0]{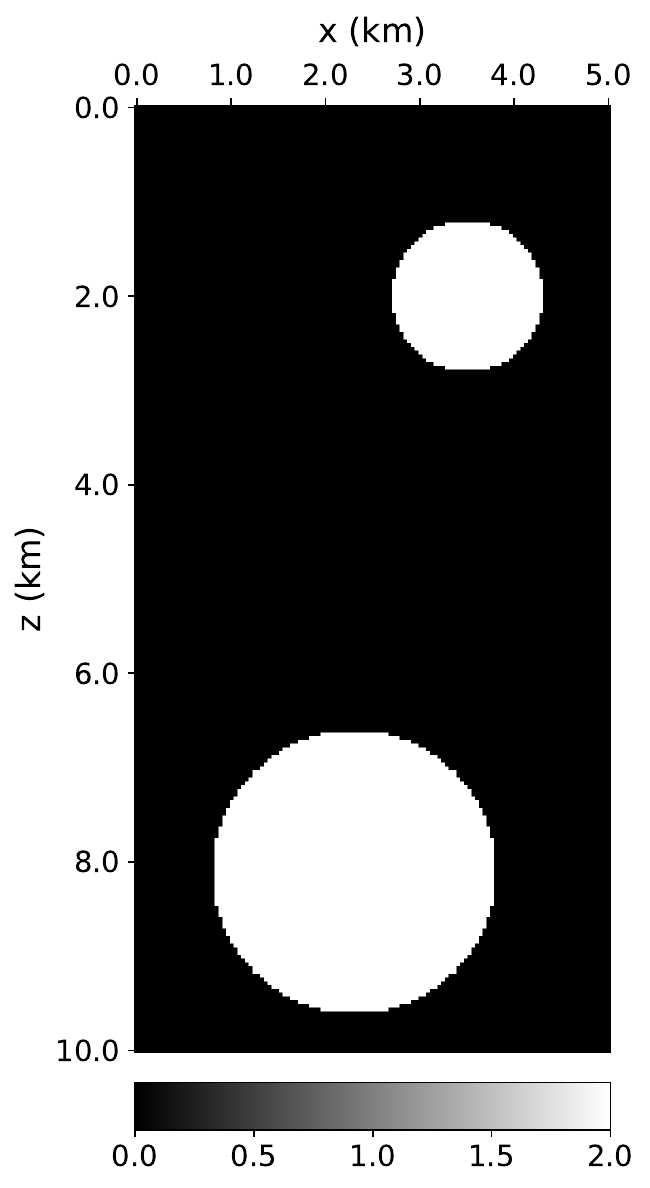}}}
(b){{\includegraphics[scale=0.24,angle=0]{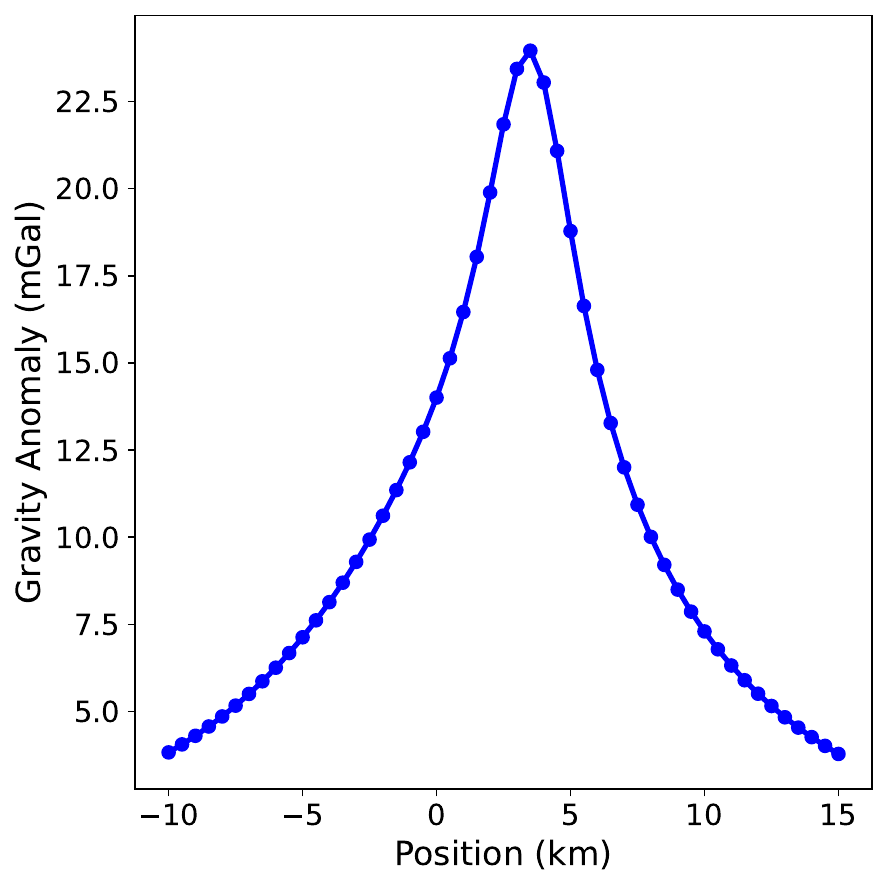}}}
{\correct (c){{\includegraphics[scale=0.24,angle=0]{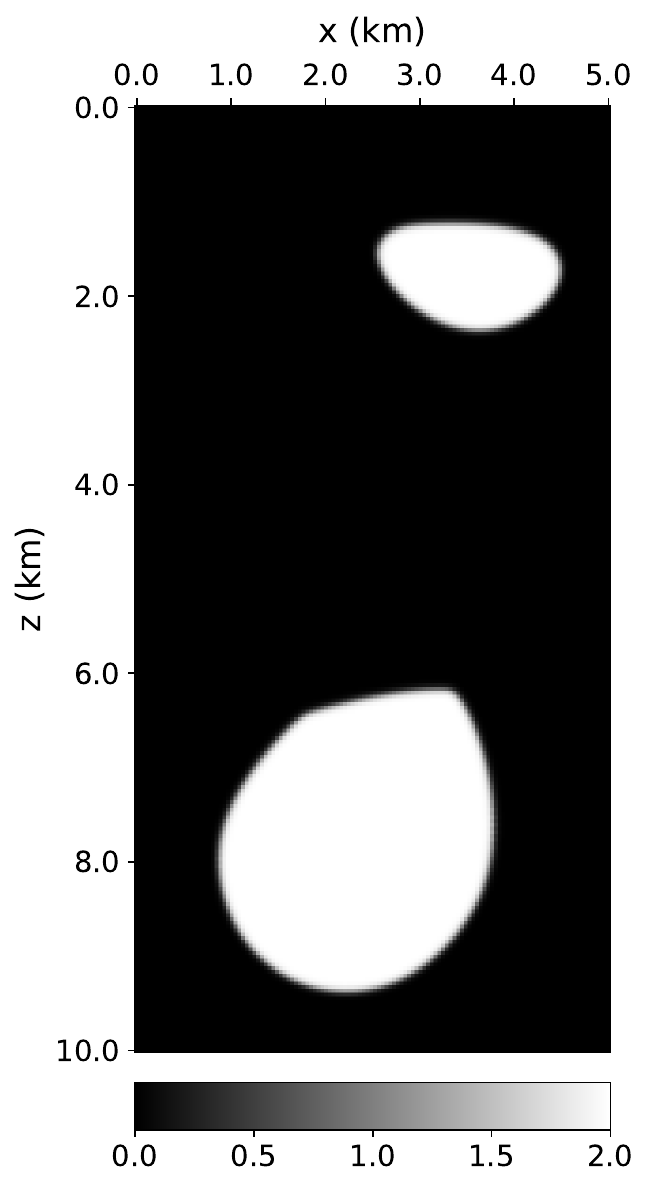}}}}
\caption{Example 1: density model and gravity inversion result. {\correct The initial density guess is shown in Figure \ref{Fig4}\,(b).} (a) True density; (b) {\correct gravity data $g_z$}; {\correct (c) recovered solution of gravity inversion.}}
\label{Fig3}
\end{figure}

Then we integrate gravity for joint inversion. Figure \ref{Fig3} shows the density model and the gravity data $g_z$. The data are measured along $z=-0.1$\,km, and there are 51 measurements with $x$-coordinates $x=-10:0.5:15$\,km. {\correct Figure \ref{Fig3}\,(c) shows the recovered solution of gravity inversion, where we use an initial density guess from Figure \ref{Fig4}\,(b), and impose the density-contrast value $f(\mathbf{r})=2.0$\,g/cm$^3$ as a priori information. Although the circular shapes appear imperfect in the density solution, it successfully recovers the deep structure. This illustrates that by appropriately using gravity data and extracting unambiguous information, the gravity approach can recover deep structures that are undetectable by waveform data.}

\begin{figure}[htbp!] % !ht 表示尽量放在当前位置或页顶
%    \captionsetup{font={color=red}}%\correct 
    \centering
    \includegraphics[width=0.40\textwidth]{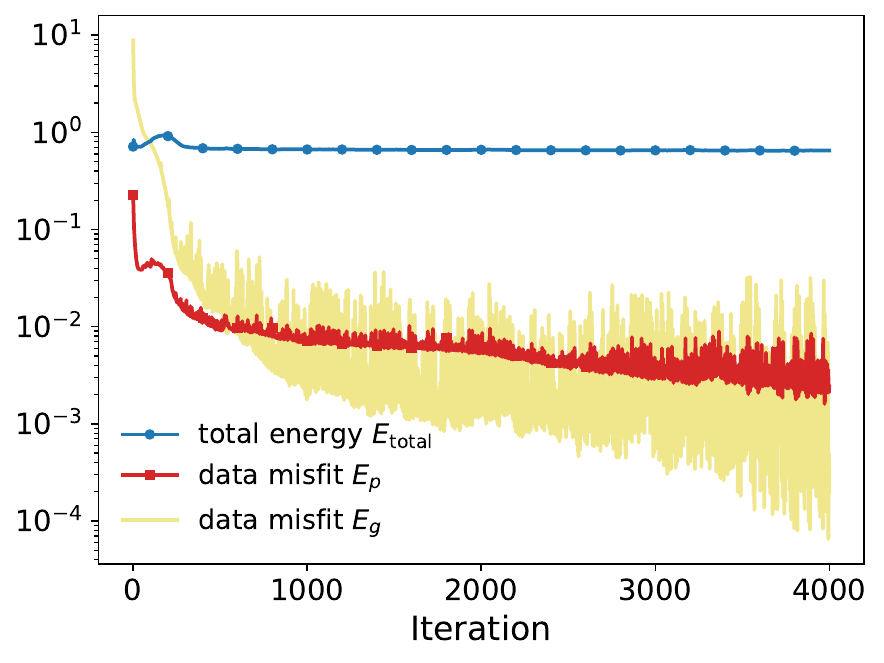} % 控制宽度为正文宽度的60%
    \caption{Example 1: Convergence history of the joint inversion. The blue line plots the total energy $E_{\mathrm{total}}$, the red line plots the seismic data misfit $E_p$, and the yellow line plots the gravity data misfit $E_g$.}
    \label{Fig_mismatch}
\end{figure}

\begin{figure}[htbp!]
\centering
(a){{\includegraphics[scale=0.24,angle=0]{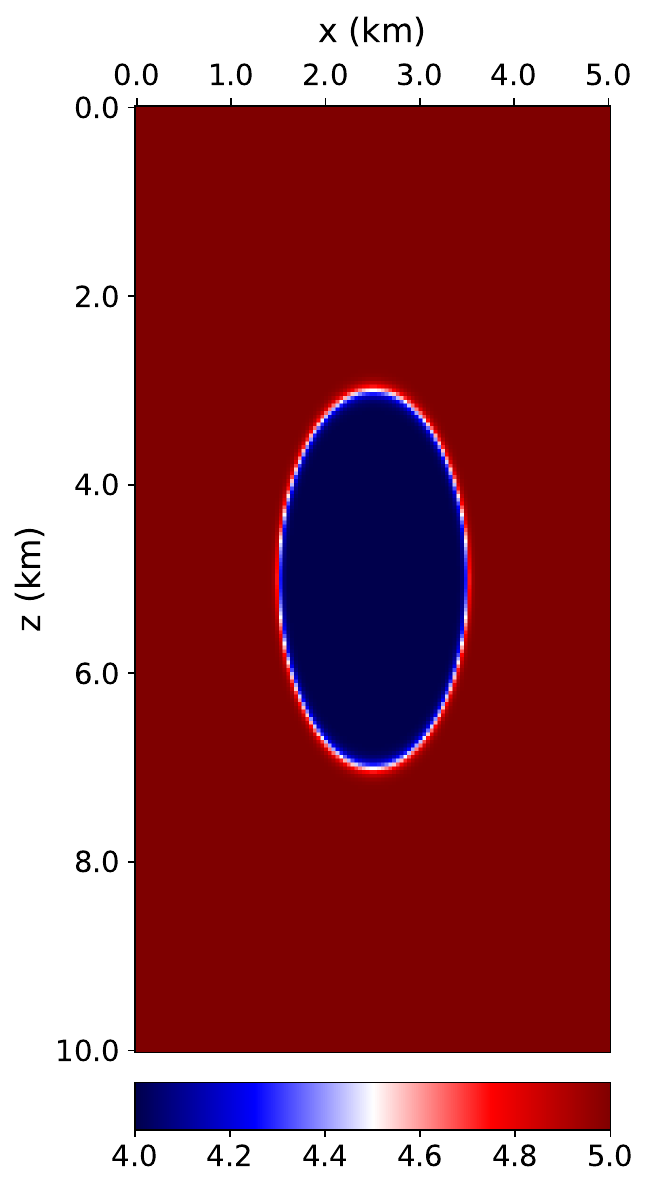}}}
(b){{\includegraphics[scale=0.24,angle=0]{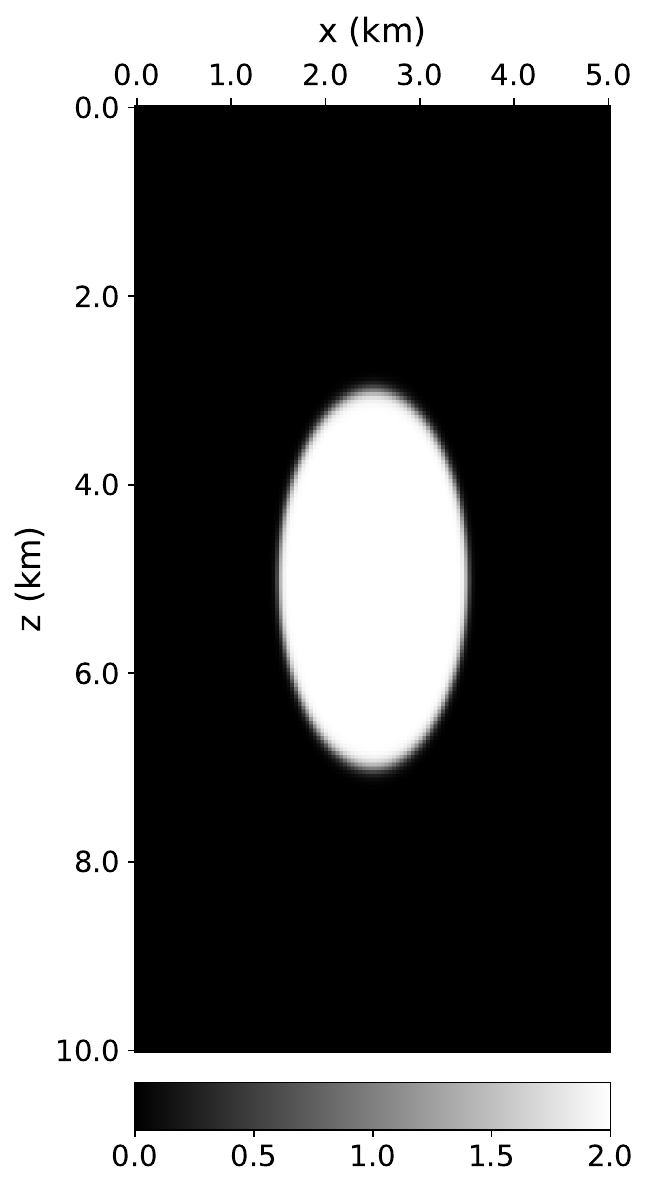}}}
(c){{\includegraphics[scale=0.24,angle=0]{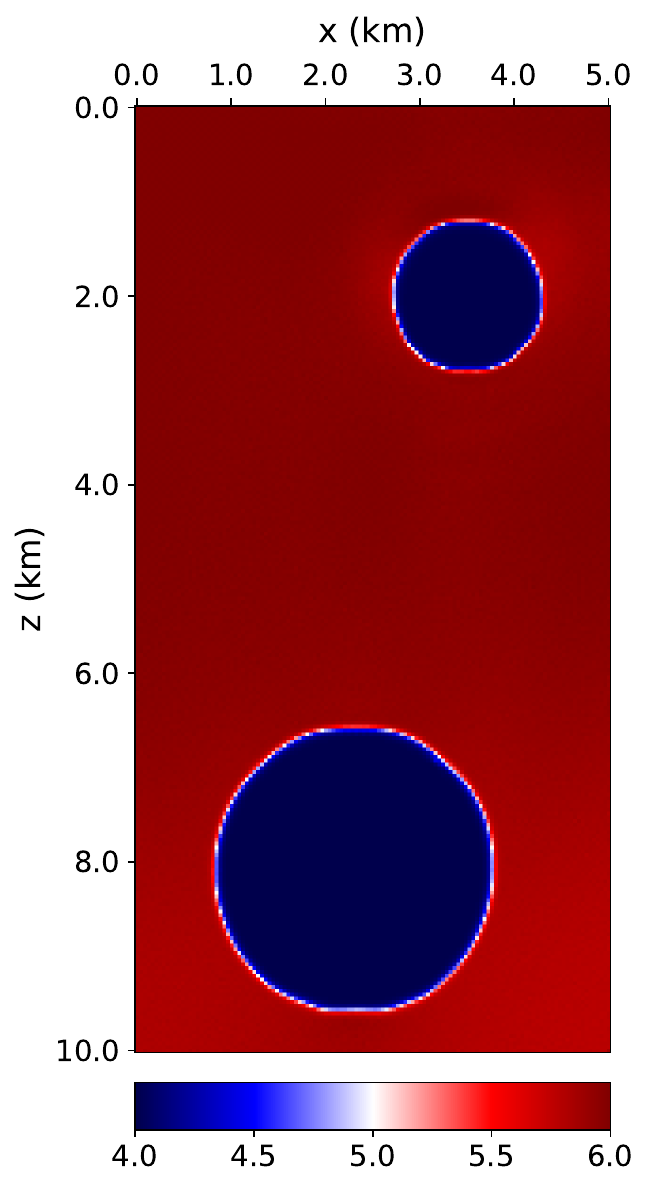}}}
(d){{\includegraphics[scale=0.24,angle=0]{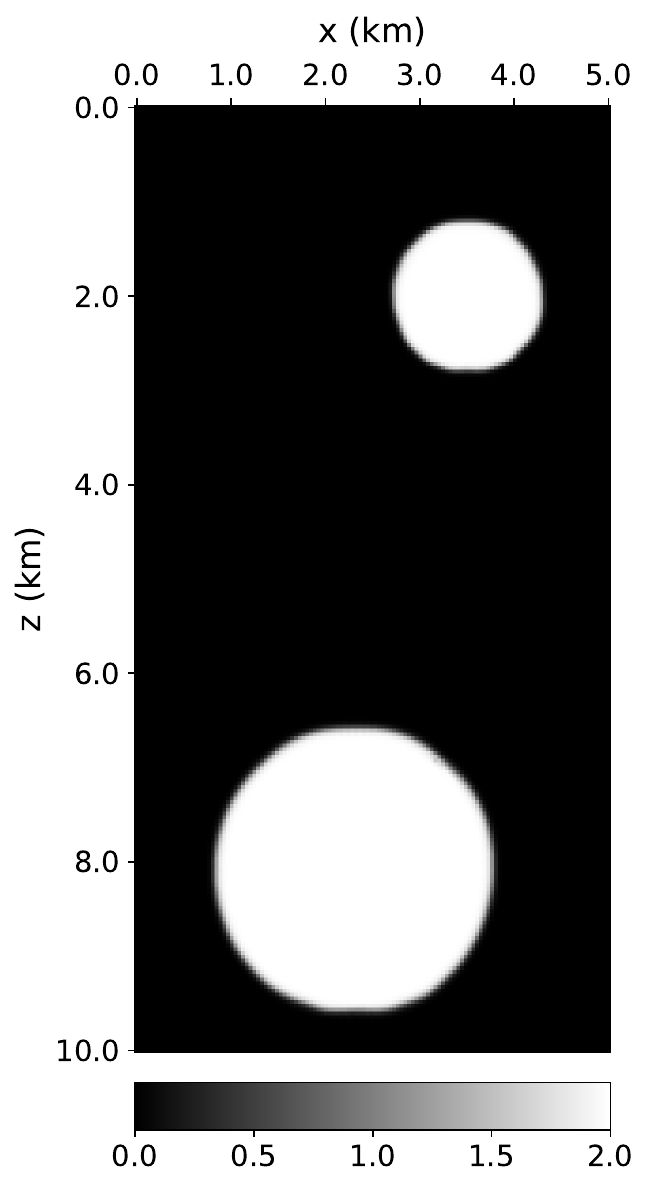}}}
\caption{Example 1: joint inversion result. (a) Initial model for velocity; (b) initial model for density; (c) recovered velocity after 4000 iterations; (d) recovered density after 4000 iterations.}
\label{Fig4}
\end{figure}

In the joint inversion, the density-contrast value $f(\mathbf{r})=2.0$\,g/cm$^3$ is imposed as a priori information; for simplicity, we further freeze the velocity value within the circles, $c_1(\mathbf{r})=4.0$\,km/s. The level-set formulation facilitates the imposition of the prior information. Table \ref{Tab2} lists the values of algorithm parameters used in the joint inversion.
{\correct Figure \ref{Fig_mismatch} shows the convergence history of the joint inversion; since we introduced a balanced and decaying weight $\omega$ in the total energy, the convergence is indicated by the data misfits $E_p$ and $E_g$.}
Figure \ref{Fig4} provides the results, where \ref{Fig4}\,(a) and \ref{Fig4}\,(b) illustrate the initial models for the joint inversion, and \ref{Fig4}\,(c) and \ref{Fig4}\,(d) plot the recovered solutions. We conclude that with a very general initial guess, the level-set joint inversion provides almost accurate reconstructions for the velocity and density models. {\correct Compared} to the FWI solution as shown in Figure \ref{Fig2}, the joint inversion integrating gravity helps to improve the imaging results.

\begin{table}[h!]
		\centering
		\begin{tabular}{ccccccc} 
			\toprule
			$\epsilon$      &  $\alpha_\phi$  &  $\alpha_{v_2}$   &  $\omega_0$    &  $\lambda$  & $\lambda_\phi$ &  $\lambda_{c_2}$\\
			\midrule
			0.03       & 1 & 3  & 5  &  $\displaystyle \frac{\mathrm{ln}5}{n_{\mathrm{max}}}$ ($n_{\mathrm{max}}=2\times10^4$)  &  $5\times10^{-5}$  & $10^{-4}$               \\
			\bottomrule
		\end{tabular}\caption{Example 1: values of algorithm parameters used in the joint inversion.}
		\label{Tab2}
\end{table}
\vspace{-8pt}

\subsubsection{Example 2.}
\begin{figure}[htbp!]
\centering
(a){{\includegraphics[scale=0.24,angle=0]{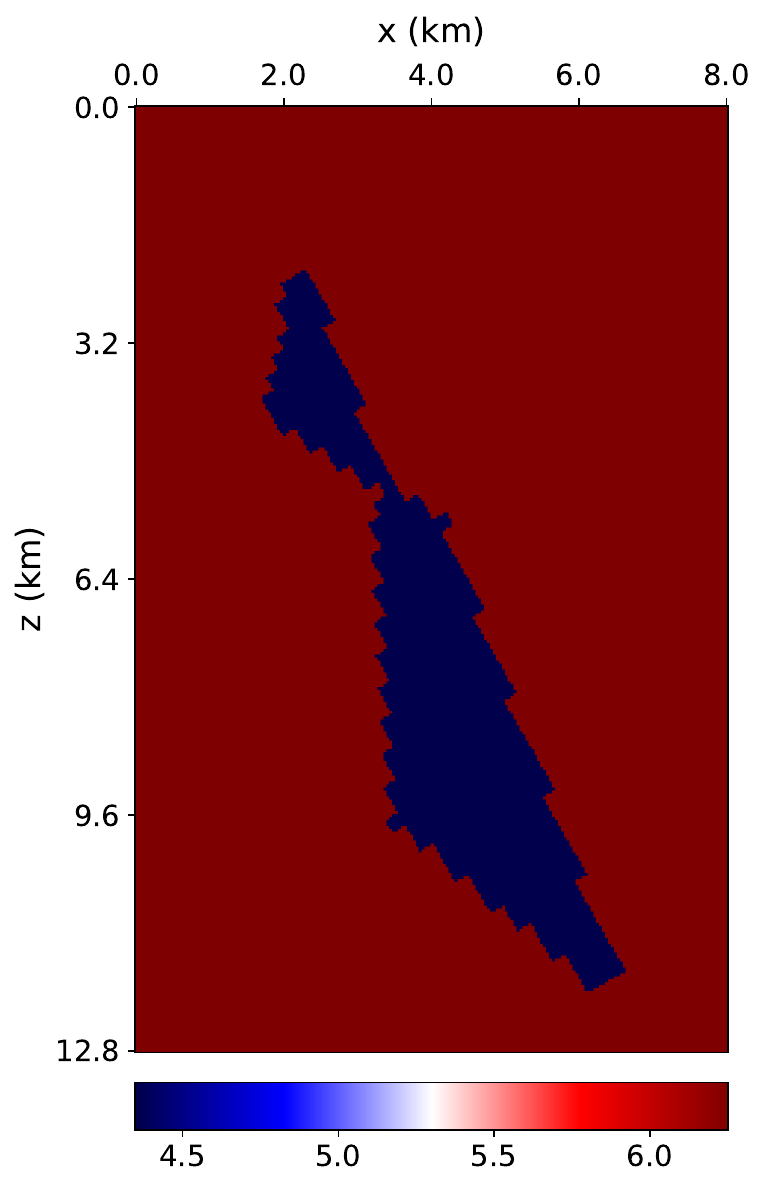}}}
(c){{\includegraphics[scale=0.24,angle=0]{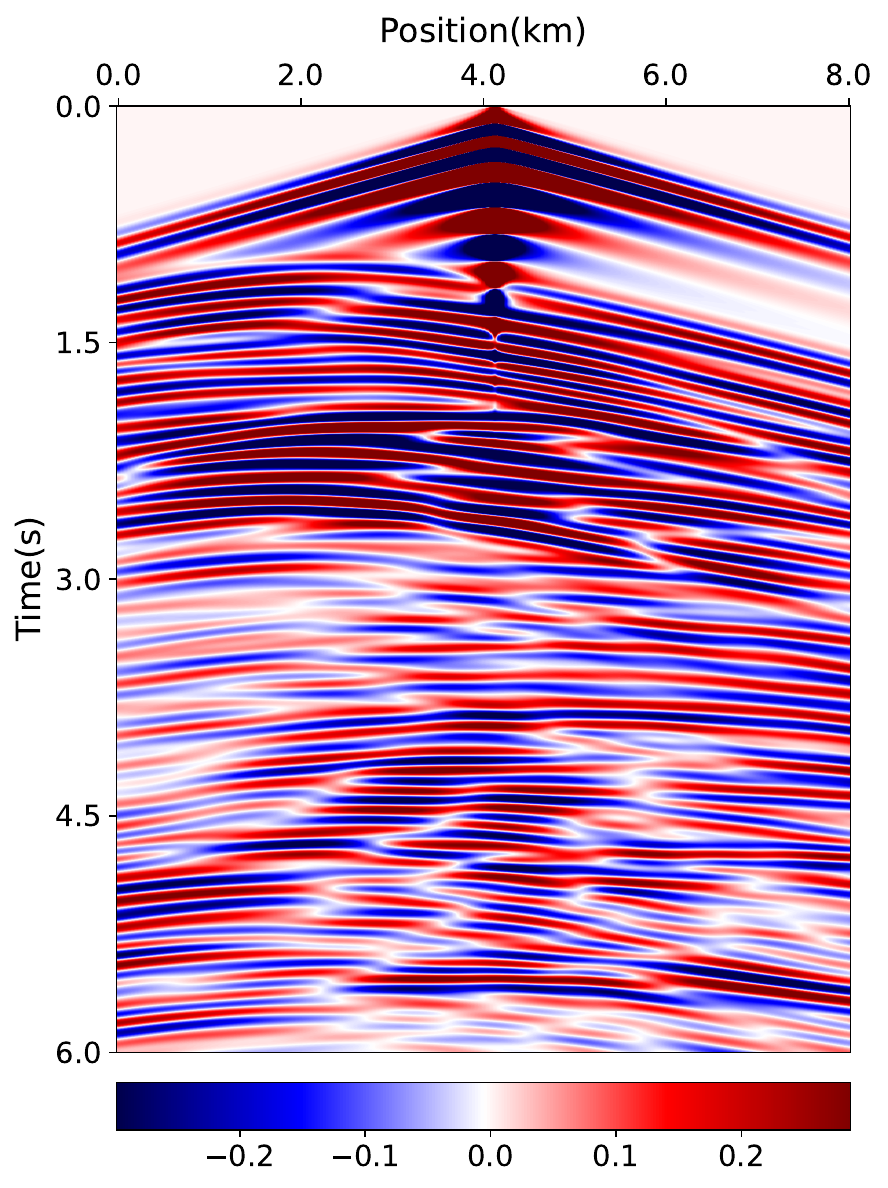}}}
(e){{\includegraphics[scale=0.24,angle=0]{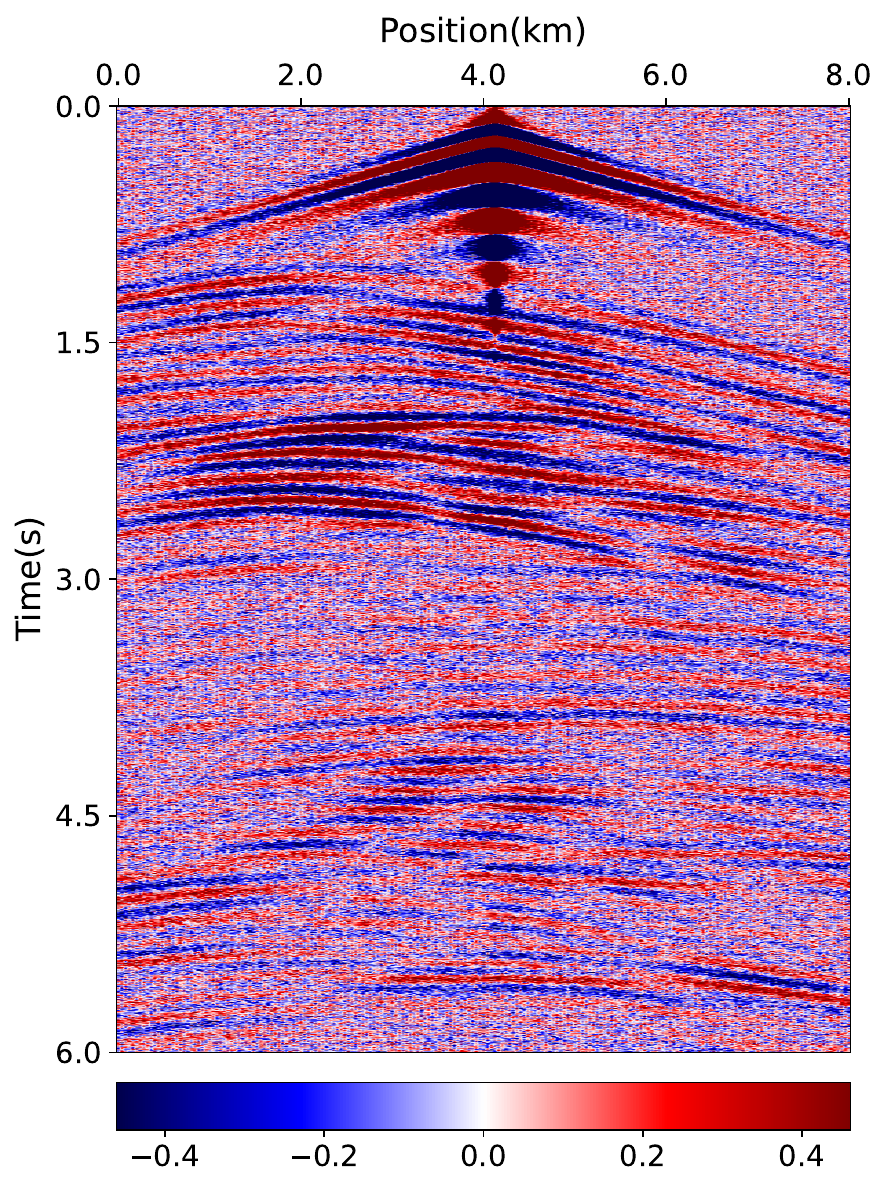}}}
{\correct (g){{\includegraphics[scale=0.24,angle=0]{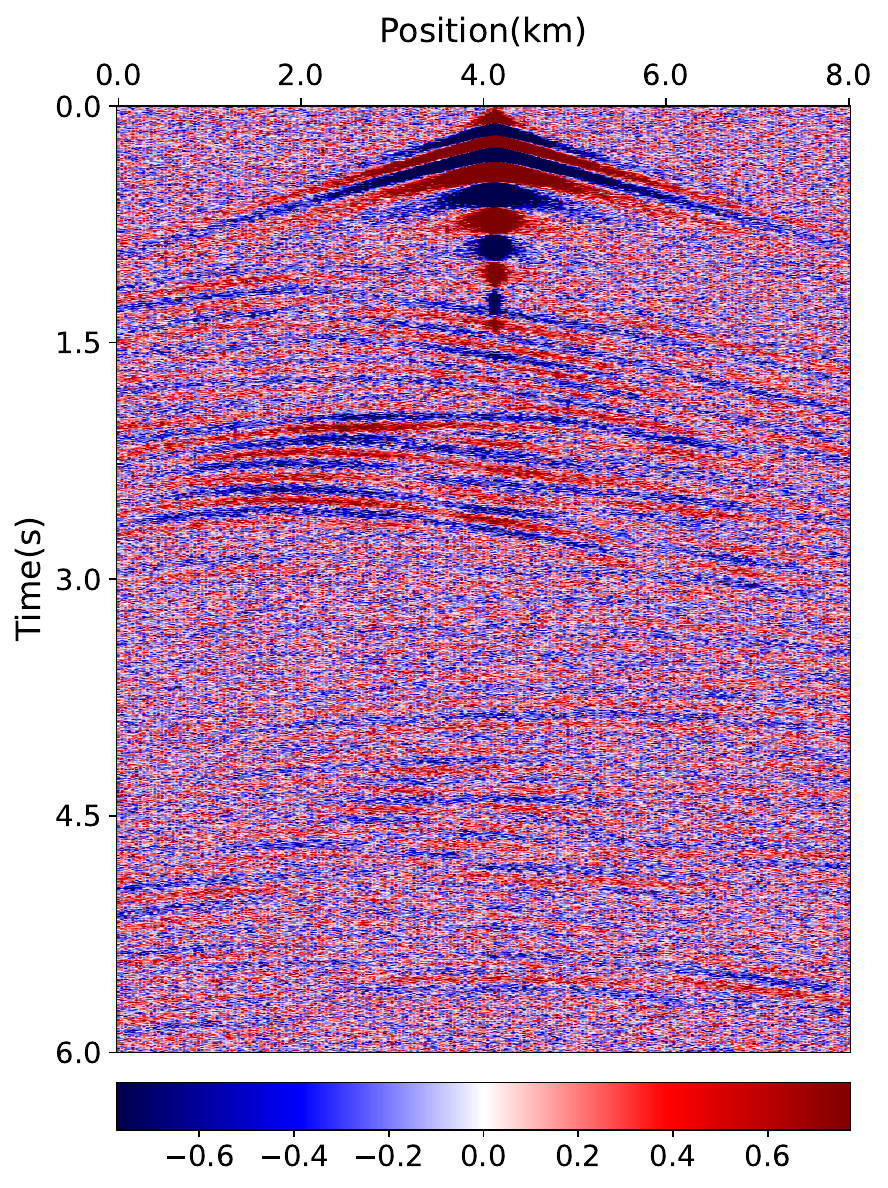}}}} \\
(b){{\includegraphics[scale=0.24,angle=0]{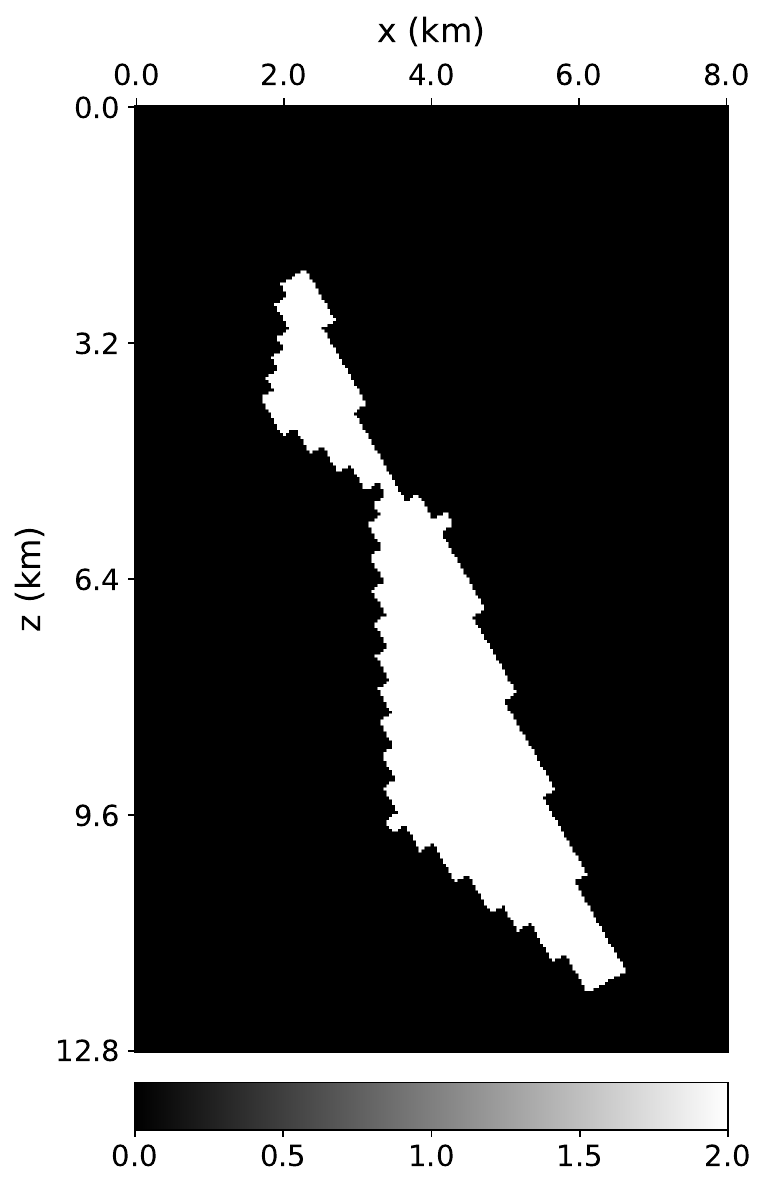}}}
(d){{\includegraphics[scale=0.24,angle=0]{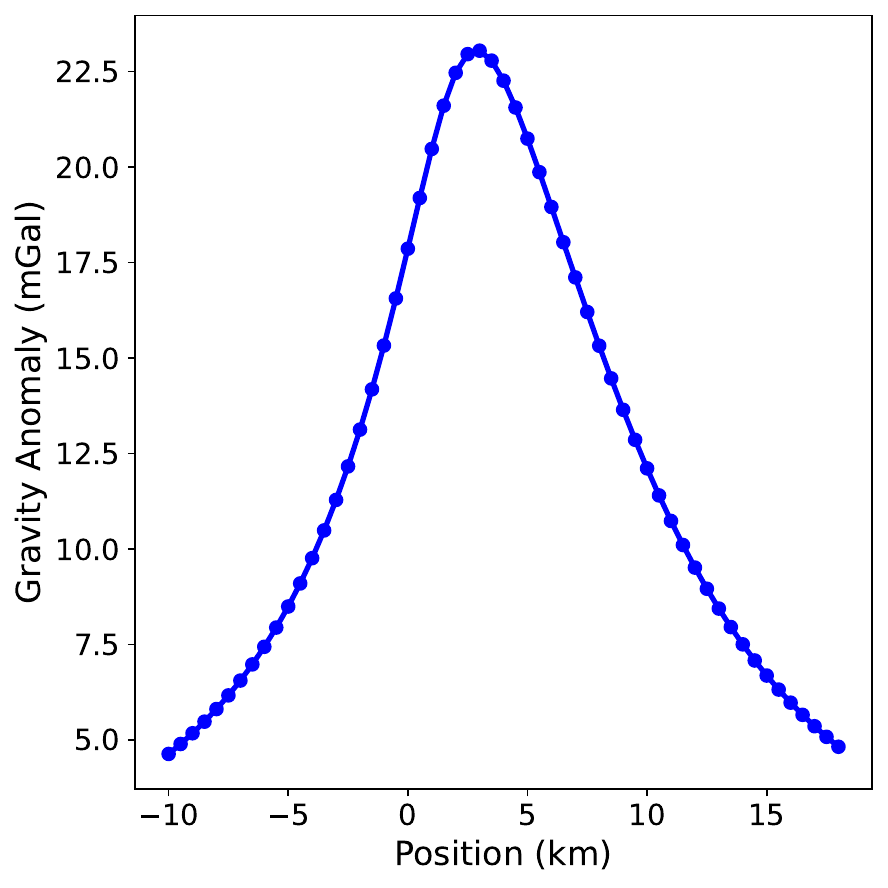}}}
(f){{\includegraphics[scale=0.24,angle=0]{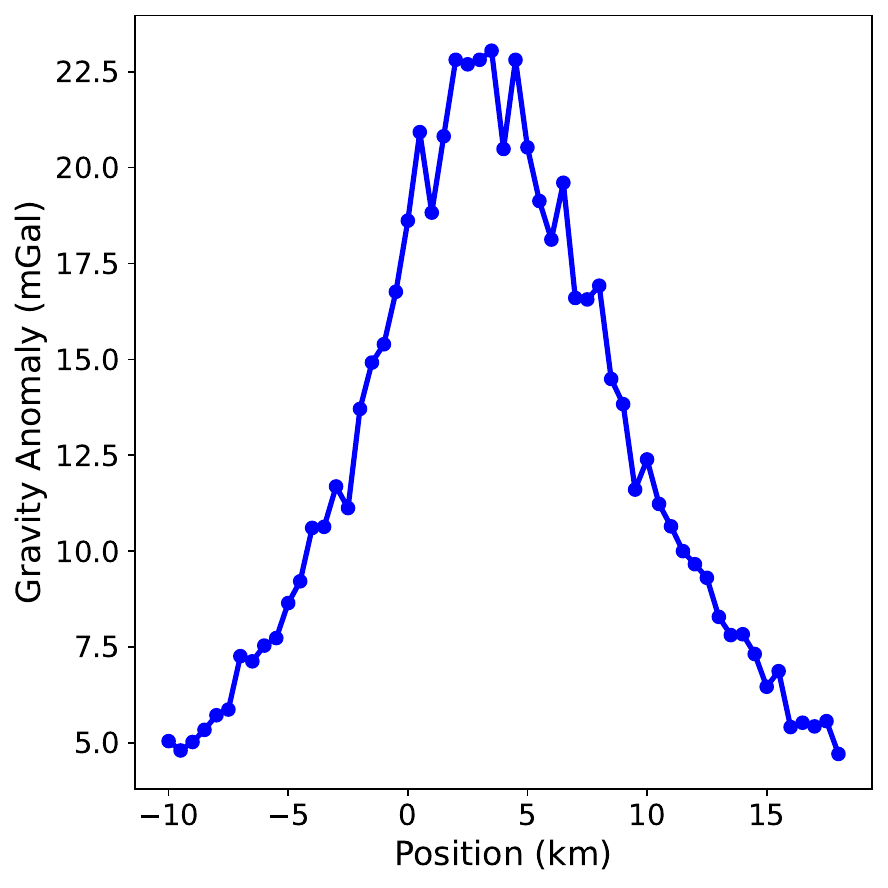}}}
{\correct (h){{\includegraphics[scale=0.24,angle=0]{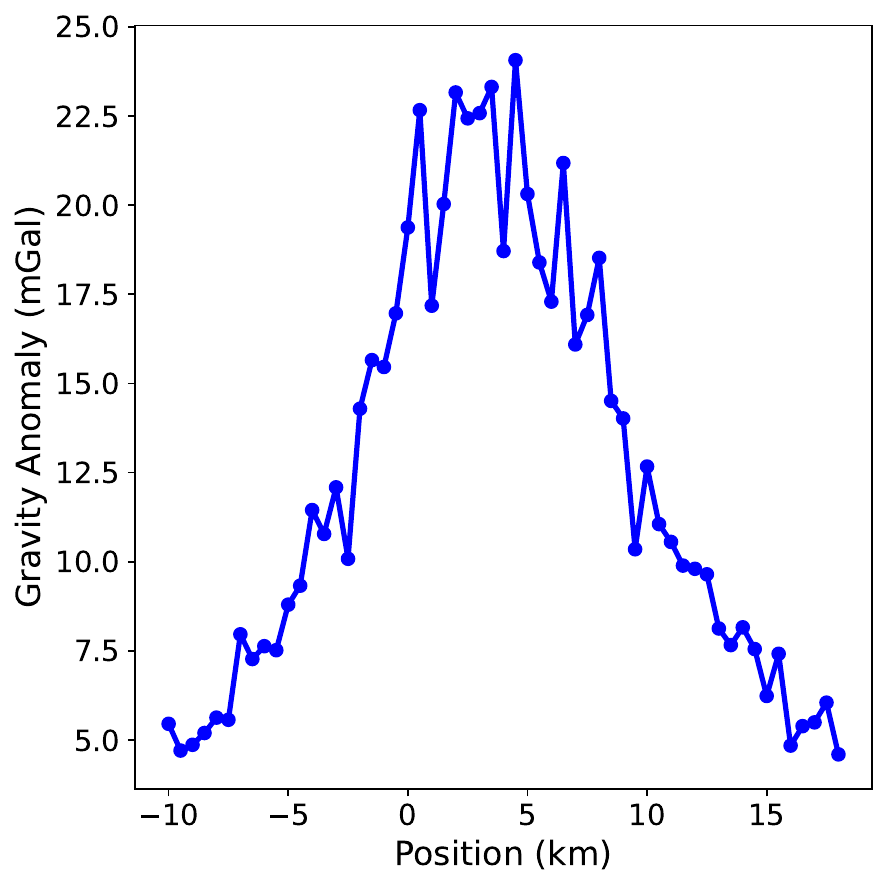}}}}
\caption{Example 2: a dipping structure; models and data. The color scale of waveform data is clipped between its $5\%$ and $95\%$ quantiles for enhanced visualization. (a) True velocity model; (b) true density model; (c) waveform data for the 16\,th source of 30; (d) gravity data $g_z$; (e) waveform data with $5\%$ Gaussian noise; (f) gravity data with $5\%$ Gaussian noise; {\correct (g) waveform data with $10\%$ Gaussian noise; (h) gravity data with $10\%$ Gaussian noise.}}
\label{Fig5}
\end{figure}

We consider a dipping structure as shown in Figure \ref{Fig5}; Figure \ref{Fig5}\,(a) shows the velocity model, and Figure \ref{Fig5}\,(b) shows the density model. The domain is $\Omega=[0,8]\times[0,12.8]$\,km, {\correct where we denote the 2D spatial coordinate as $\mathbf{r}=(x,z)$.}
The seismic sources and receivers are located along $z=0$\,km: 30 point sources with $x$-coordinates $x=0:0.276:8$\,km, and 201 receivers with $x$-coordinates $x=0:0.04:8$\,km. {\correct Again the source wavelet is the high-pass Ricker wavelet, as shown in Figure \ref{Fig_RickerWavelet} by the black solid line. The waveform data are simulated by solving equation (\ref{eqn1}) with a time step size of $\Delta t=0.003$\,s and a spatial mesh size of $h=0.04$\,km; it satisfies the CFL condition for the 2D wave equation, i.e.,  $\Delta t \le \frac{1}{c_{\mathrm{max}}}\frac{h}{\sqrt{2}}$.} The total recording time is $6.0$\,s, with a sampling interval of $0.003$\,s. Figure \ref{Fig5}\,(c) illustrates the waveform data for the 16\,th source of 30. The gravity data are measured along $z=-0.1$\,km, and there are 56 measurements with $x$-coordinates $x=-10:0.5:17.5$\,km; Figure \ref{Fig5}\,(d) plots the gravity data $g_z$. To test the algorithm's robustness, {\correct we further add $5\%$ and $10\%$ Gaussian noise to the measurement data, in separate tests.} The noisy data are simulated as follows:
\begin{eqnarray}
\tilde{p}^*_{i,j,k}= p^*_{i,j,k}+ \eta E \cdot\mathcal{N}(0,1) \ \ \mathrm{with}\ \ E=\sqrt{\frac{1}{N_r N_t}\sum_{j=1}^{N_r}\sum_{k=1}^{N_t}\left(p^*_{i,j,k}\right)^2}\,,  \label{eqn34}\\
\tilde{g}^*_{z,j}=g^*_{z,j}\,(1+\eta\cdot\mathcal{N}(0,1))\,, \label{eqn35}
\end{eqnarray}
{\correct where $\eta$ controls the noise level, e.g. $\eta=5\%$ or $10\%$,} and $\mathcal{N}(0,1)$ denotes the Gaussian noise with a mean of 0 and a standard deviation of 1. Figures {\correct \ref{Fig5}\,(e)\,-\,\ref{Fig5}\,(h)} illustrate the data with noise.

\begin{figure}[htbp!]
\centering
(a){{\includegraphics[scale=0.24,angle=0]{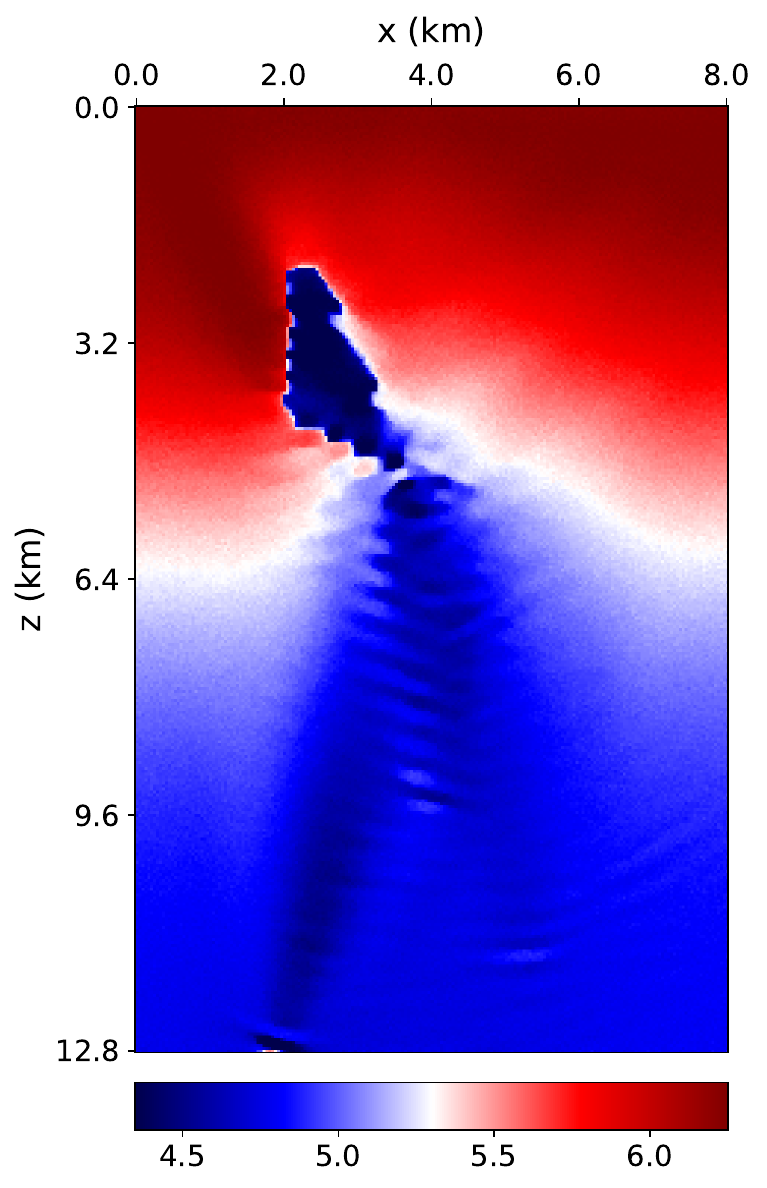}}}
(b){{\includegraphics[scale=0.24,angle=0]{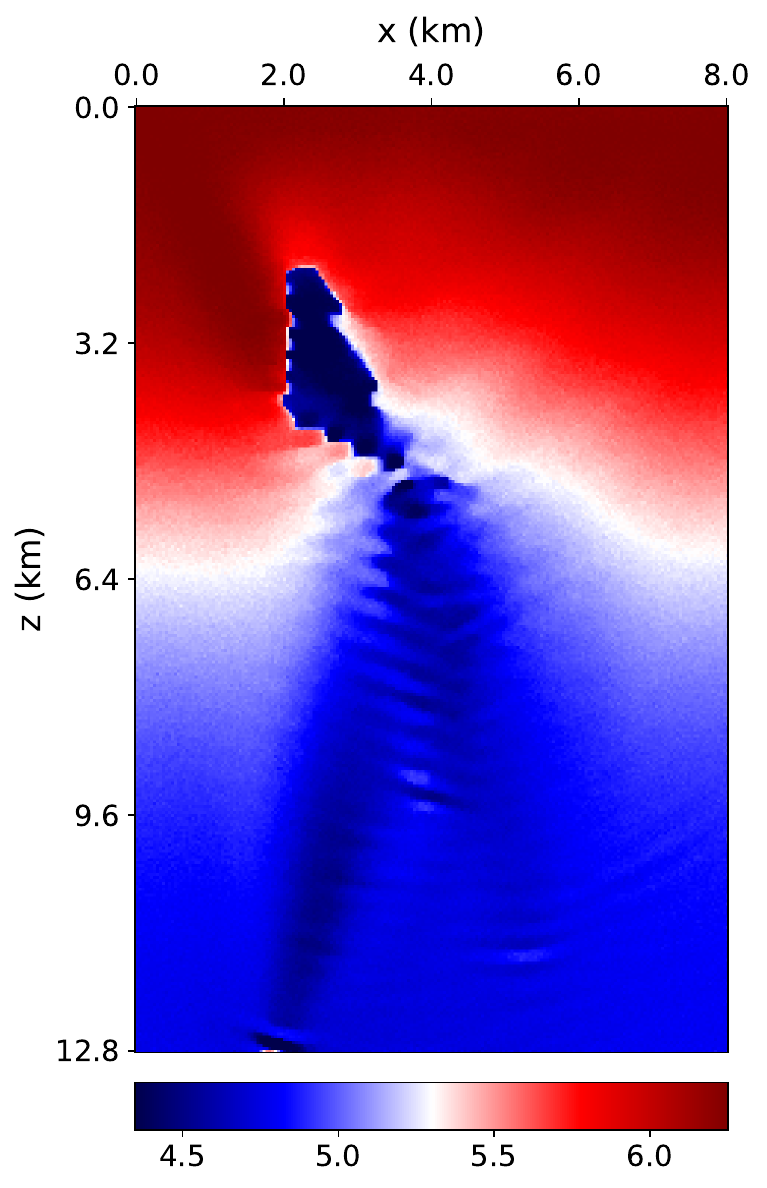}}}
(c){{\includegraphics[scale=0.24,angle=0]{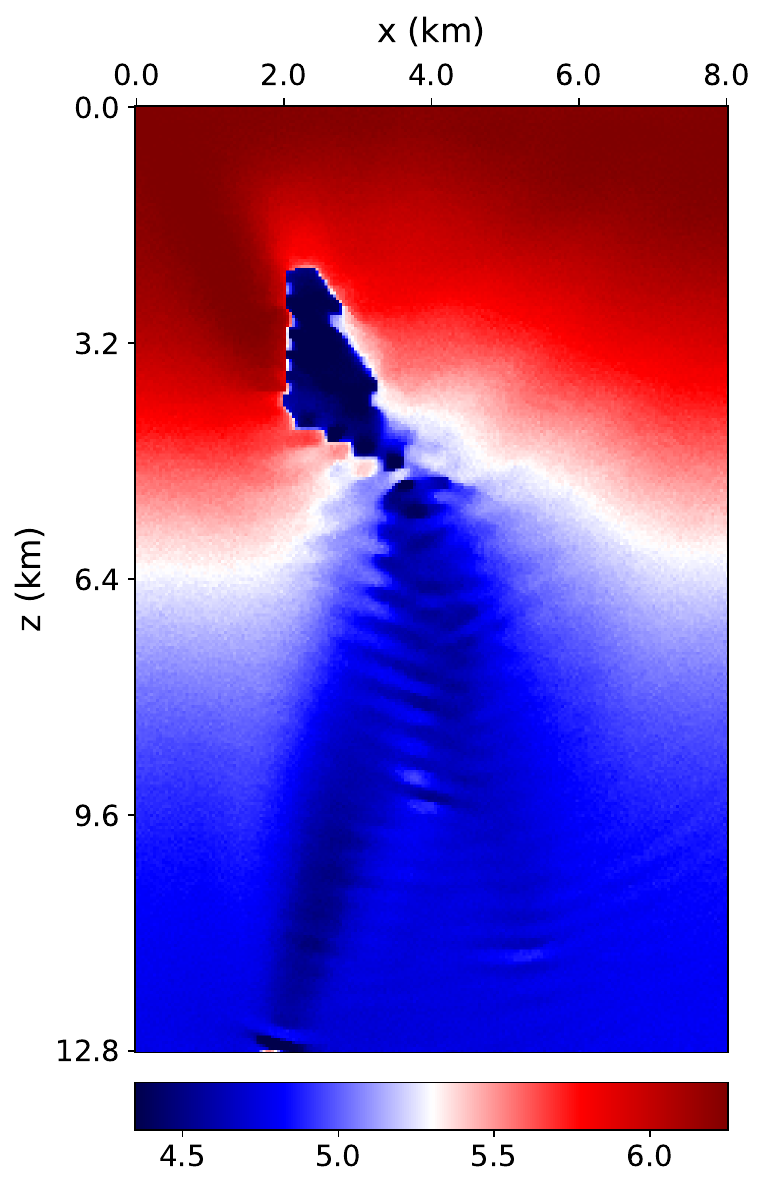}}}
\caption{Example 2: full-waveform inversion results. {\correct The initial velocity guess is shown in Figure \ref{Fig7}\,(a).} (a) FWI solution from clean data; (b) FWI solution from the data with $5\%$ Gaussian noise; \correct{(c) FWI solution from the data with $10\%$ Gaussian noise. }}
\label{Fig6}
\end{figure}

As a benchmark, we firstly consider the solution by full-waveform inversion. {\correct Figure \ref{Fig6} shows the results, where we use an initial velocity guess from Figure \ref{Fig7}\,(a).  Figure \ref{Fig6}\,(a) shows the FWI solution from clean data, and  Figures \ref{Fig6}\,(b) and \ref{Fig6}\,(c) show the solutions from data with $5\%$ and $10\%$ Gaussian noise, respectively.} For this dipping model, FWI adequately recovers its shallow structure, but poorly images the extensive part in the deep region. 

\begin{figure}[htbp!]
%\captionsetup{font={color=red}}%\correct 
\centering
\correct
(a){{\includegraphics[scale=0.24,angle=0]{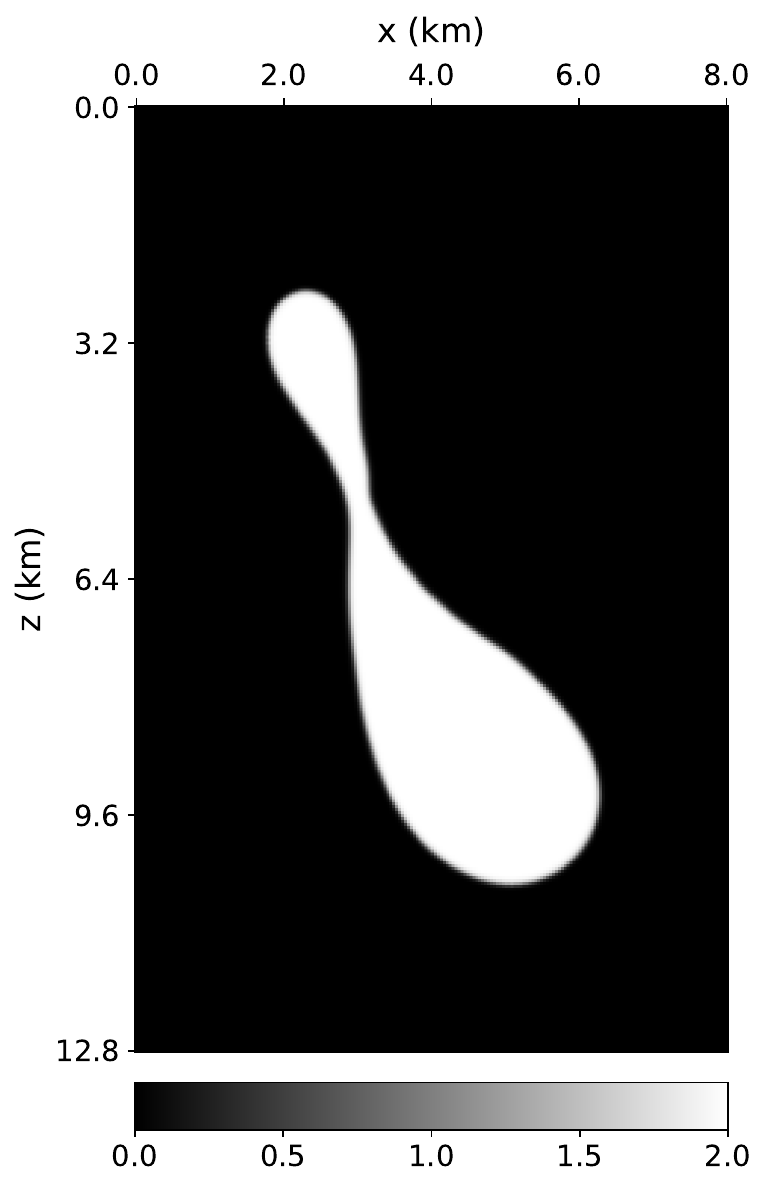}}}
(b){{\includegraphics[scale=0.24,angle=0]{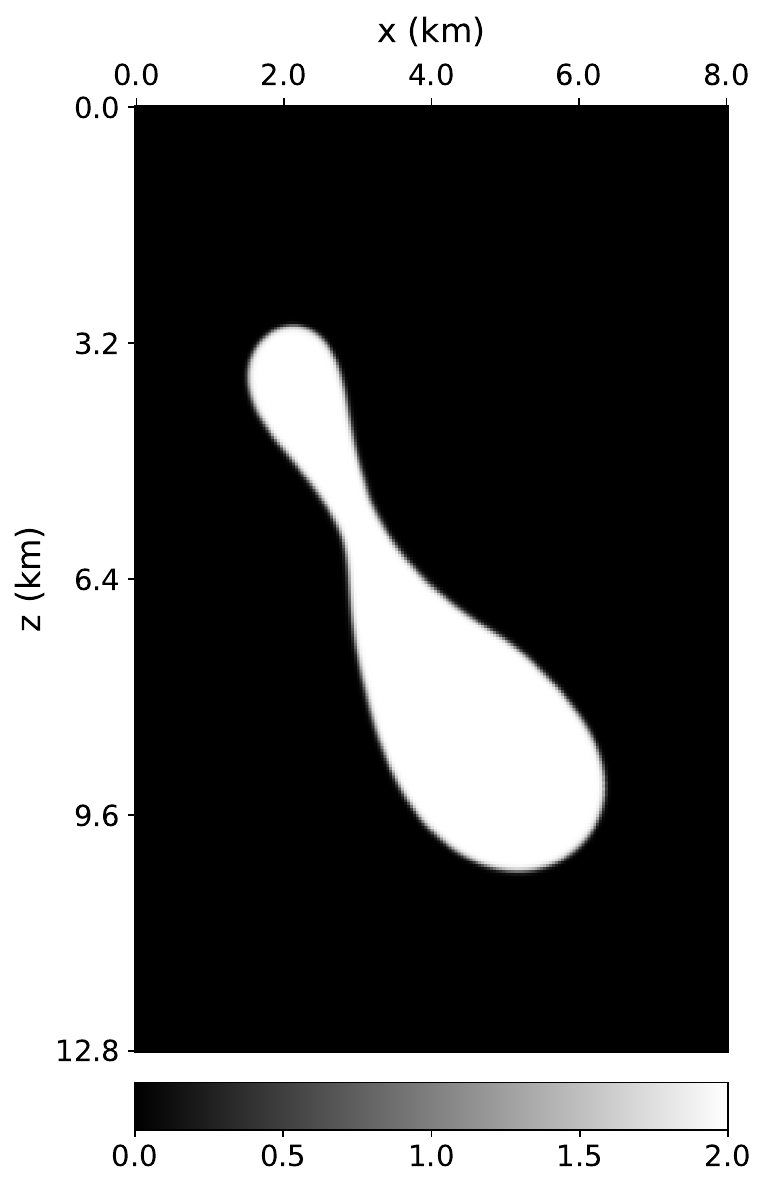}}}
(c){{\includegraphics[scale=0.24,angle=0]{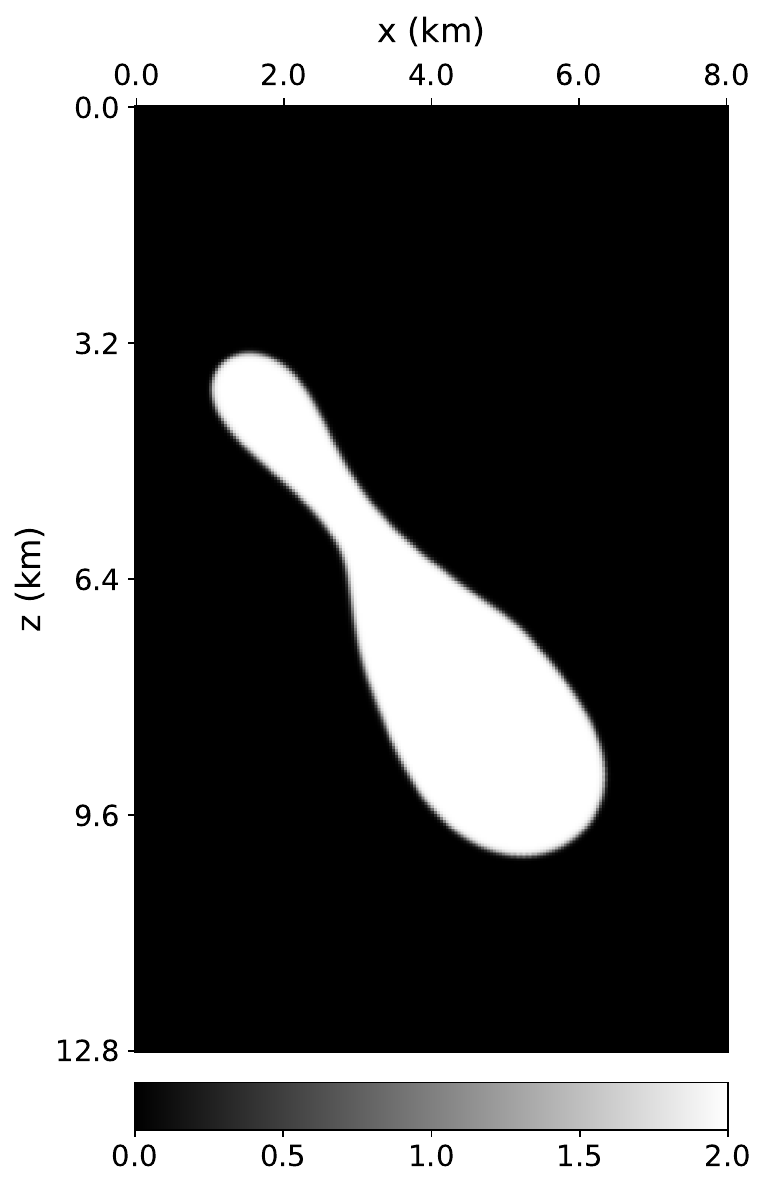}}}
\caption{ Example 2: gravity inversion results. The initial density guess is shown in Figure \ref{Fig7}\,(b). (a) Solution of gravity inversion from clean data; (b) solution of gravity inversion from the data with $5\%$ Gaussian noise; (c) solution of gravity inversion from the data with $10\%$ Gaussian noise. }
\label{Fig_EX2_gravity}
\end{figure}

\begin{figure}[htbp!]
\centering
(a){{\includegraphics[scale=0.24,angle=0]{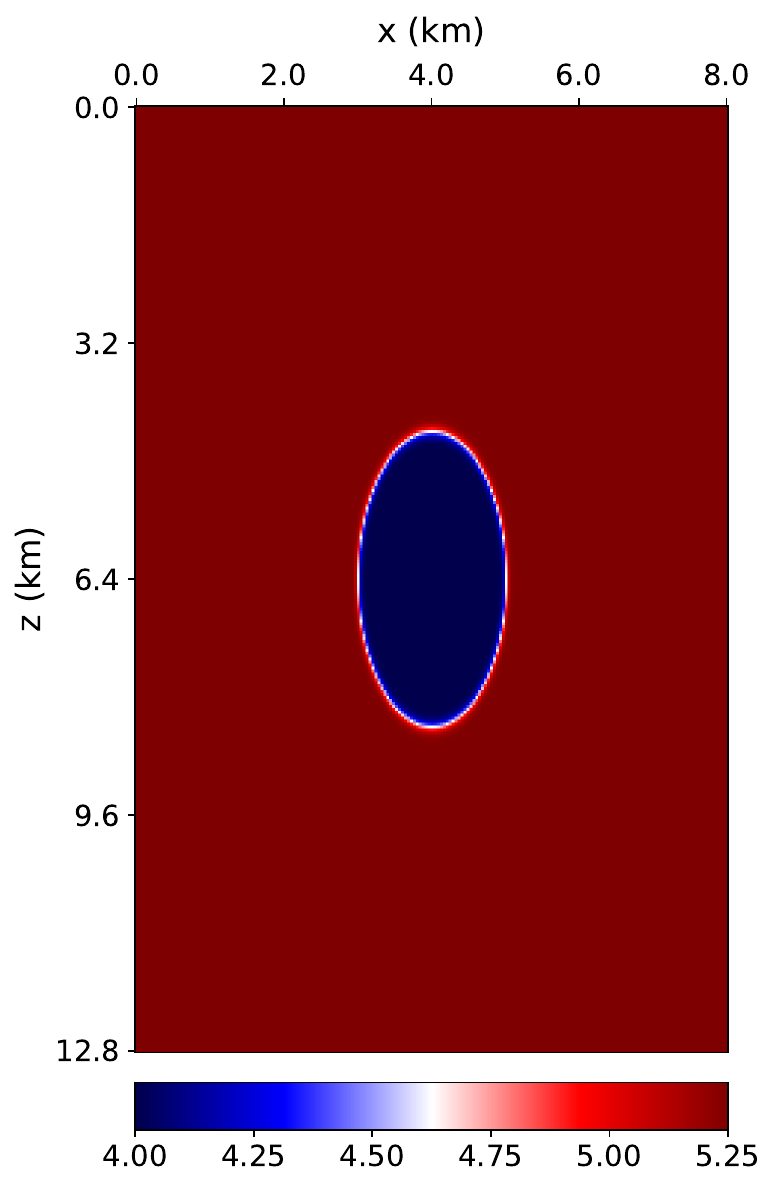}}}
(c){{\includegraphics[scale=0.24,angle=0]{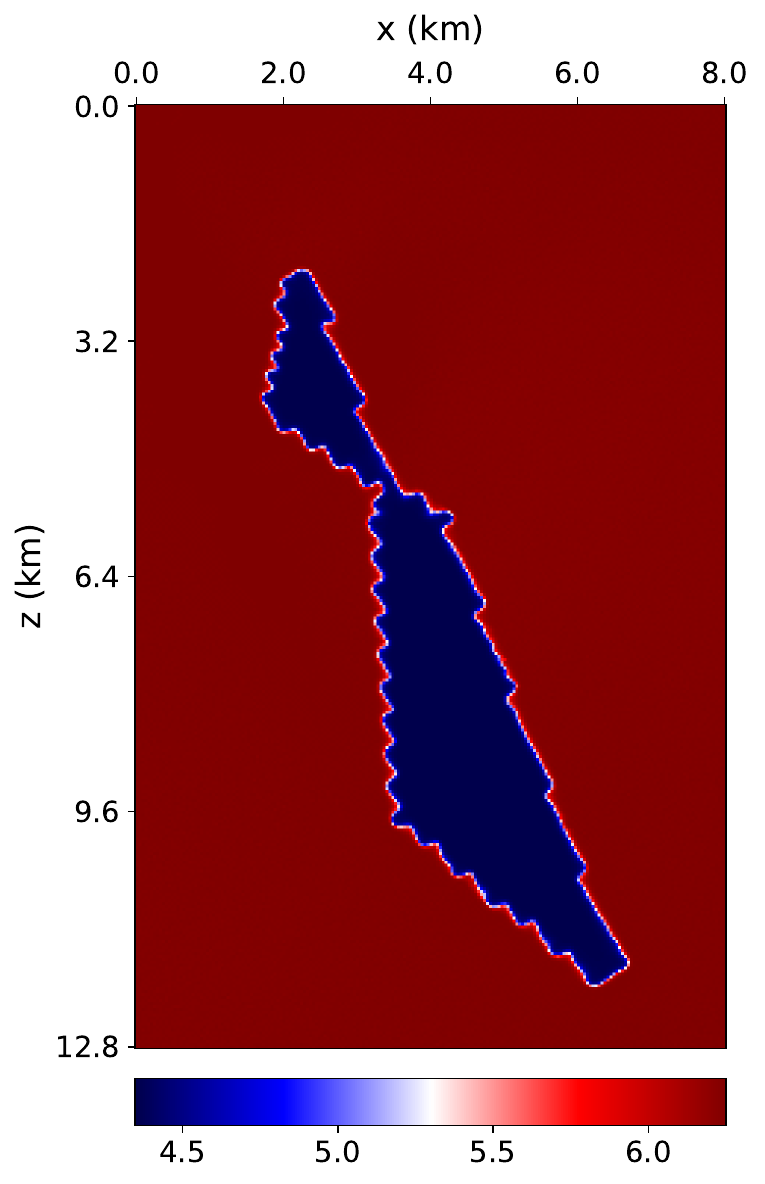}}}
(e){{\includegraphics[scale=0.24,angle=0]{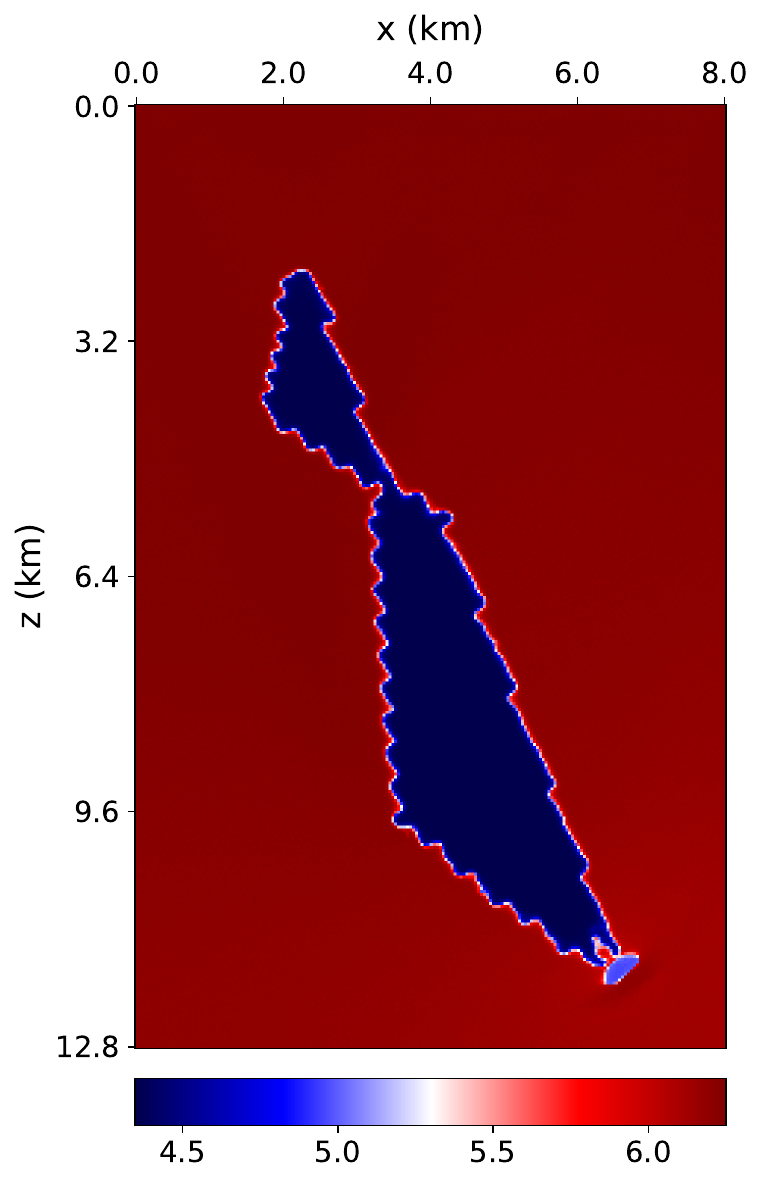}}}
{\correct (g){{\includegraphics[scale=0.24,angle=0]{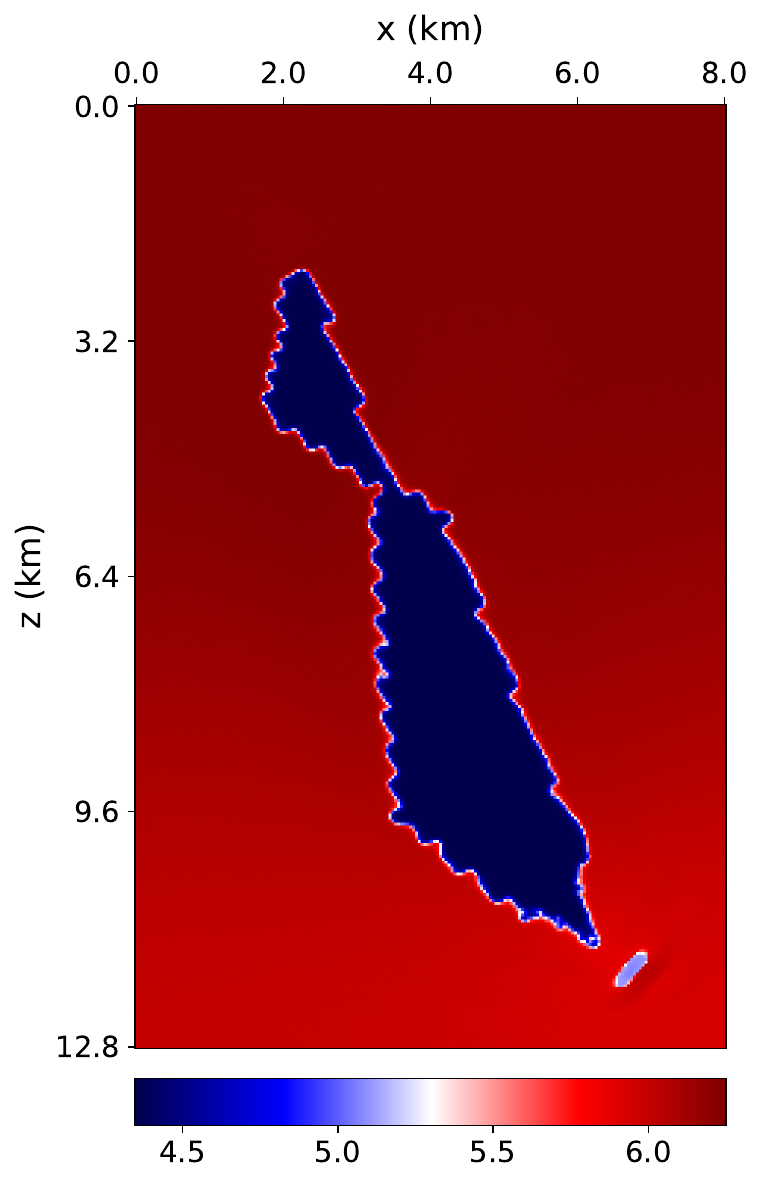}}}} \\
(b){{\includegraphics[scale=0.24,angle=0]{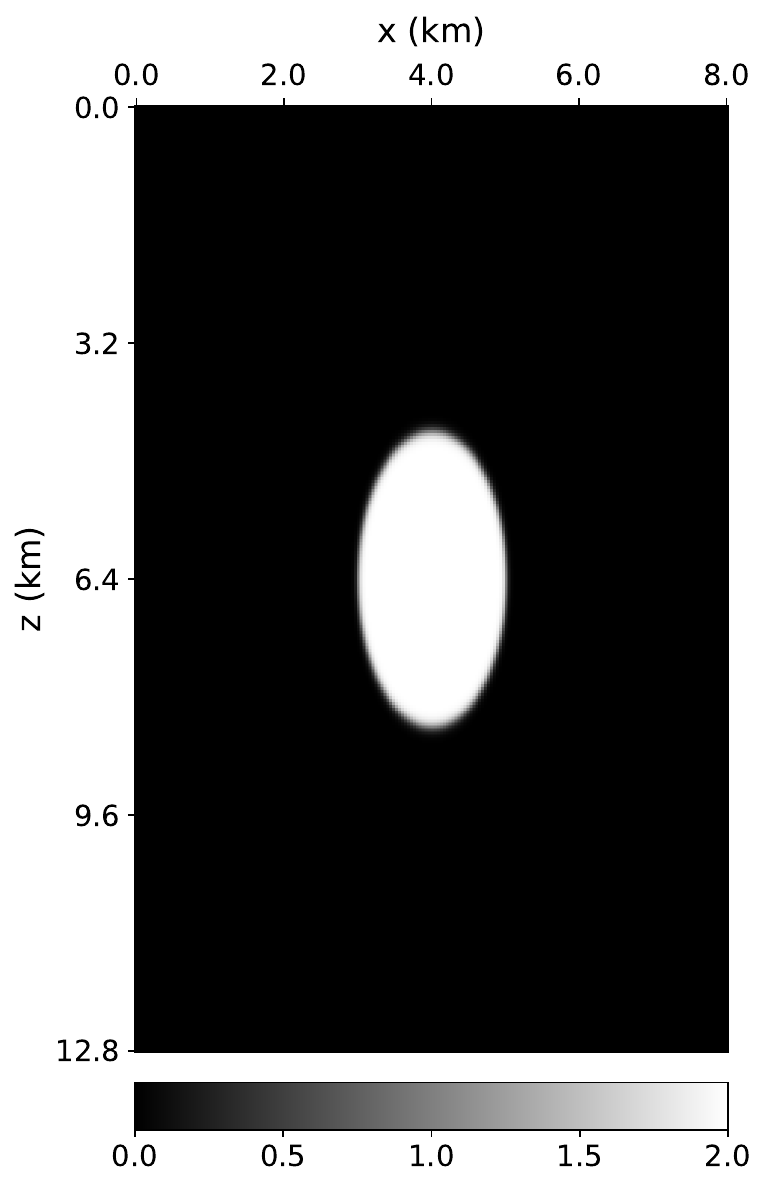}}}
(d){{\includegraphics[scale=0.24,angle=0]{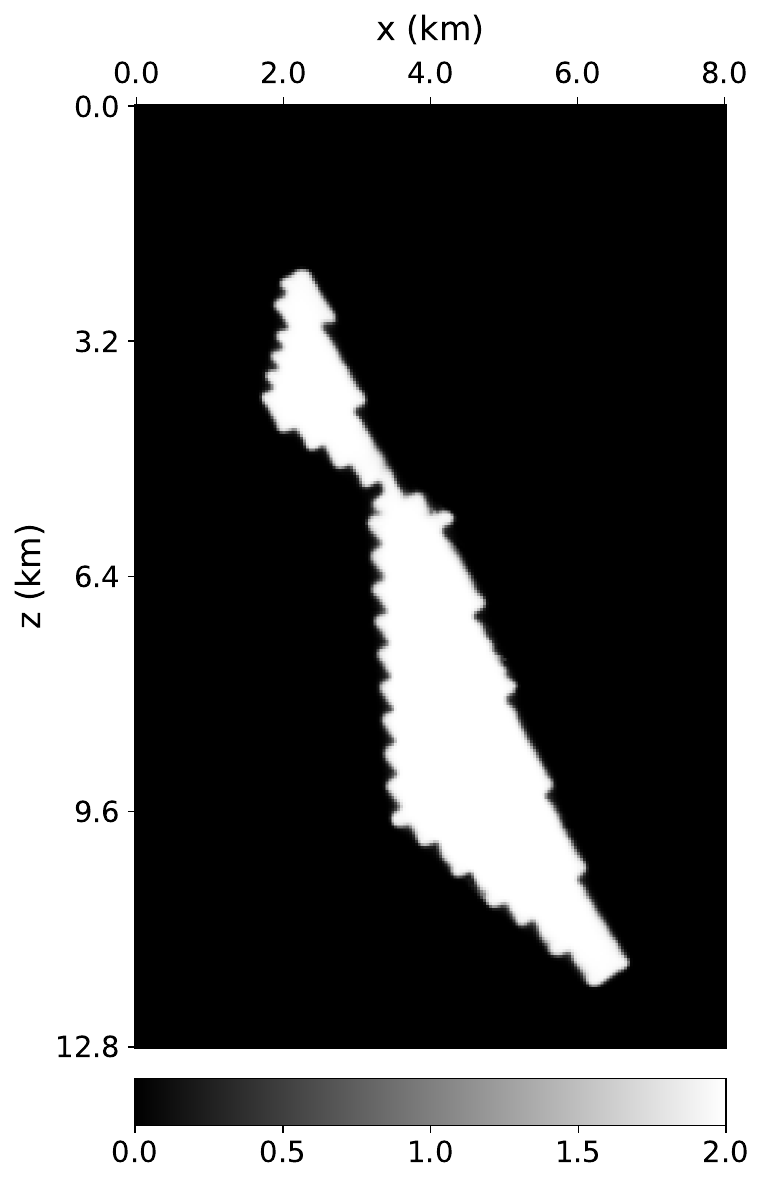}}}
(f){{\includegraphics[scale=0.24,angle=0]{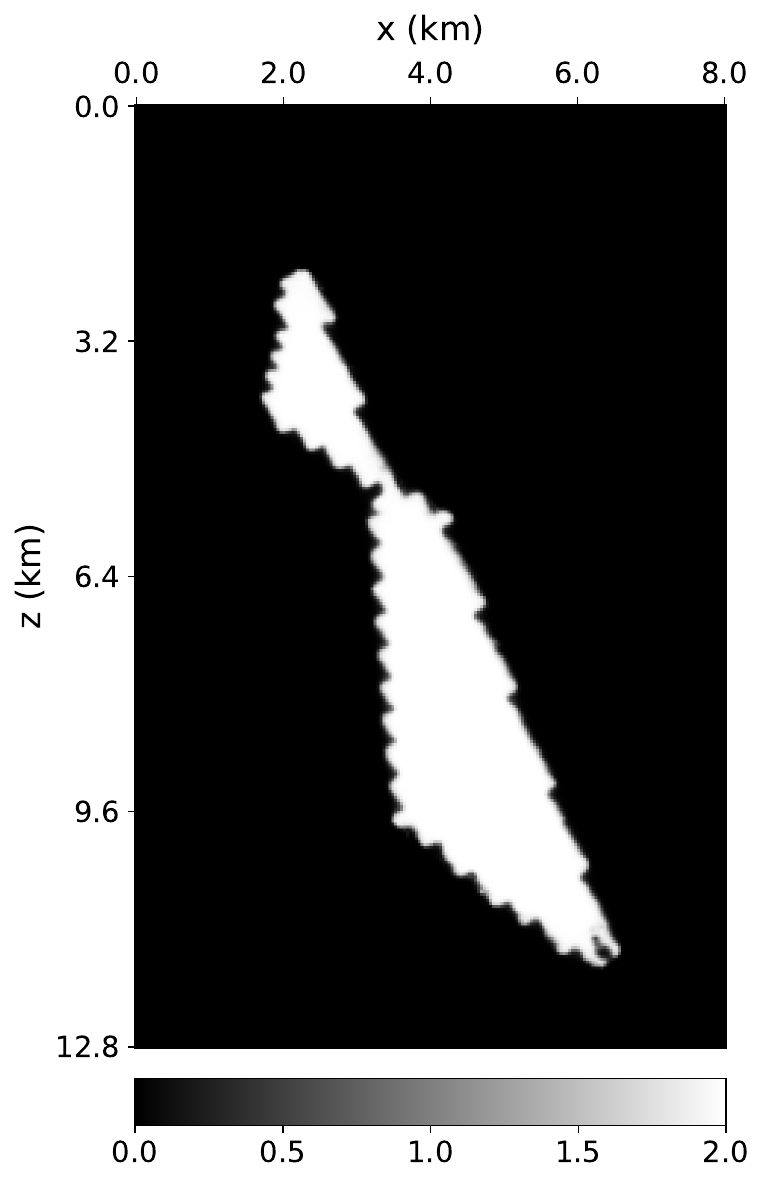}}}
{\correct (h){{\includegraphics[scale=0.24,angle=0]{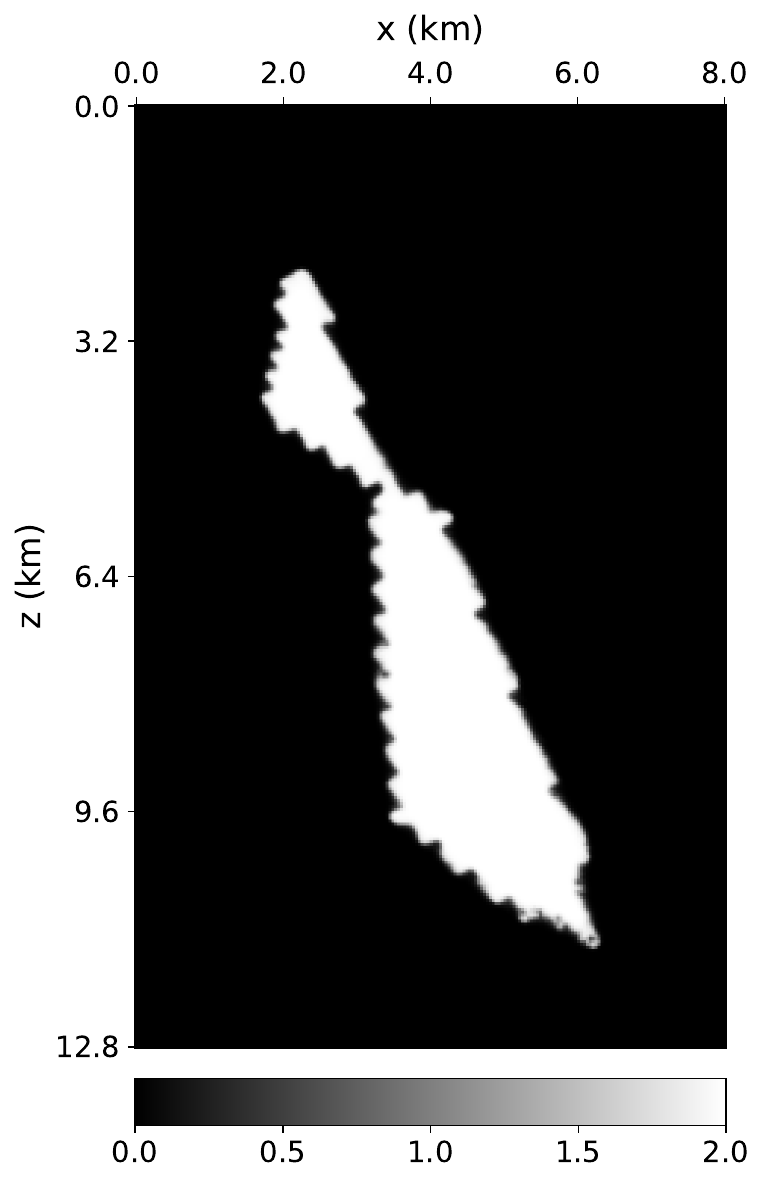}}}}
\caption{Example 2: joint inversion results. (a) Initial model for velocity; (b) initial model for density; (c) recovered velocity of joint inversion from clean data; (d) recovered density of joint inversion from clean data; (e) recovered velocity from the data with $5\%$ Gaussian noise; (f)  recovered density from the data with $5\%$ Gaussian noise; {\correct (g) recovered velocity from the data with $10\%$ Gaussian noise; (h) recovered density from the data with $10\%$ Gaussian noise.}}
\label{Fig7}
\end{figure}

Then we integrate gravity for joint inversion. {\correct To illustrate the effect of gravity approach, we provide the pure gravity inversion results in Figure \ref{Fig_EX2_gravity}. The initial density guess is shown in Figure \ref{Fig7}\,(b), and we impose the density-contrast value $f(\mathbf{r})=2.0$\,g/cm$^3$ as a priori information. The gravity inversion recovers the overall profile of the dipping object, including its extensive part at depth, although the shape of the recovered model is imperfect.}

In the joint inversion, the density-contrast value $f(\mathbf{r})=2.0$\,g/cm$^3$ is imposed as a priori information; for simplicity, we freeze the velocity value of the dipping object, $c_1(\mathbf{r})=4.348$\,km/s. The level-set formulation is effective for the imposition of the prior information. Table \ref{Tab3} lists the values of algorithm parameters employed in the joint inversion.
Figure \ref{Fig7} shows the joint inversion results, where \ref{Fig7}\,(a) and \ref{Fig7}\,(b) illustrate the initial guesses for velocity and density models, \ref{Fig7}\,(c) and \ref{Fig7}\,(d) plot the recovered solutions from clean data, and {\correct \ref{Fig7}\,(e)\,-\,\ref{Fig7}\,(h) plot the solutions from the data with $5\%$ and $10\%$ Gaussian noise.} The joint inversions from both clean data and noisy data yield excellent solutions for the velocity and density models. {\correct The extensive dipping object is successfully recovered, and the shape of interfaces closely matches the true model. Although $10\%$ noise is considerable for gravity data, the joint inversion demonstrates robustness to noise contamination. In our level-set based joint inversion algorithm, the gravity inversion and FWI effectively complement each other.}

\begin{table}[h!]
		\centering
		\begin{tabular}{ccccccc} 
			\toprule
			$\epsilon$      &  $\alpha_\phi$  &  $\alpha_{v_2}$   &  $\omega_0$    &  $\lambda$  & $\lambda_\phi$ &  $\lambda_{c_2}$\\
			\midrule
			0.05       & 1 & 1  & 5  &  $\displaystyle \frac{\mathrm{ln}50}{n_{\mathrm{max}}}$ ($n_{\mathrm{max}}=2\times10^4$)  &  $10^{-6}$  & $2 \times 10^{-5}$               \\
			\bottomrule
		\end{tabular}\caption{Example 2: values of algorithm parameters used in the joint inversion. The values are set the same for the inversions with and without noise.}
		\label{Tab3}
\end{table}
\vspace{-8pt}

% -------------------------
\subsubsection{{\correct Example 3.}}

\begin{figure}[htbp!]
%\captionsetup{font={color=red}}%\correct 
\centering
\correct
(a){{\includegraphics[scale=0.24,angle=0]{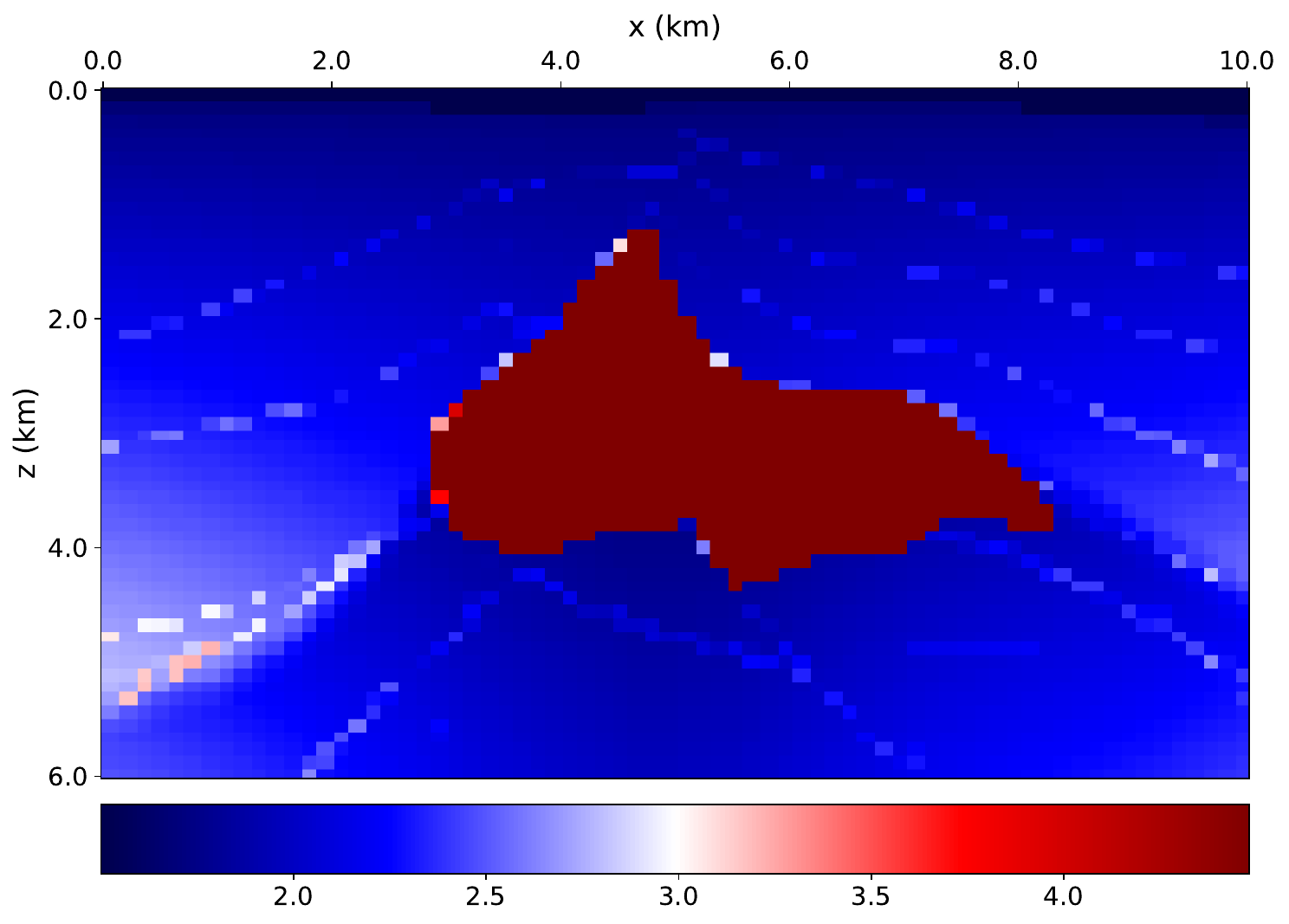}}}
(b){{\includegraphics[scale=0.24,angle=0]{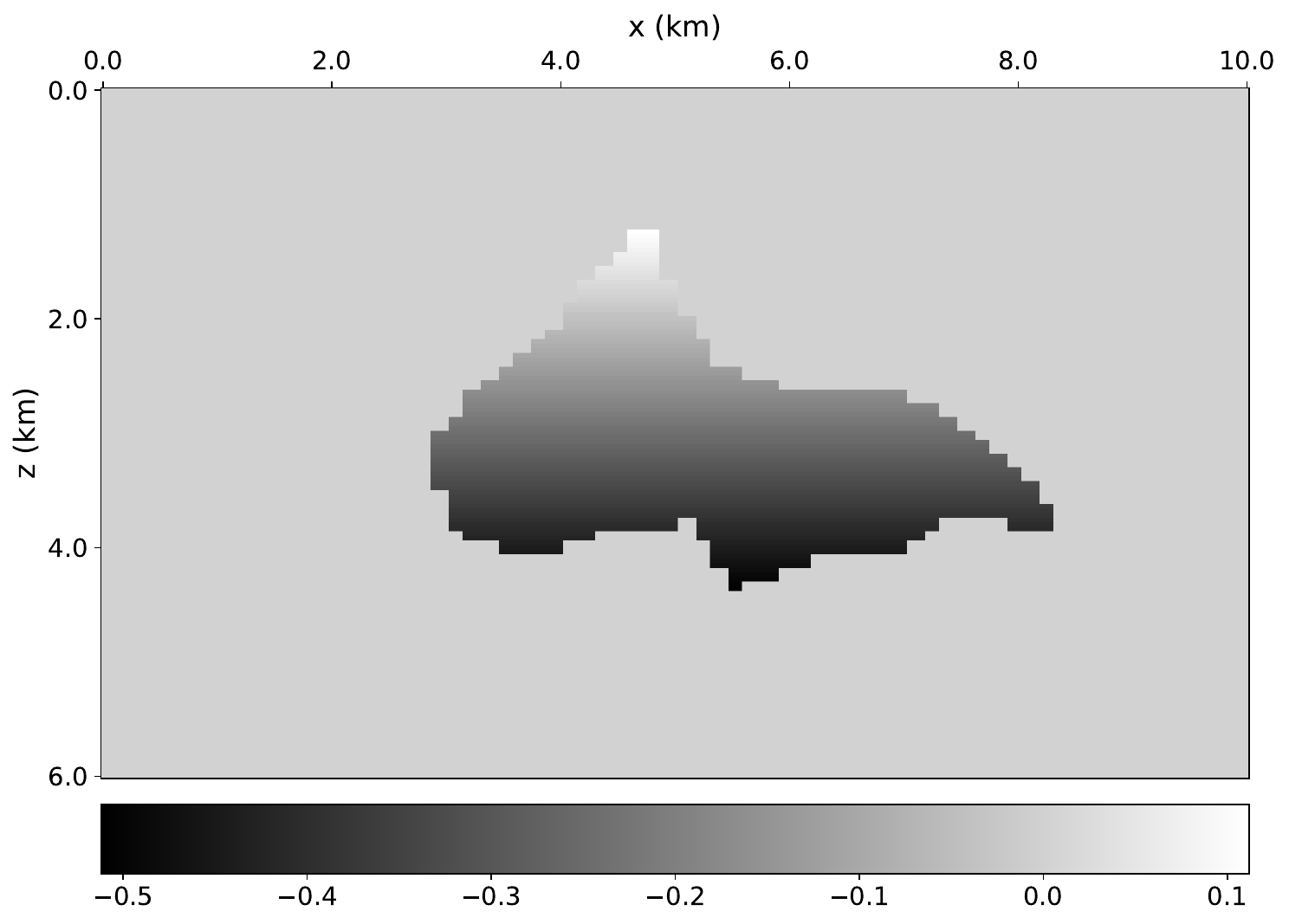}}}
\caption{Example 3: a salt model. (a) True velocity model; (b) true density model.}
\label{Figexp31}
\end{figure}

\begin{figure}[htbp!]
%\captionsetup{font={color=red}}%\correct 
\centering
\correct
(a){{\includegraphics[scale=0.24,angle=0]{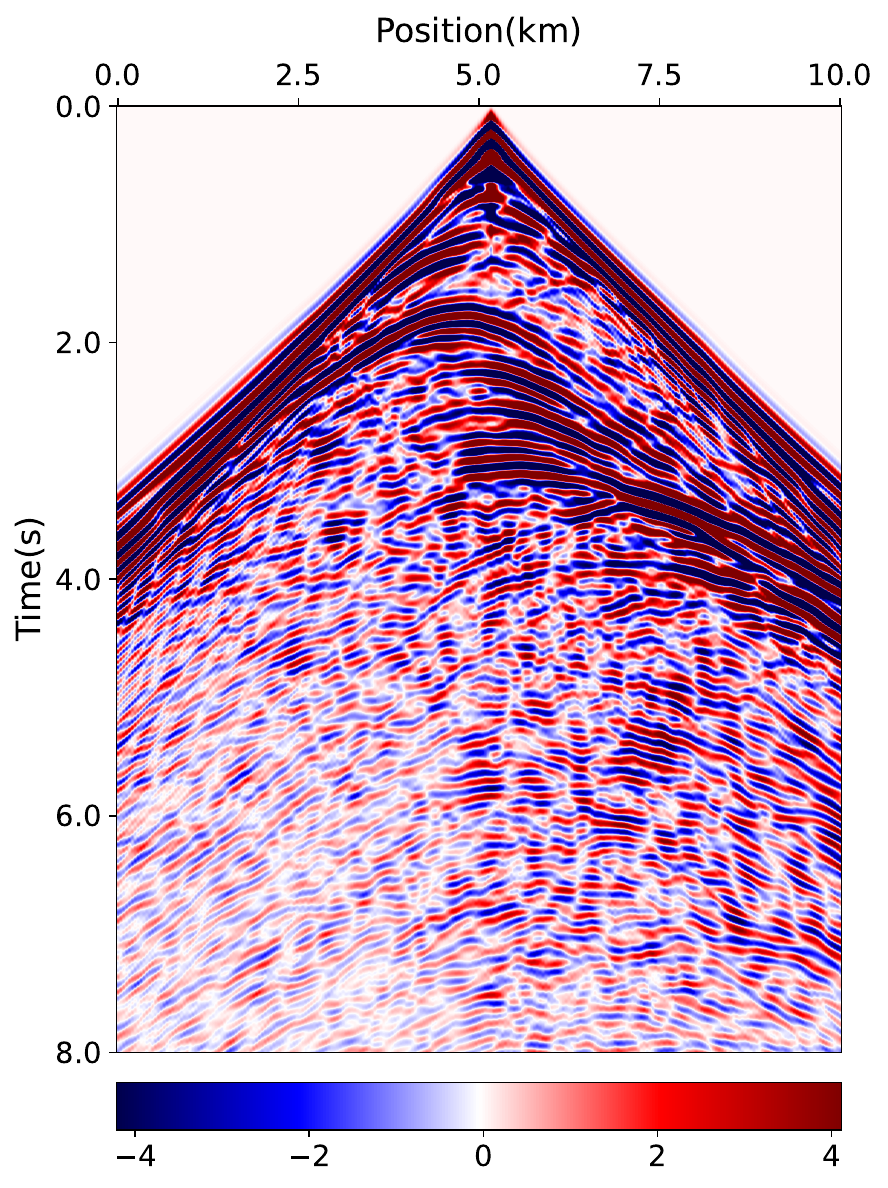}}} 
(c){{\includegraphics[scale=0.24,angle=0]{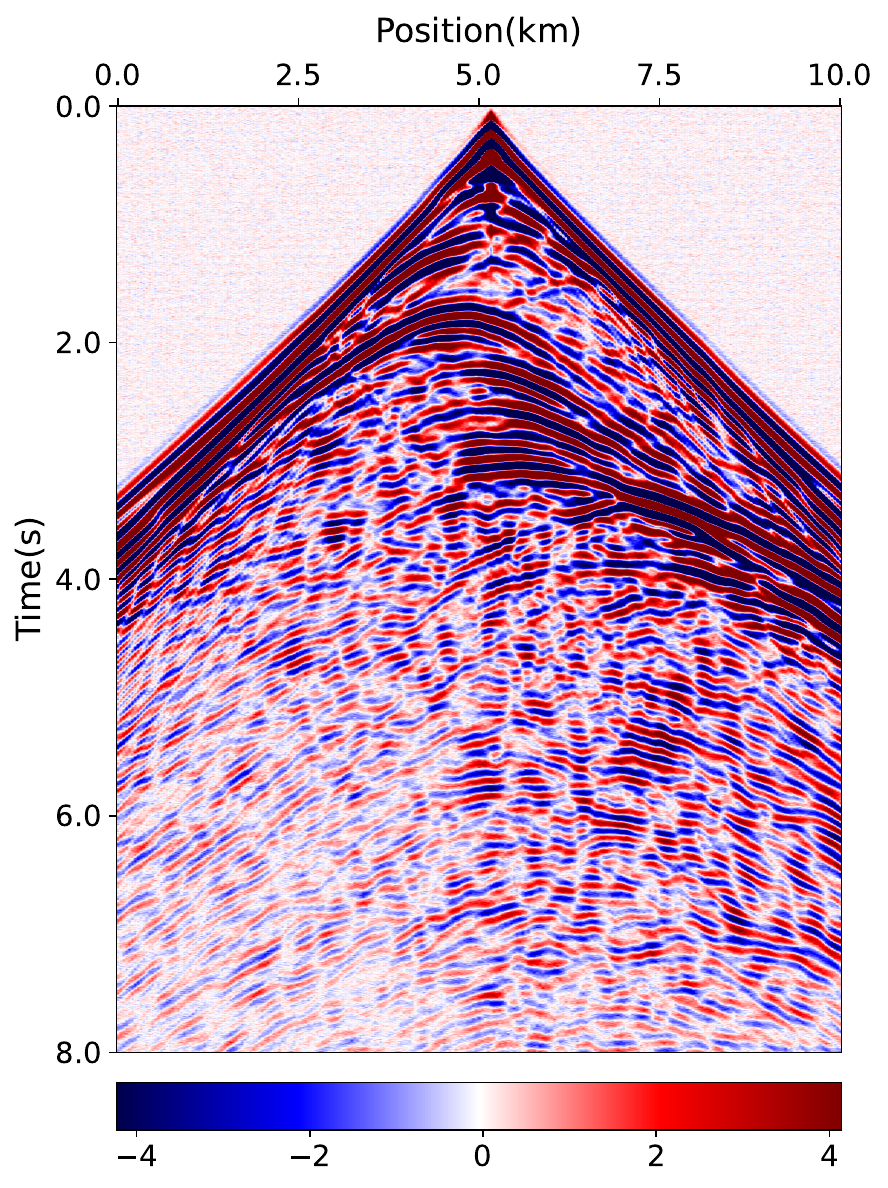}}}
(e){{\includegraphics[scale=0.24,angle=0]{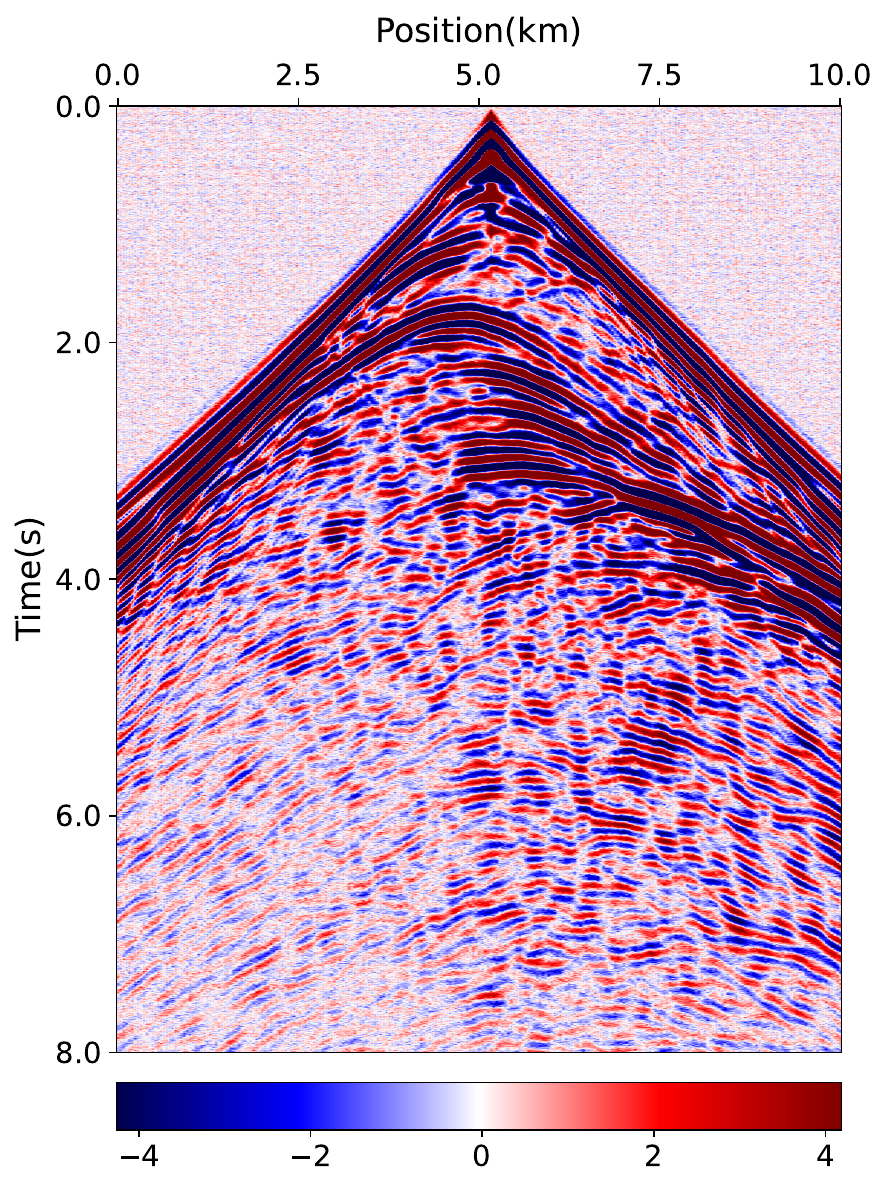}}} \\
(b){{\includegraphics[scale=0.24,angle=0]{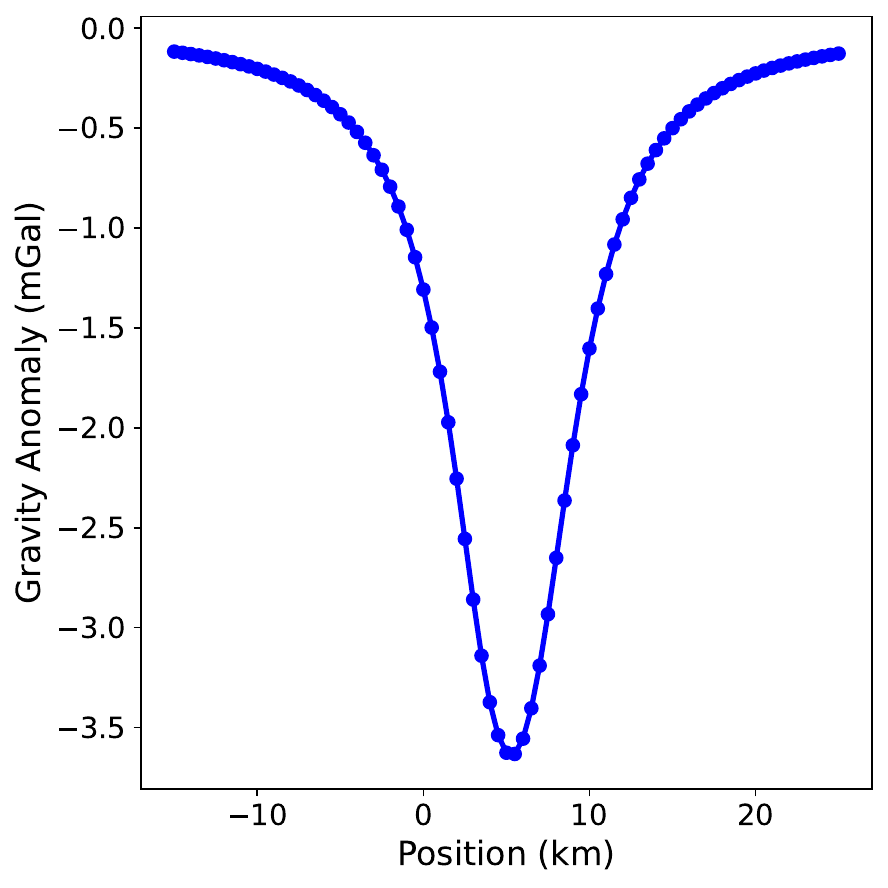}}}
(d){{\includegraphics[scale=0.24,angle=0]{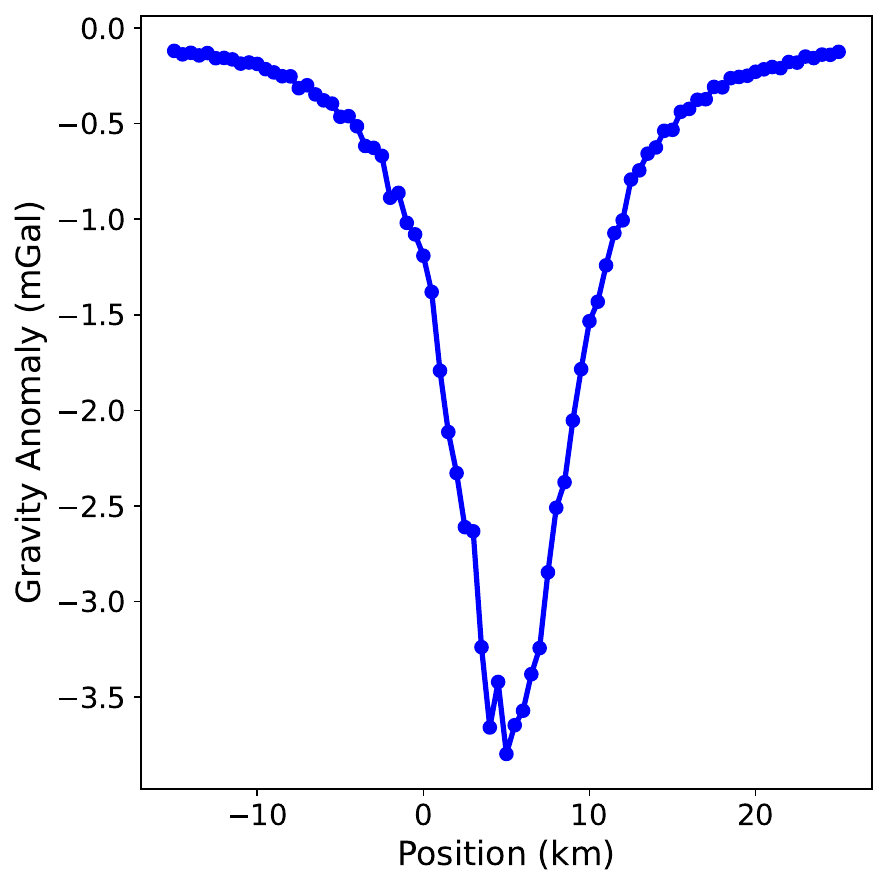}}}
(f){{\includegraphics[scale=0.24,angle=0]{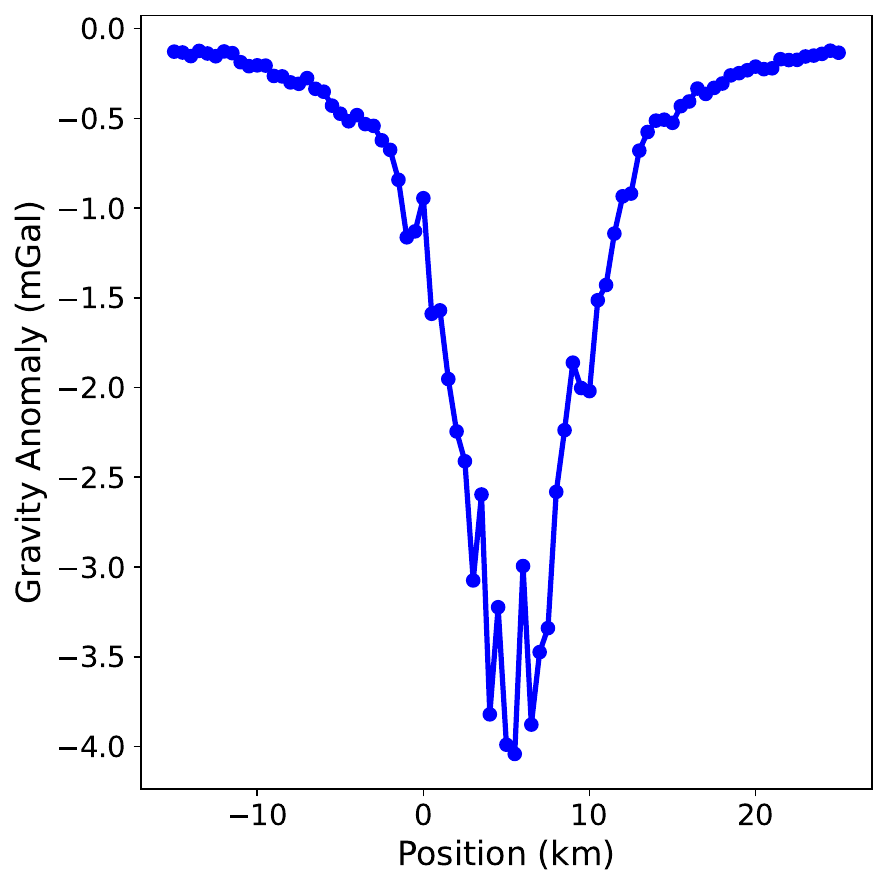}}}
\caption{Example 3: measurement data. The color scale of waveform data is clipped between its $5\%$ and $95\%$ quantiles for enhanced visualization. (a) waveform data for the 16\,th source of 30; (b) gravity data $g_z$; (c) waveform data with $5\%$ Gaussian noise; (d) gravity data with $5\%$ Gaussian noise; (e) waveform data with $10\%$ Gaussian noise; (f) gravity data with $10\%$ Gaussian noise.}
\label{Figexp32}
\end{figure}

{\correct We consider a salt model shown in Figure \ref{Figexp31}, where \ref{Figexp31}\,(a) and  \ref{Figexp31}\,(b) show the velocity and density models, respectively. The density-contrast value is set as $f(\mathbf{r})=(1.8-z)\times 0.2$\,g/cm$^3$ \cite{liluqia16}. The velocity and density share a common interface, which delineates the boundary of the salt region. The computational domain is $\Omega=[0,10]\times[0,6]$\,km, with the 2D spatial coordinate denoted by $\mathbf{r}=(x,z)$. The seismic sources and receivers are located along $z=0$\,km; there are 30 point sources with $x$-coordinates $x=0:0.345:10$\,km, and 251 receivers with $x$-coordinates $x=0:0.04:10$\,km. 
The source wavelet is the high-pass Ricker wavelet, as shown in Figure \ref{Fig_RickerWavelet} by the black solid line. When solving the wave equation, we use a time step size of $\Delta t=0.004$\,s and a spatial mesh size of $h=0.04$\,km; it satisfies the CFL condition for the 2D wave equation, i.e., $\Delta t \le \frac{1}{c_{\mathrm{max}}}\frac{h}{\sqrt{2}}$. The total recording time is $8$\,s, with a sampling interval of $0.004$\,s. Figure \ref{Figexp32}\,(a) shows the waveform data for the 16\,th source of 30. The gravity data are acquired along $z=-0.1$\,km, where we have 81 measurements with $x$-coordinates $x=-15:0.5:25$\,km. Figure \ref{Figexp32}\,(b) plots the gravity data $g_z$. To test the algorithm's robustness, we add $5\%$ and $10\%$ Gaussian noise to the measurement data. Figures \ref{Figexp32}\,(c)\,-\,\ref{Figexp32}\,(f) plot the data with noise.}

\begin{figure}[htbp!]
%\captionsetup{font={color=red}}%\correct 
\centering
\correct
(a){{\includegraphics[scale=0.24,angle=0]{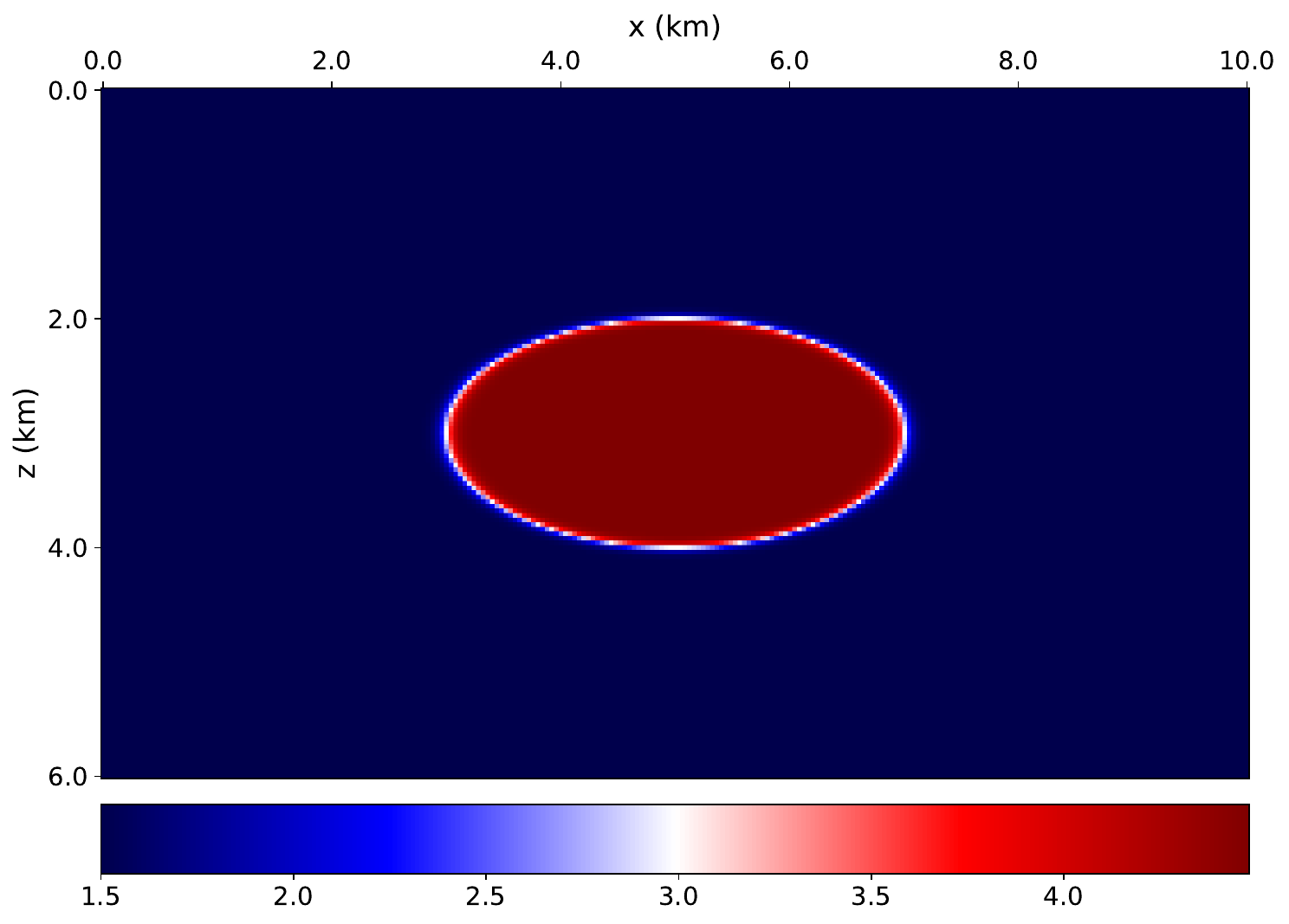}}}
(b){{\includegraphics[scale=0.24,angle=0]{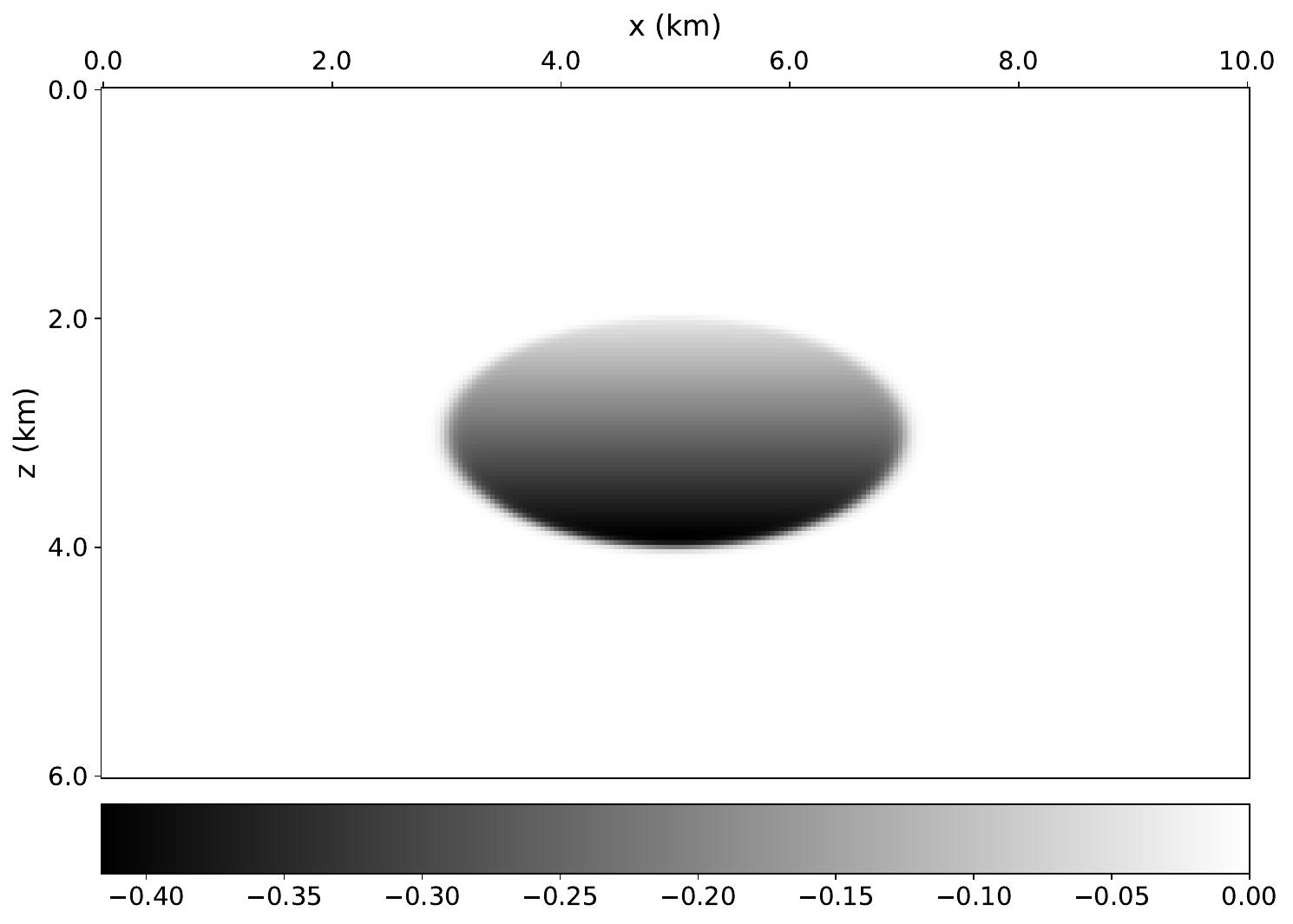}}}
(c){{\includegraphics[scale=0.24,angle=0]{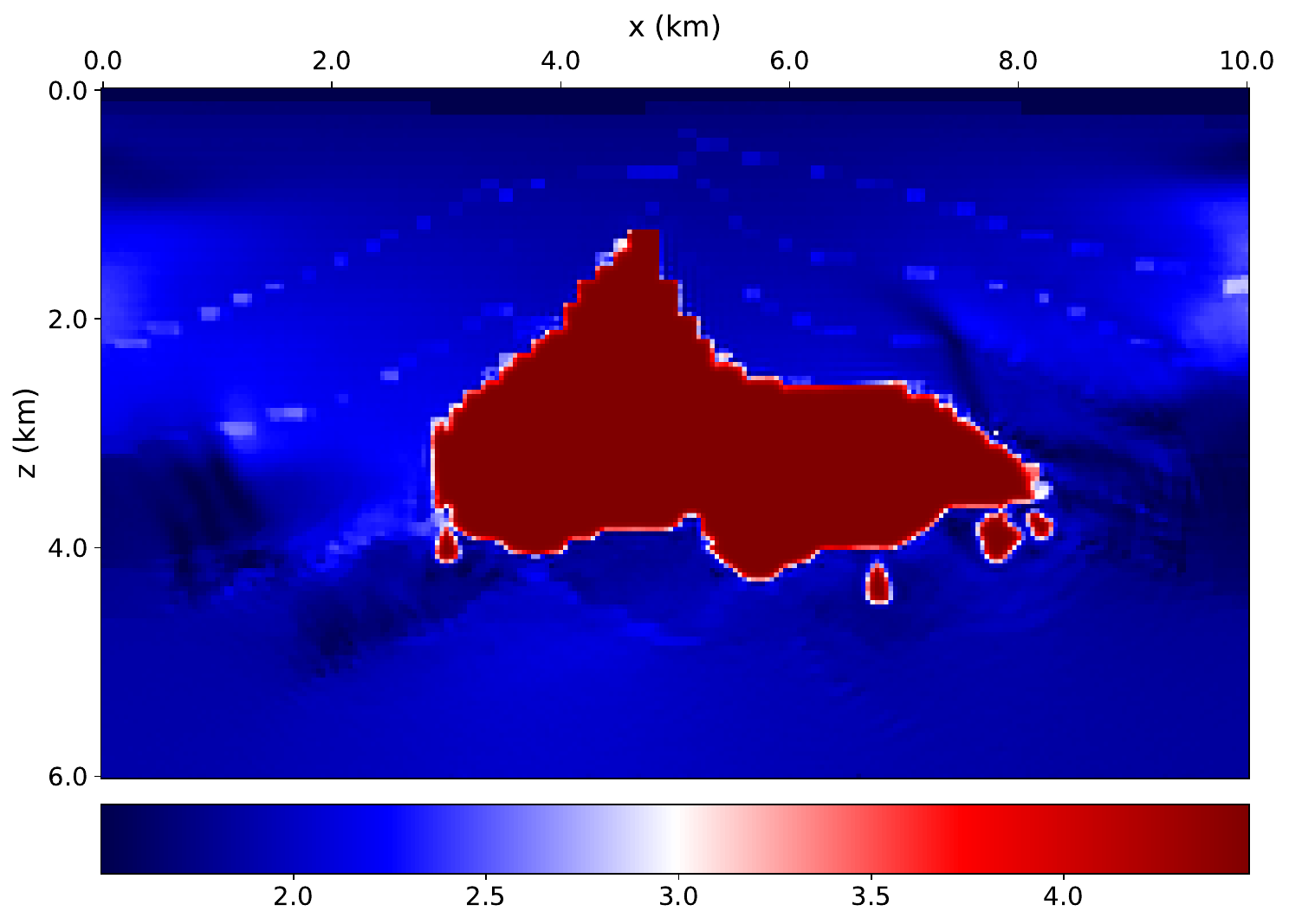}}}
(d){{\includegraphics[scale=0.24,angle=0]{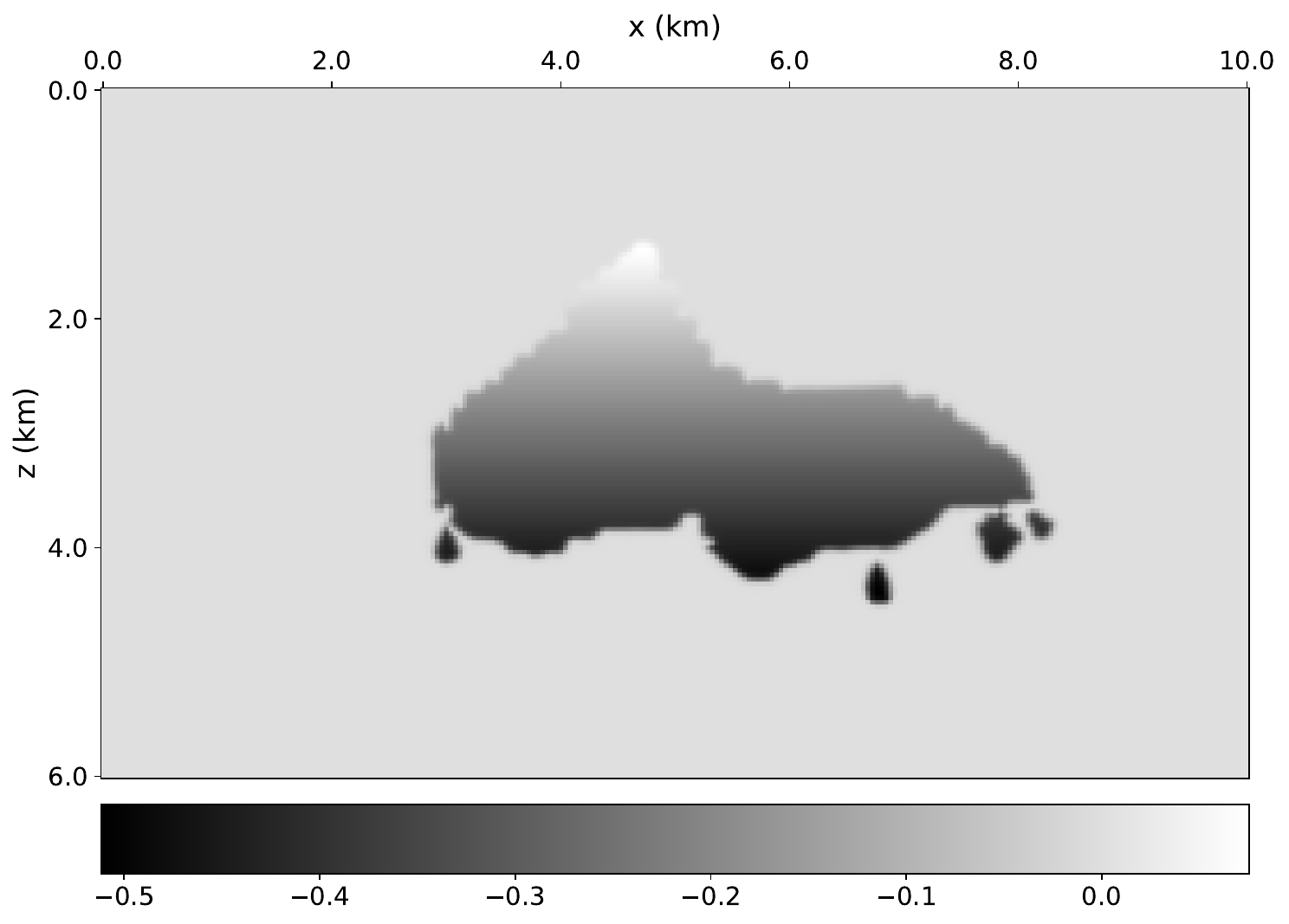}}}
(e){{\includegraphics[scale=0.24,angle=0]{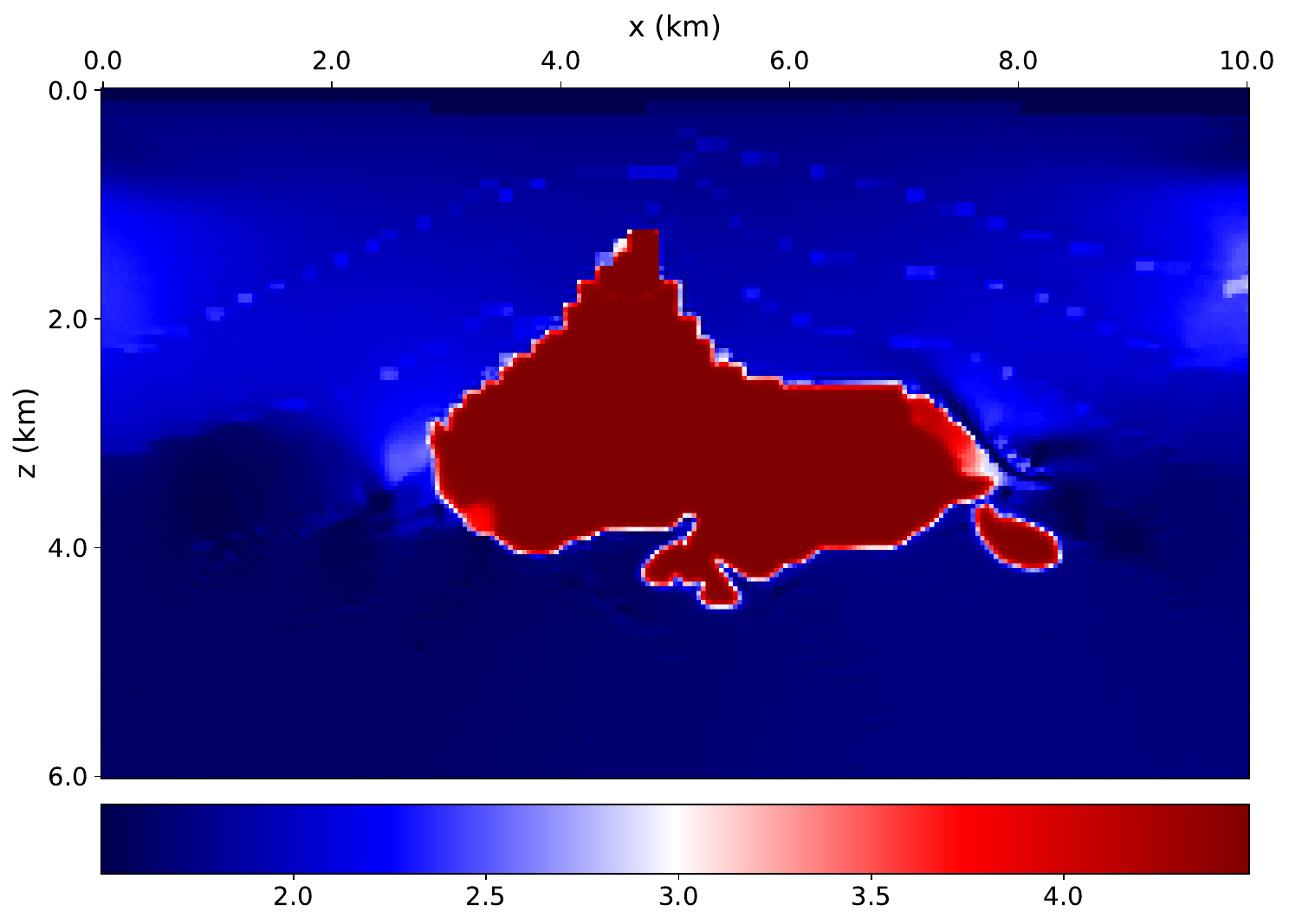}}}
(f){{\includegraphics[scale=0.24,angle=0]{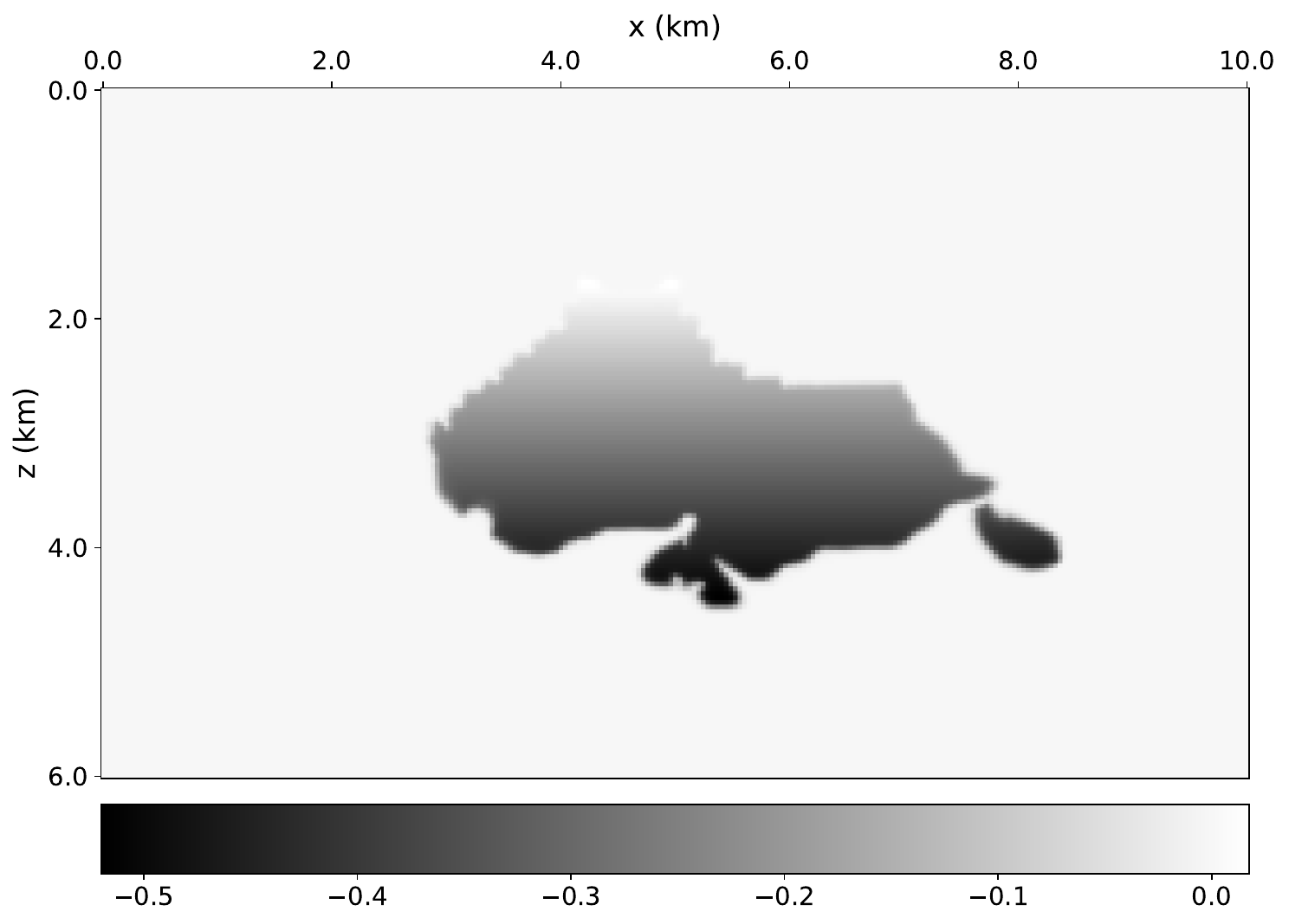}}}
(g){{\includegraphics[scale=0.24,angle=0]{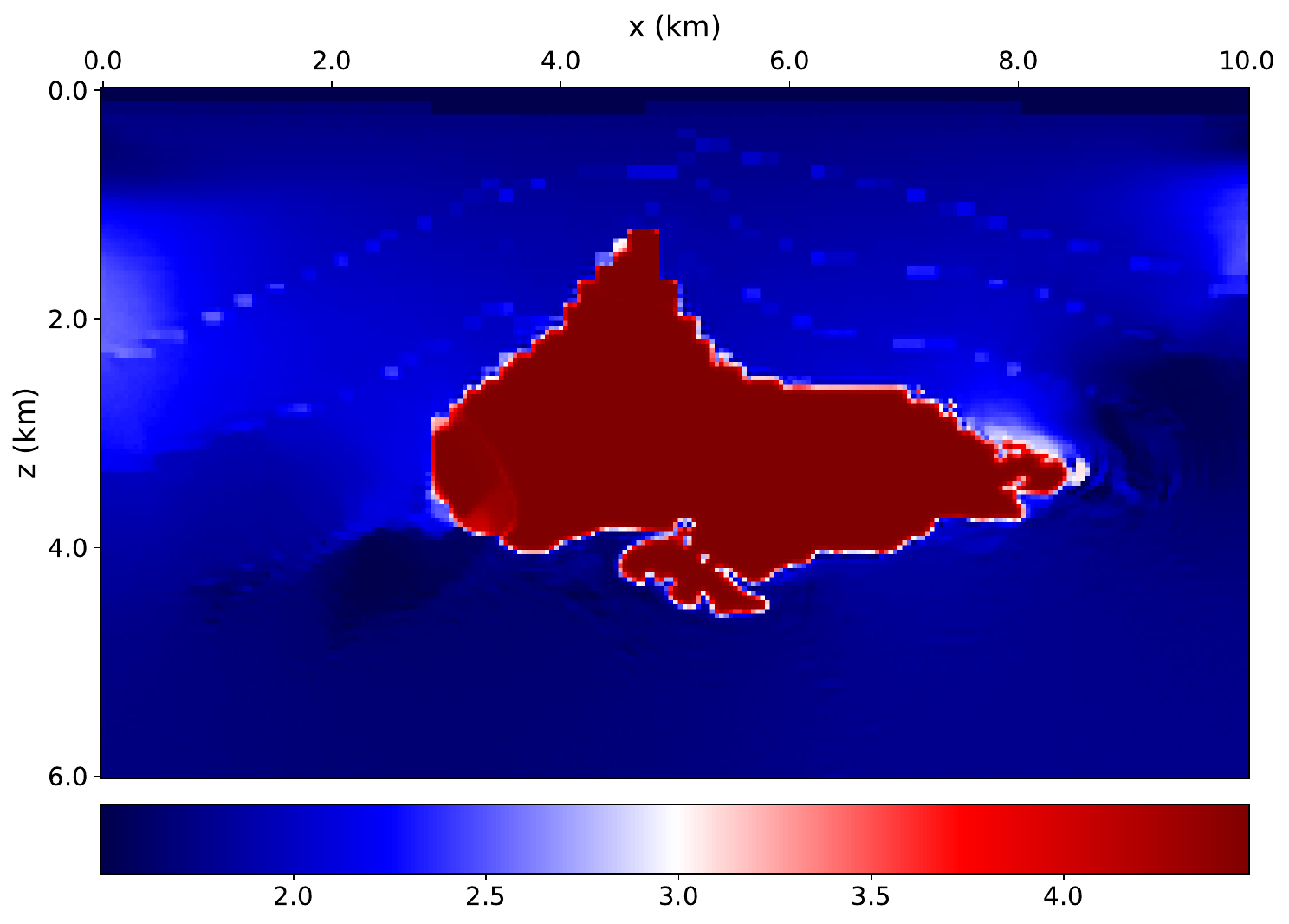}}}
(h){{\includegraphics[scale=0.24,angle=0]{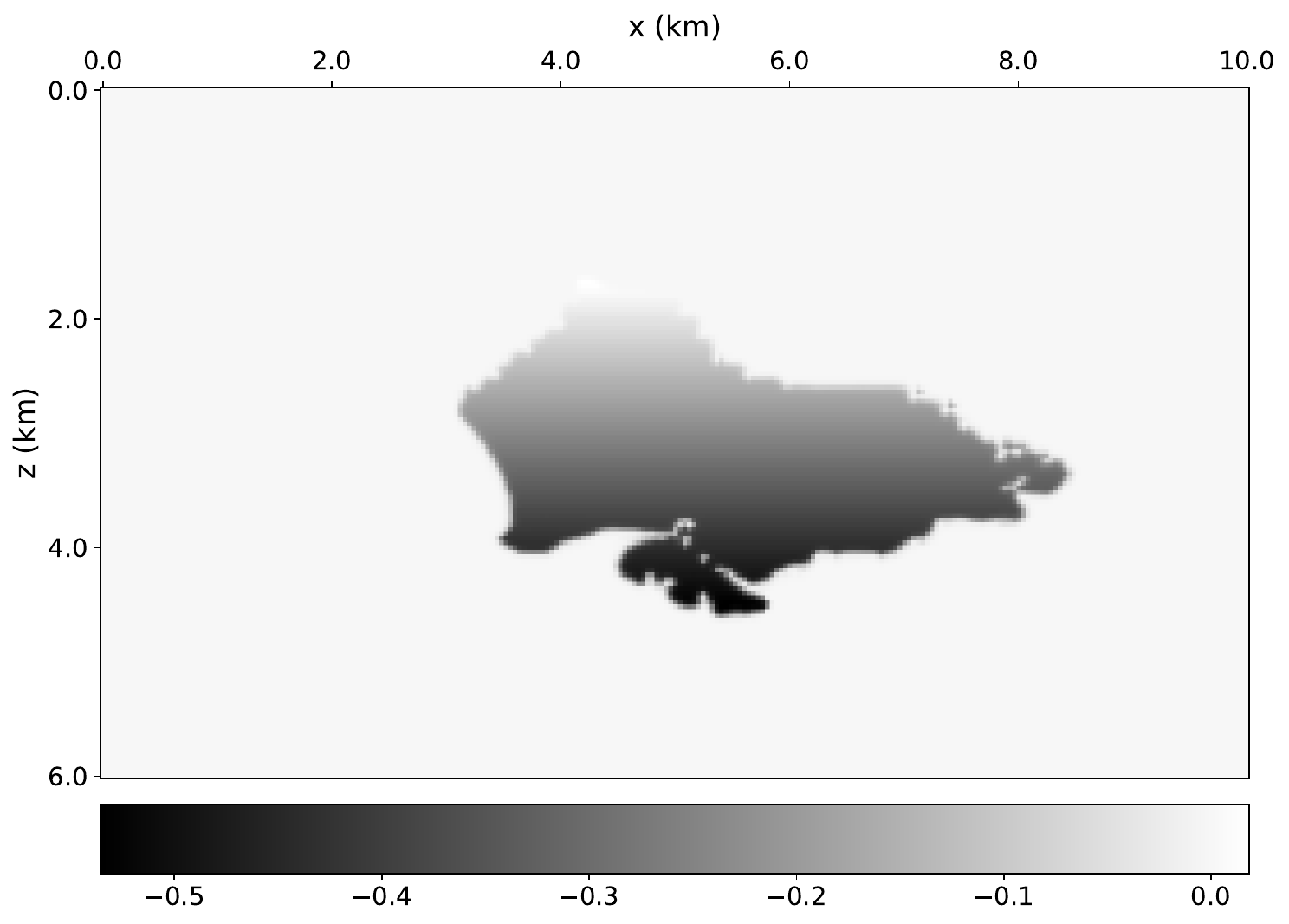}}}
\caption{Example 3: joint inversion results. (a) Initial model for velocity; (b) initial model for density; (c) recovered velocity of joint inversion from clean data; (d) recovered density of joint inversion from clean data; (e) recovered velocity from the data with $5\%$ Gaussian noise; (f)  recovered density from the data with $5\%$ Gaussian noise; (g) recovered velocity from the data with $10\%$ Gaussian noise; (h)  recovered density from the data with $10\%$ Gaussian noise.}
\label{Figexp33}
\end{figure}

\begin{figure}[htbp!]
%\captionsetup{font={color=red}}%\correct 
\centering
\correct
(a){{\includegraphics[scale=0.24,angle=0]{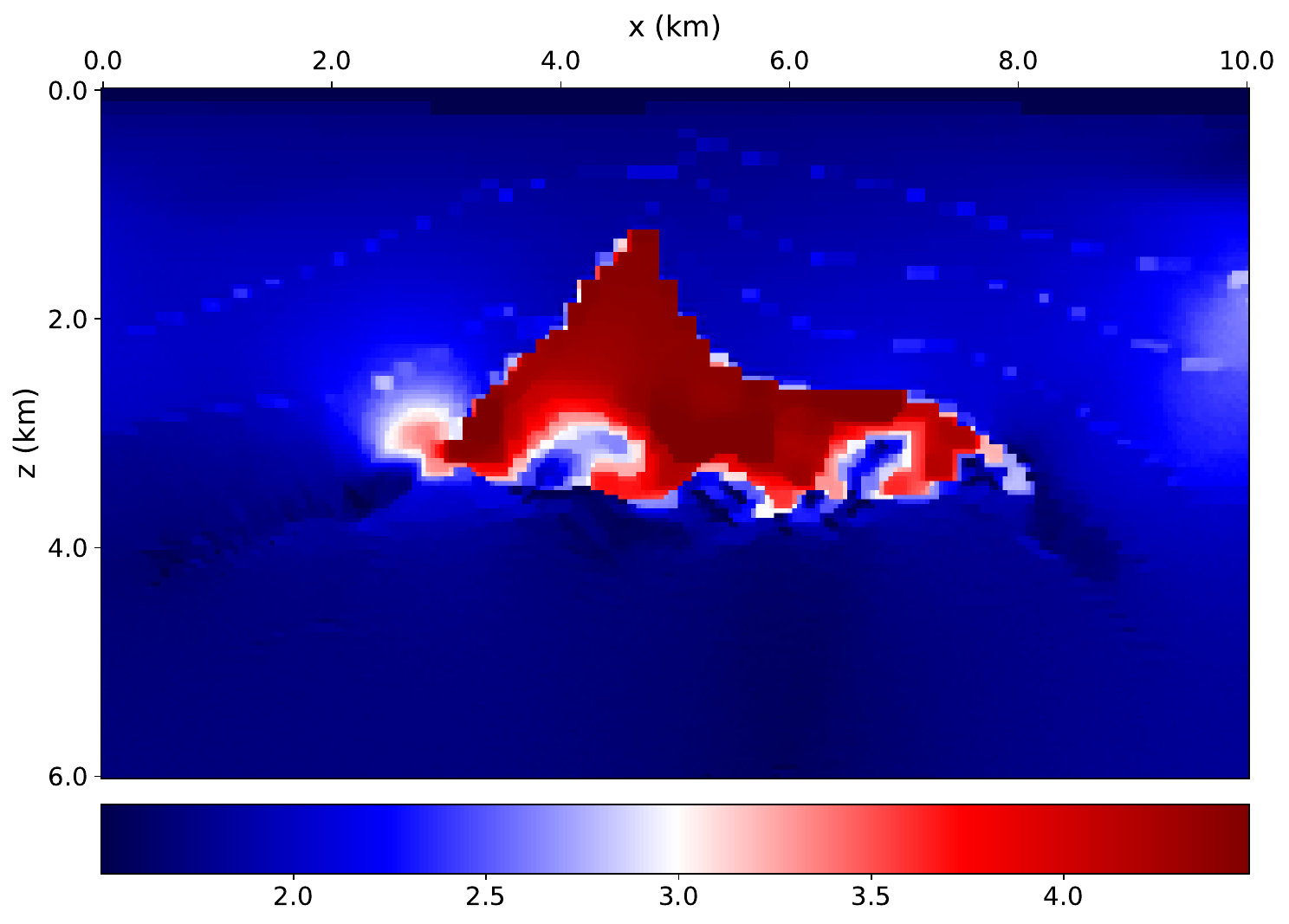}}} 
(b){{\includegraphics[scale=0.24,angle=0]{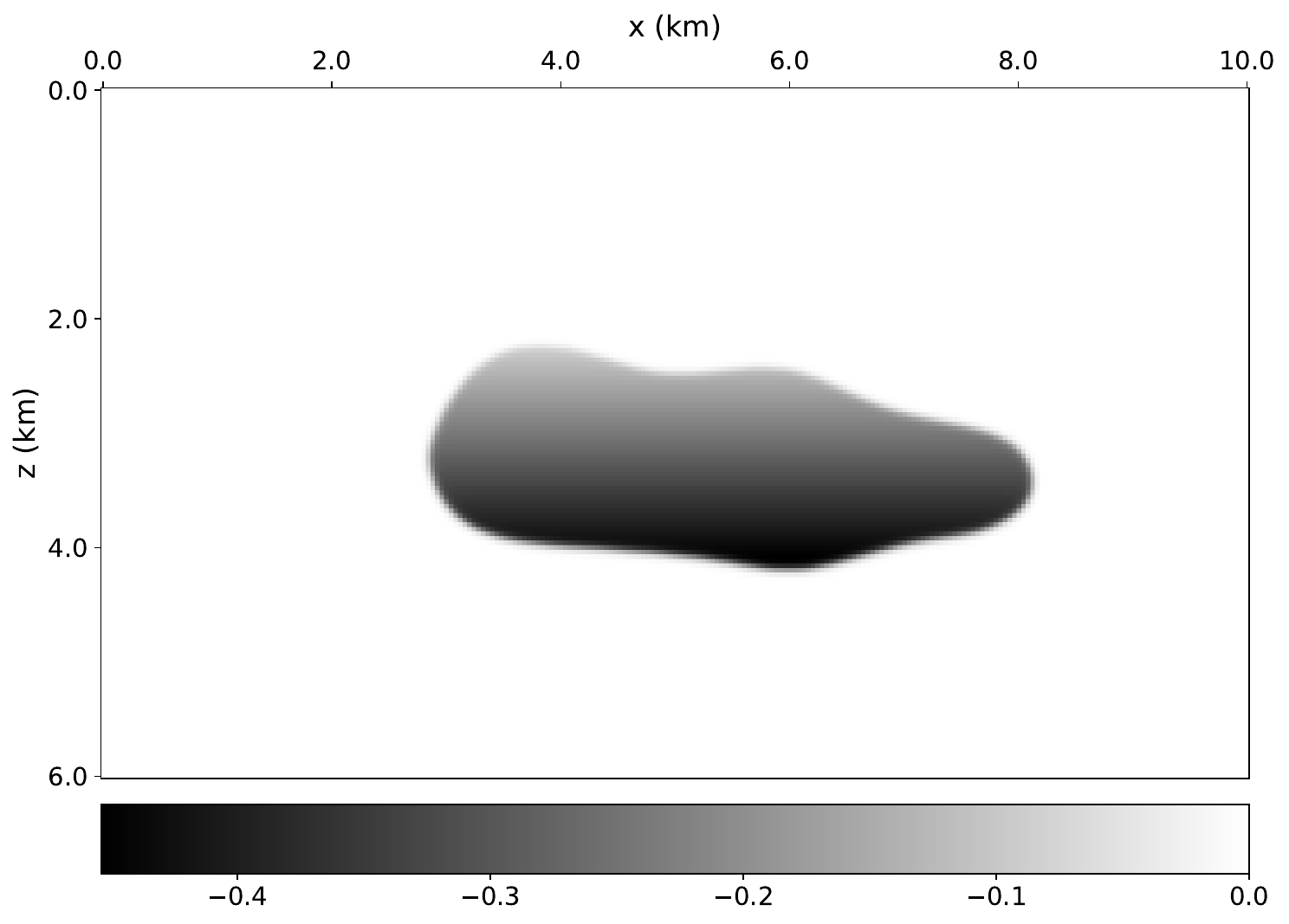}}}
(c){{\includegraphics[scale=0.24,angle=0]{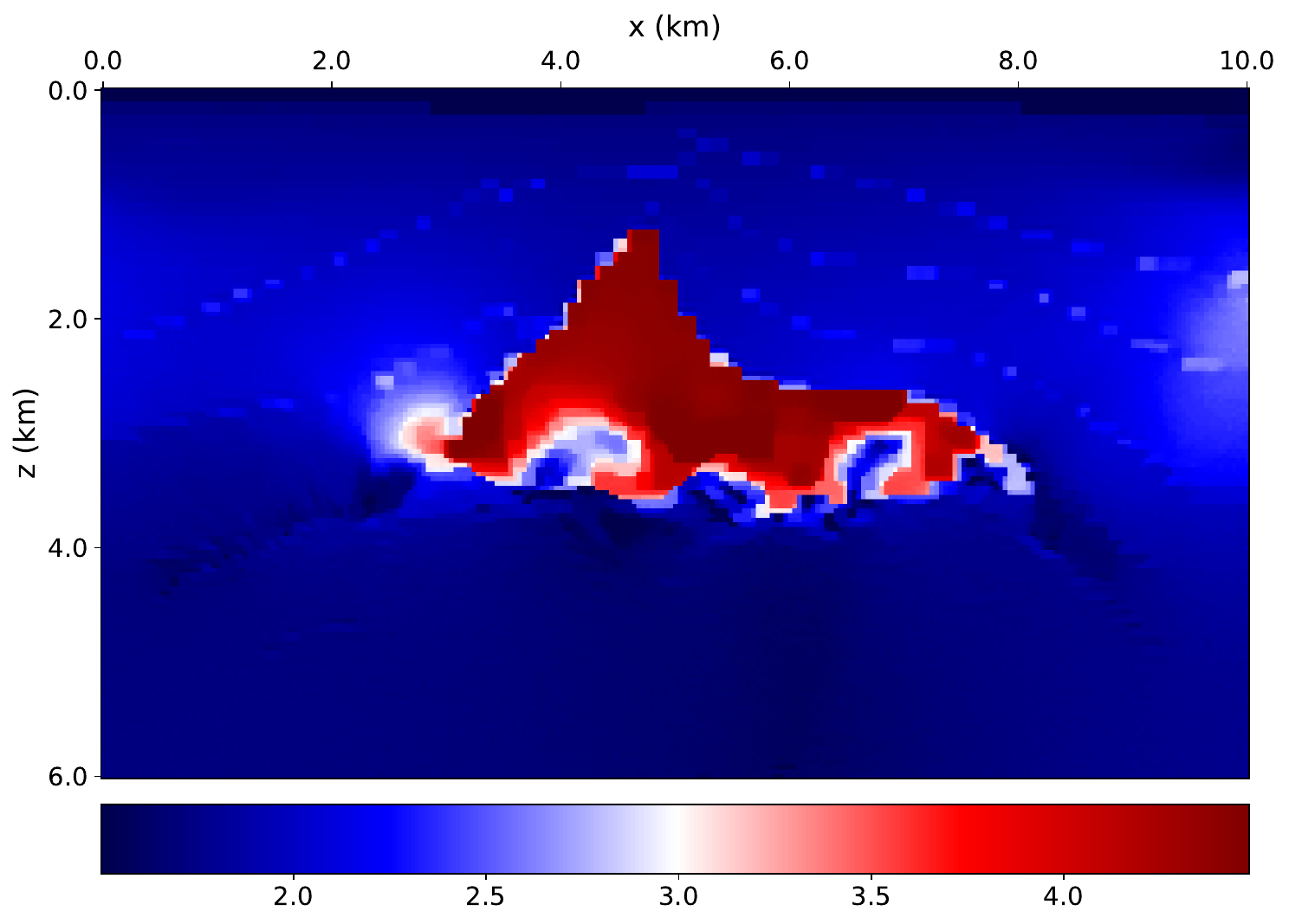}}}
(d){{\includegraphics[scale=0.24,angle=0]{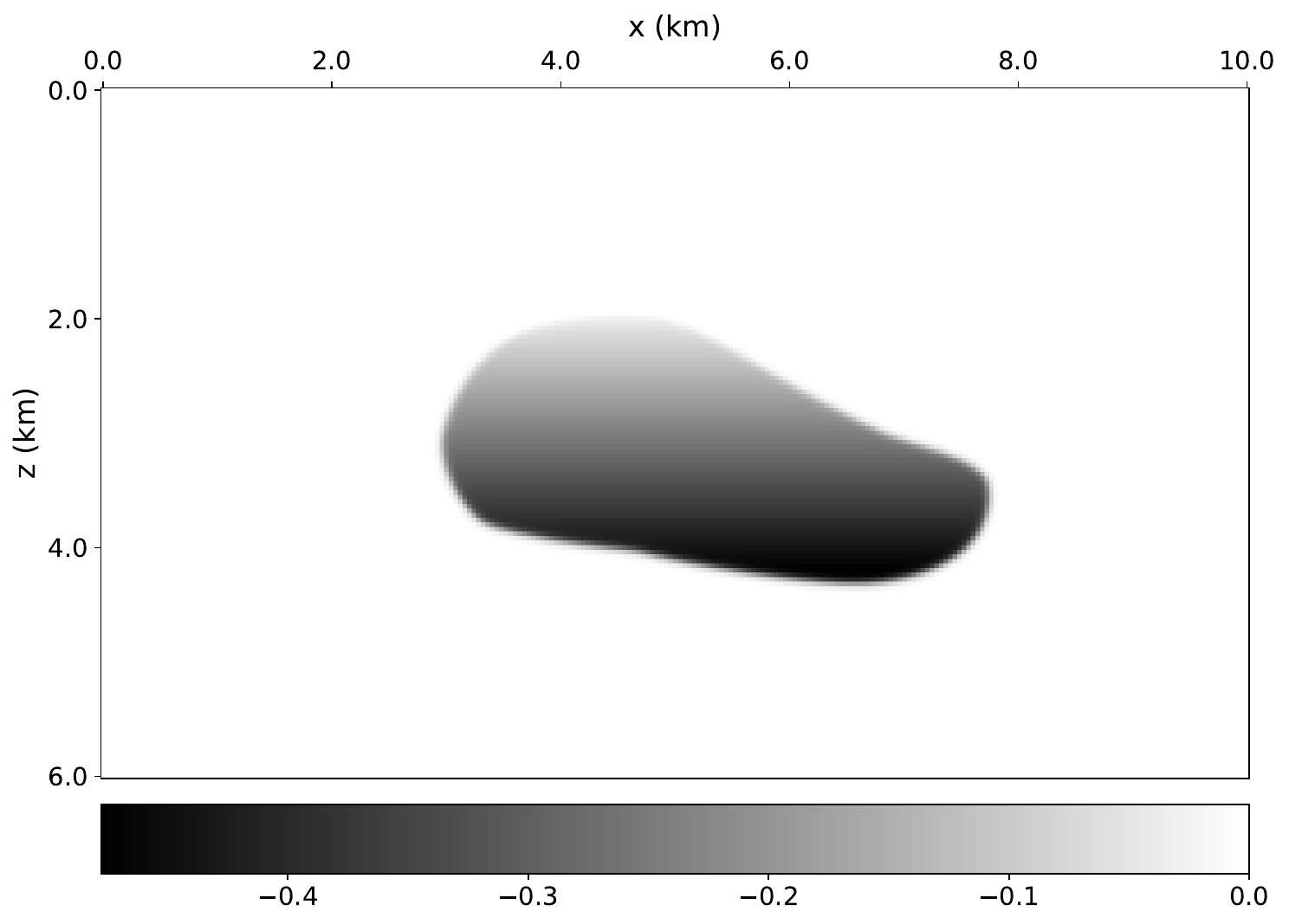}}}
(e){{\includegraphics[scale=0.24,angle=0]{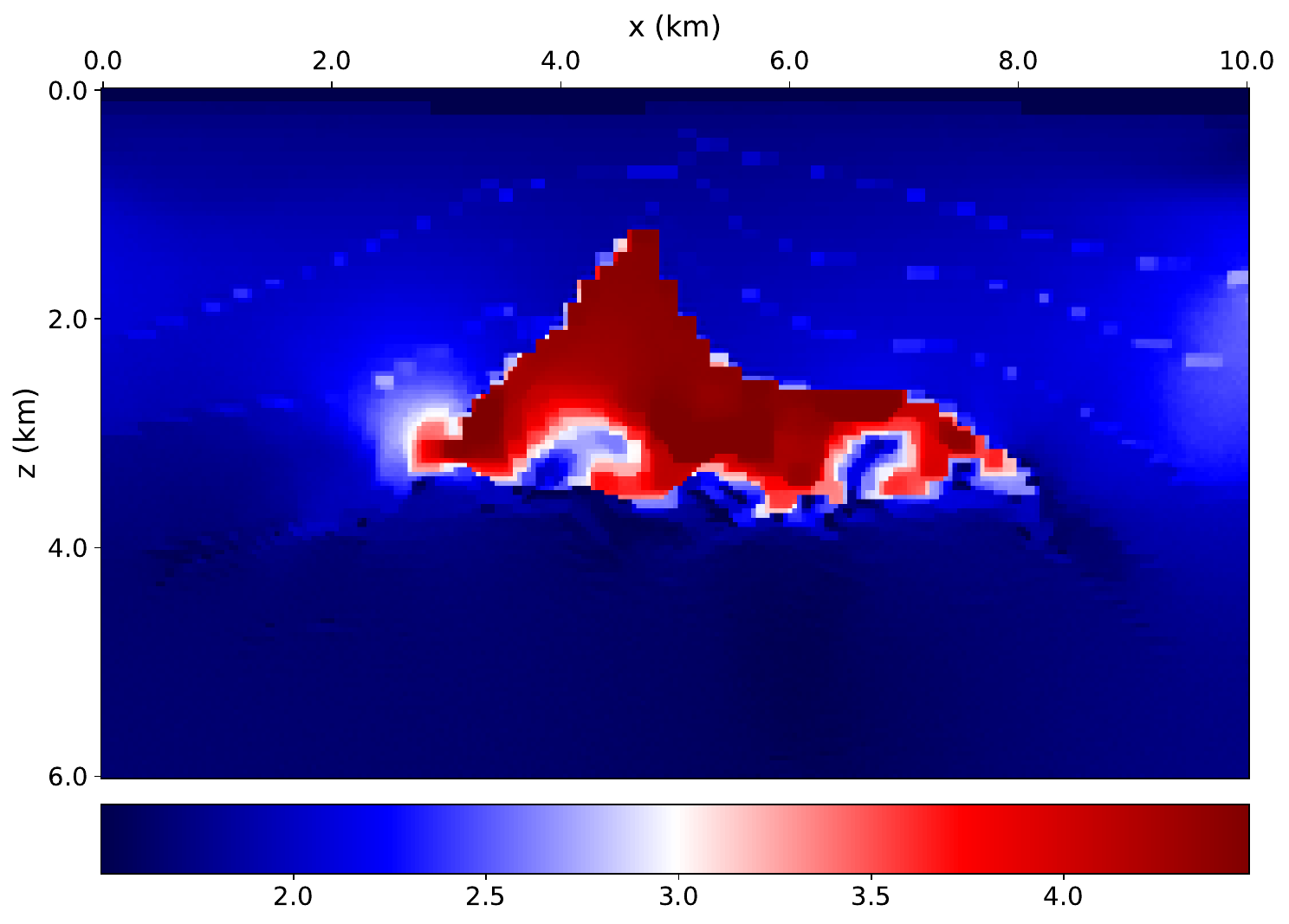}}}
(f){{\includegraphics[scale=0.24,angle=0]{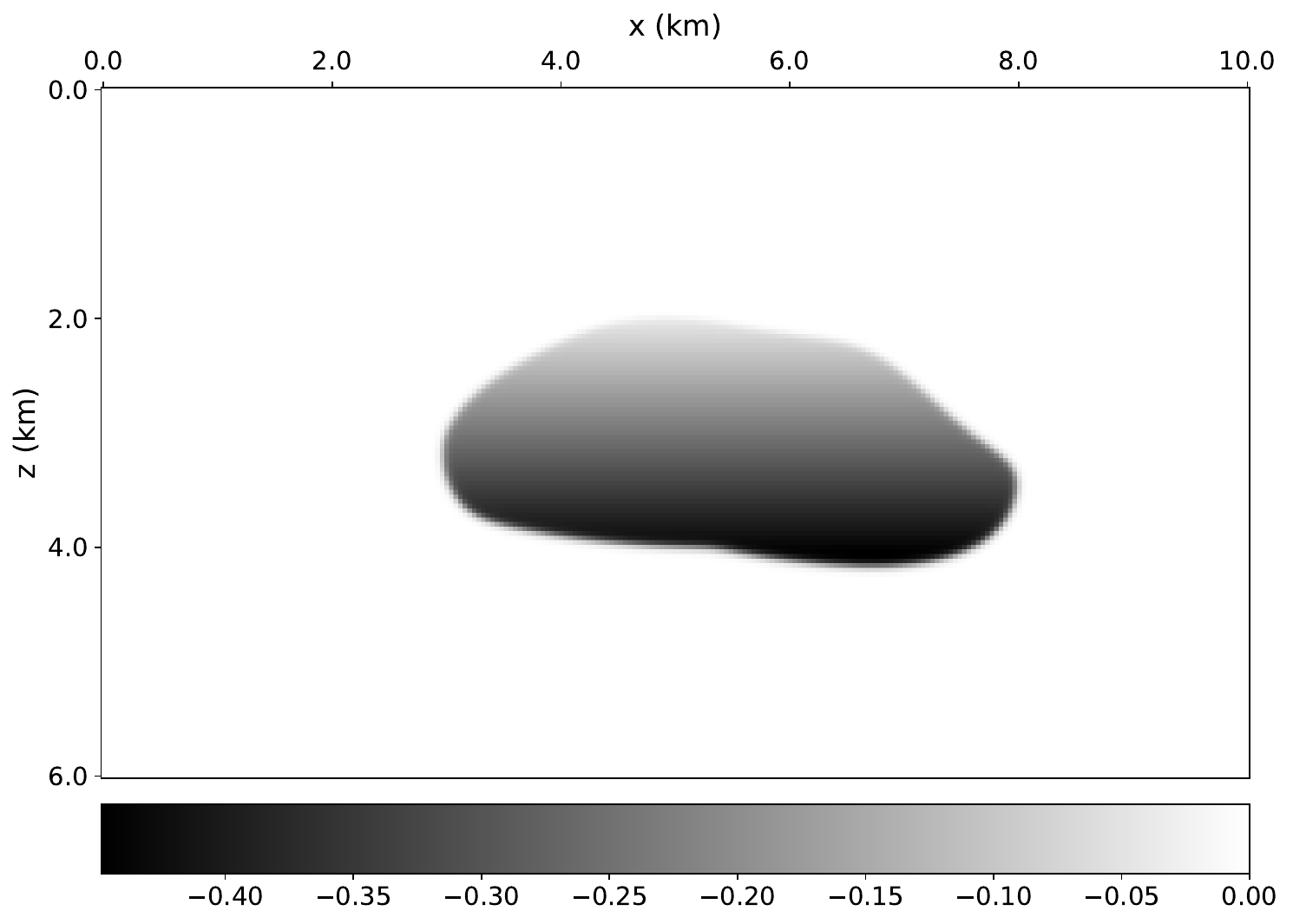}}}
\caption{Example 3: solutions of pure full-waveform inversion (FWI) and pure gravity inversion, respectively. The initial velocity guess is shown in Figure \ref{Figexp33}\,(a), and the initial density guess is shown in Figure \ref{Figexp33}\,(b). (a) FWI solution from clean data; (b) solution of gravity inversion from clean data; (c)  FWI solution from the data with $5\%$ Gaussian noise; (d) solution of gravity inversion from the data with $5\%$ Gaussian noise; (e) FWI solution from the data with $10\%$ Gaussian noise.}
\label{Figexp34}
\end{figure}

{\correct In the level-set joint inversion, we impose the density-contrast value $f(\mathbf{r})$ as a priori information, and freeze the velocity value within the salt, $c_1(\mathbf{r})=4.482$\,km/s. The level-set formulation is effective for the imposition of the prior information.
Table \ref{Tabexp3} lists the values of algorithm parameters used in the joint inversion. Figure \ref{Figexp33} provides the results, where \ref{Figexp33}\,(a) and \ref{Figexp33}\,(b) illustrate the initial guesses for velocity and density models, \ref{Figexp33}\,(c) and \ref{Figexp33}\,(d) plot the recovered solutions from clean data, and \ref{Figexp33}\,(e)\,-\,\ref{Figexp33}\,(h) plot the solutions from the noisy data ($5\%$ and $10\%$ Gaussian noise, respectively). For comparison,  Figure \ref{Figexp34} presents the solutions of pure full-waveform inversion and pure gravity inversion.}

{\correct The full-waveform inversion accurately delineates the top of the salt model but fails to resolve its wide body at depth. Conversely, the gravity inversion successfully recovers the overall profile of the salt model, although the detailed shape is inaccurate. The level-set joint inversion combines the advantages of both FWI and gravity inversion,  producing superior solutions that closely match the true model. Even with significant noise in the gravity data, while the recovered density model may appear incomplete (e.g. Figure \ref{Figexp33}\,(h)), the velocity model remains nearly correct. Thus, in the joint inversion, the velocity and density solutions provide complementary information for a complete understanding of the salt structure.}

\begin{table}[h]
%		\captionsetup{labelfont={color=red}} %\correct
		\centering
		\correct
		\begin{tabular}{ccccccc} 
			\toprule
			$\epsilon$      &  $\alpha_\phi$  &  $\alpha_{v_2}$   &  $\omega_0$    &  $\lambda$  & $\lambda_\phi$ &  $\lambda_{c_2}$\\
			\midrule
			0.05       & 1 & 1  & 5  &  $\displaystyle \frac{\mathrm{ln}50}{n_{\mathrm{max}}}$ ($n_{\mathrm{max}}=2\times10^4$)  &  \makecell{for clean data: $2\times10^{-6}$ \\  for noisy data: $2\times10^{-5}$} &  \makecell{for clean data: $2\times10^{-5}$ \\  for noisy data: $1\times10^{-4}$}               \\
			\bottomrule
		\end{tabular}\caption{\correct Example 3: values of algorithm parameters used in the joint inversion.}
		\label{Tabexp3}
\end{table}
\vspace{-8pt}

\subsubsection{Example 4. }
\begin{figure}[htbp!]
\centering
(a){{\includegraphics[scale=0.24,angle=0]{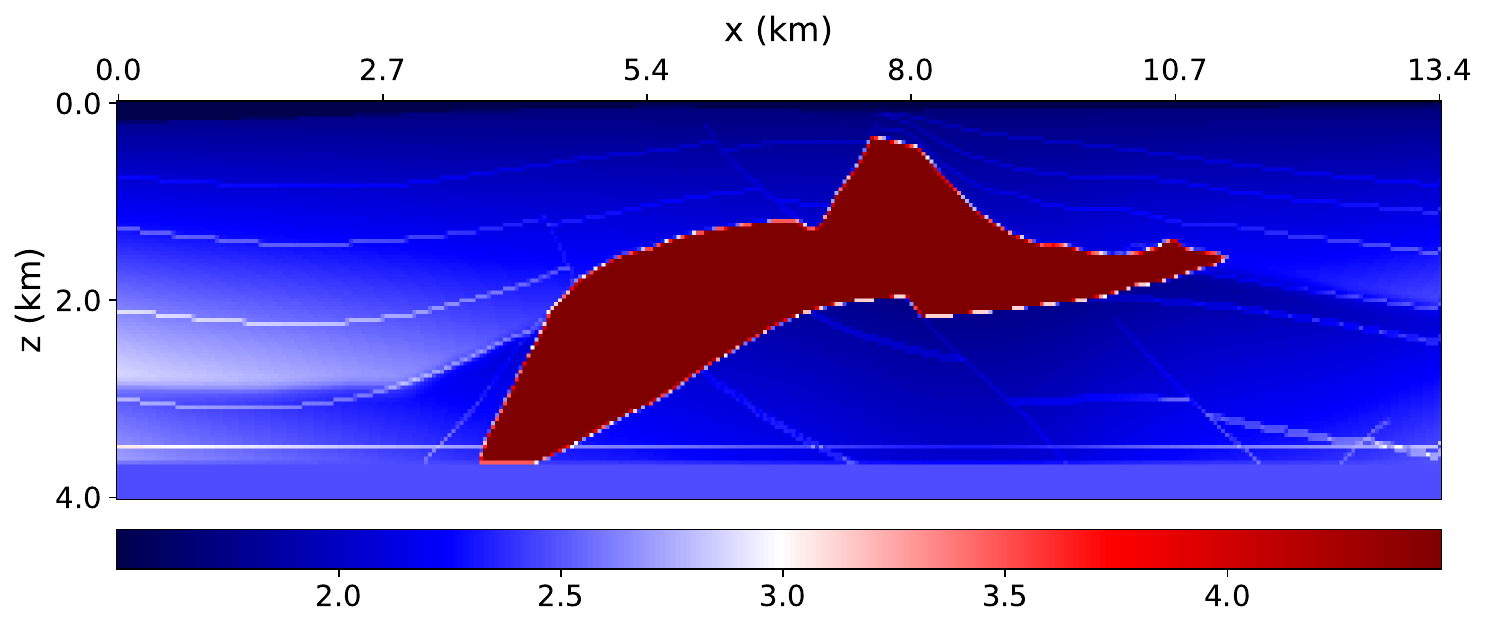}}}
(b){{\includegraphics[scale=0.24,angle=0]{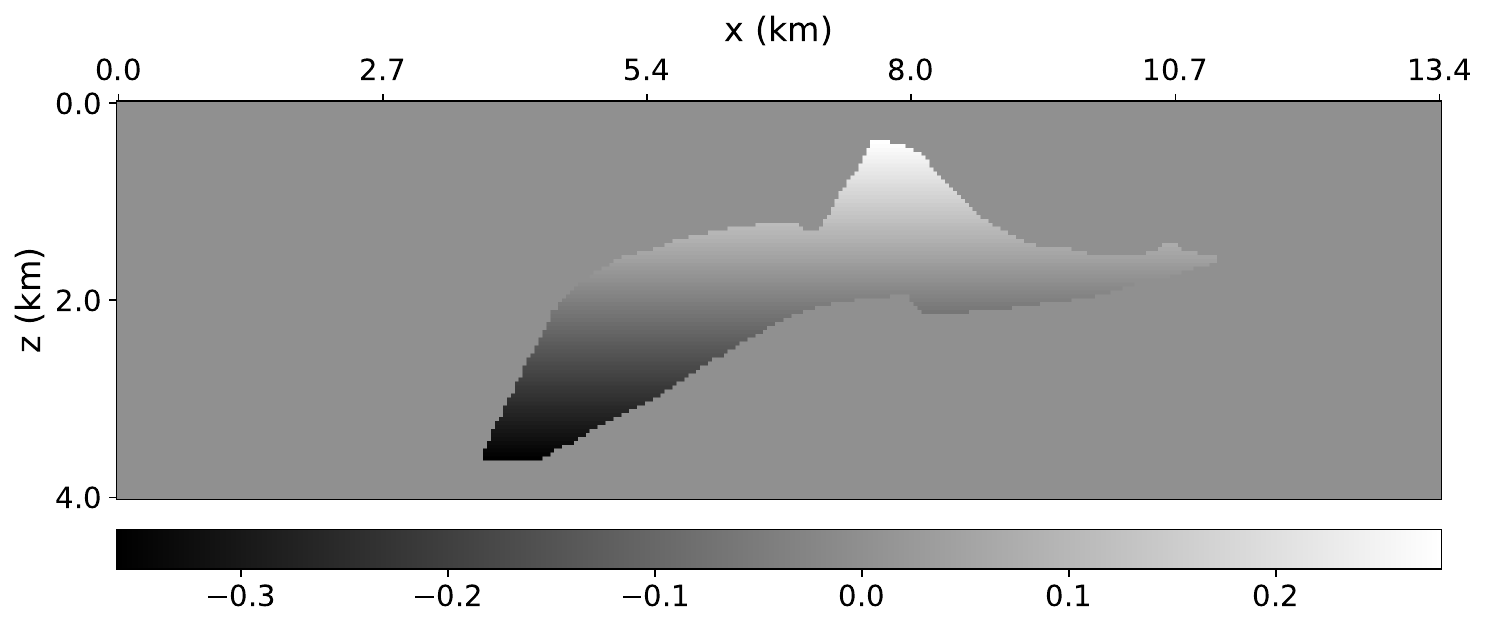}}}
\caption{Example 4: 2D SEG/EAGE salt model. (a) True velocity model; (b) true density model.}
\label{Fig8}
\end{figure}

In this example, we consider a 2D SEG/EAGE salt model. Figure \ref{Fig8}\,(a) shows the velocity model. It is a cross-section of the 3D SEG/EAGE salt model \cite{amibrakun97}, where the slice is taken along the plane defined by three points \cite{liluqia16}: $(0,2.2,0)$\,km, $(13.4,6.6,0)$\,km, $(13.4,6.6,4)$\,km. To build the density model, we extract the salt structure, and impose a density-contrast value $f(\mathbf{r})=(1.8-z)\times 0.2$\,g/cm$^3$ \cite{liluqia16}; Figure \ref{Fig8}\,(b) plots the density model.

\begin{figure}[htbp!]
\centering
(a){{\includegraphics[scale=0.24,angle=0]{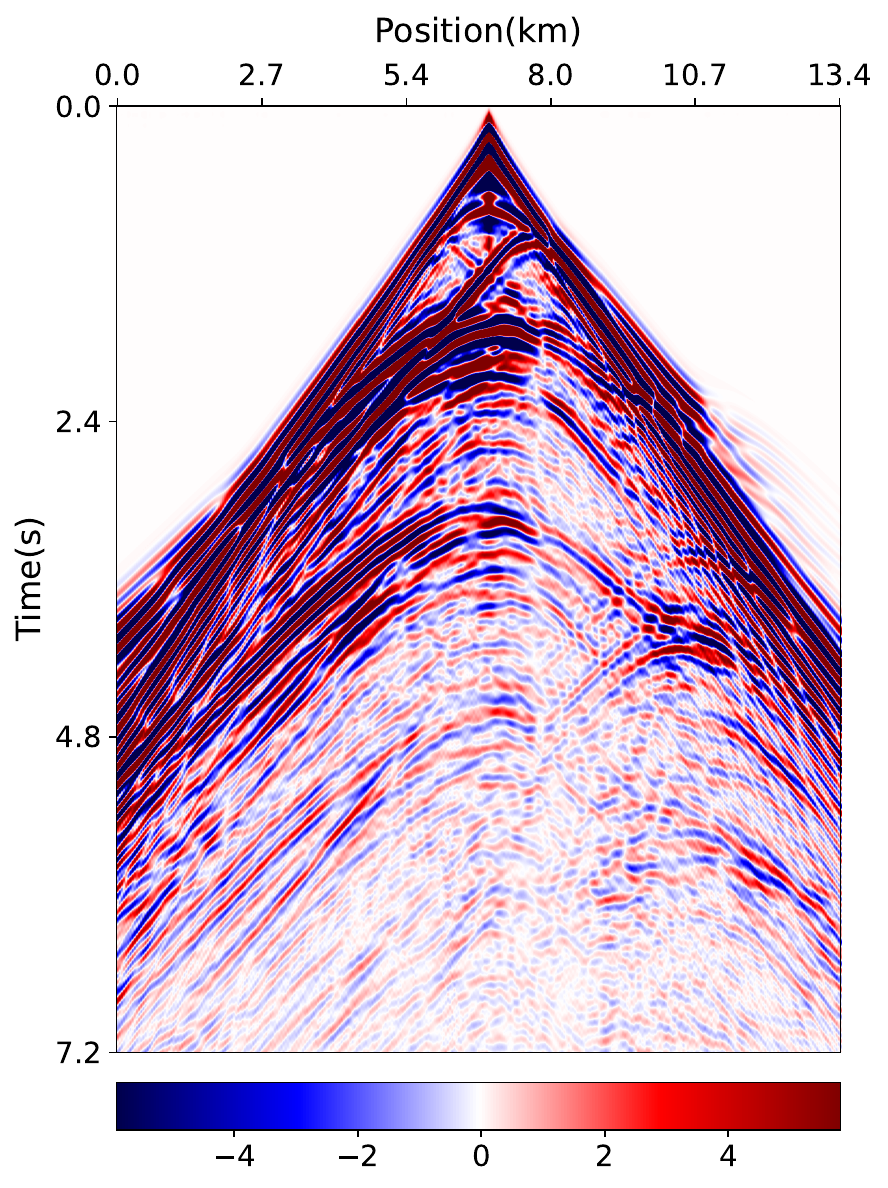}}} 
(c){{\includegraphics[scale=0.24,angle=0]{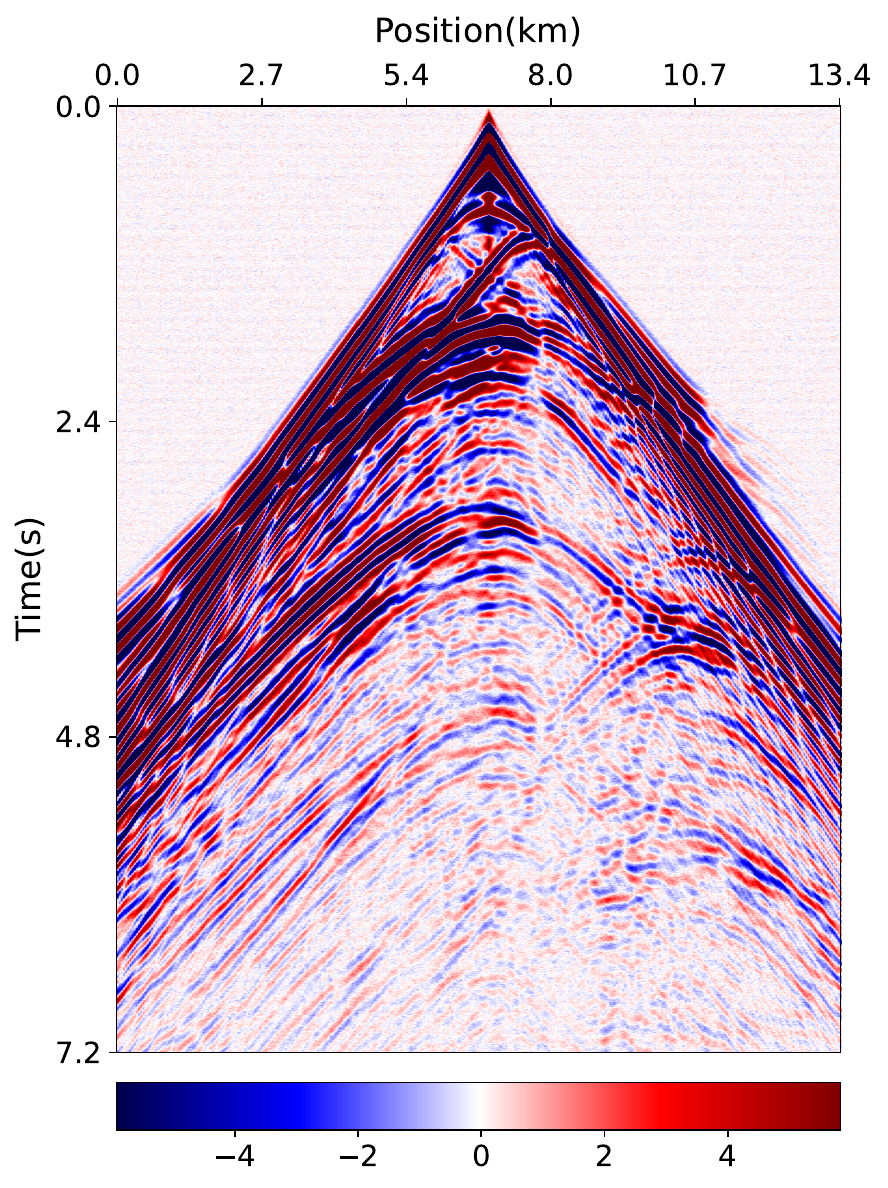}}}\\
(b){{\includegraphics[scale=0.24,angle=0]{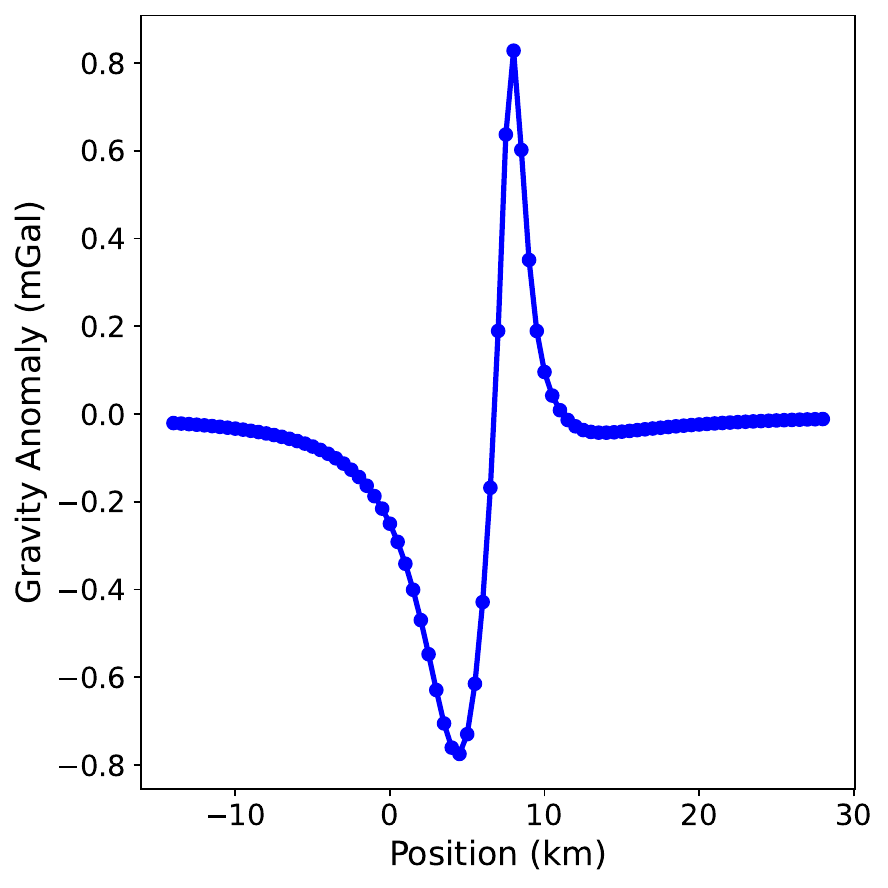}}}
(d){{\includegraphics[scale=0.24,angle=0]{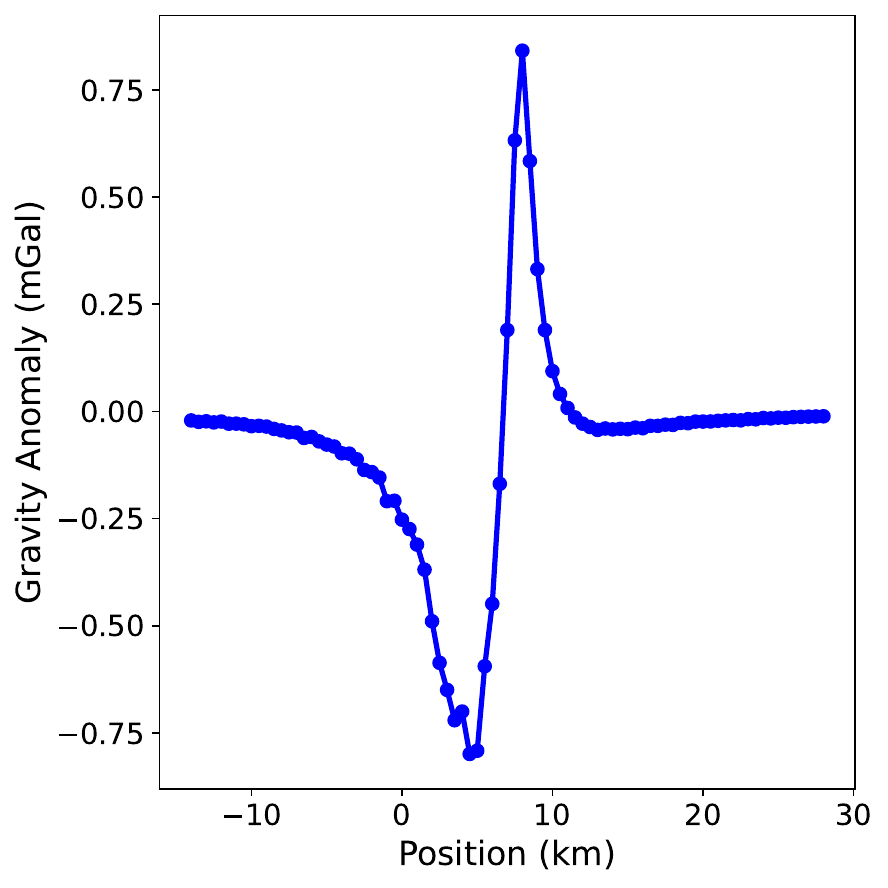}}}
\caption{Example 4: measurement data. The color scale of waveform data is clipped between its $5\%$ and $95\%$ quantiles for enhanced visualization. (a) waveform data for the 16\,th source of 30; (b) gravity data $g_z$; (c) waveform data with $5\%$ Gaussian {\correct noise}; (d) gravity data with $5\%$ Gaussian {\correct noise}.}
\label{Fig9}
\end{figure}

The computational domain is $\Omega=[0,13.4]\times[0,4]$\,km, {\correct with the 2D spatial coordinate denoted by $\mathbf{r}=(x,z)$.} The seismic sources and receivers are located along $z=0$\,km; there are 30 point sources with $x$-coordinates $x=0:0.46:13.34$\,km, and 336 receivers with $x$-coordinates $x=0:0.04:13.4$\,km.
{\correct The source wavelet is the high-pass Ricker wavelet, as shown in Figure \ref{Fig_RickerWavelet} by the black solid line. When solving the wave equation, we use a time step size of $\Delta t=0.004$\,s and a spatial mesh size of $h=0.04$\,km; it satisfies the CFL condition for the 2D wave equation, i.e., $\Delta t \le \frac{1}{c_{\mathrm{max}}}\frac{h}{\sqrt{2}}$.}
The total recording time is $7.2$\,s, with a sampling interval of $0.004$\,s. Figure \ref{Fig9}\,(a) shows the waveform data for the 16\,th source of 30, and Figure \ref{Fig9}\,(c) plots the data with $5\%$ Gaussian {\correct noise}, where the noisy data are simulated by equation (\ref{eqn34}).  The gravity data are acquired along $z=-0.1$\,km, and {\correct there are 85 measurements with $x$-coordinates $x=-14:0.5:28$\,km.} Figure \ref{Fig9}\,(b) plots the gravity data $g_z$; Figure \ref{Fig9}\,(d) plots the data with $5\%$ Gaussian {\correct noise}, where the noisy data are simulated by equation (\ref{eqn35}).

\begin{figure}[htbp!]
\centering
(a){{\includegraphics[scale=0.24,angle=0]{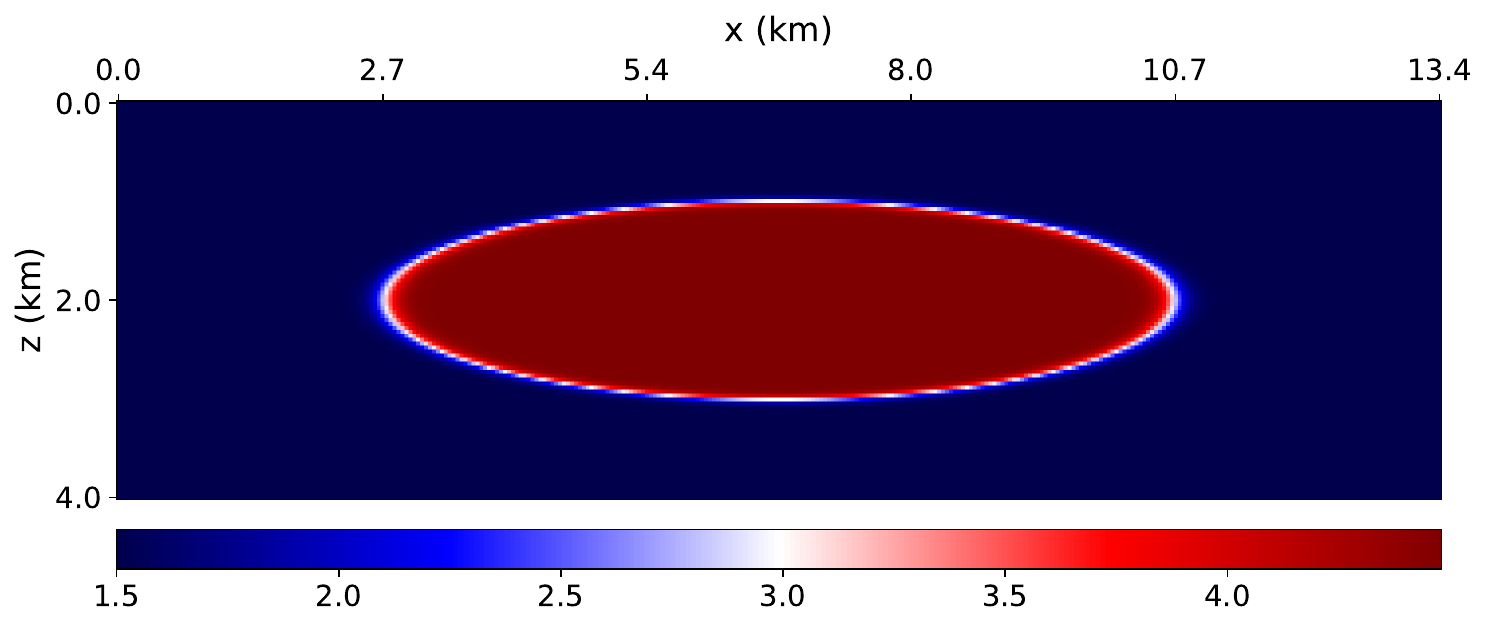}}}
(b){{\includegraphics[scale=0.24,angle=0]{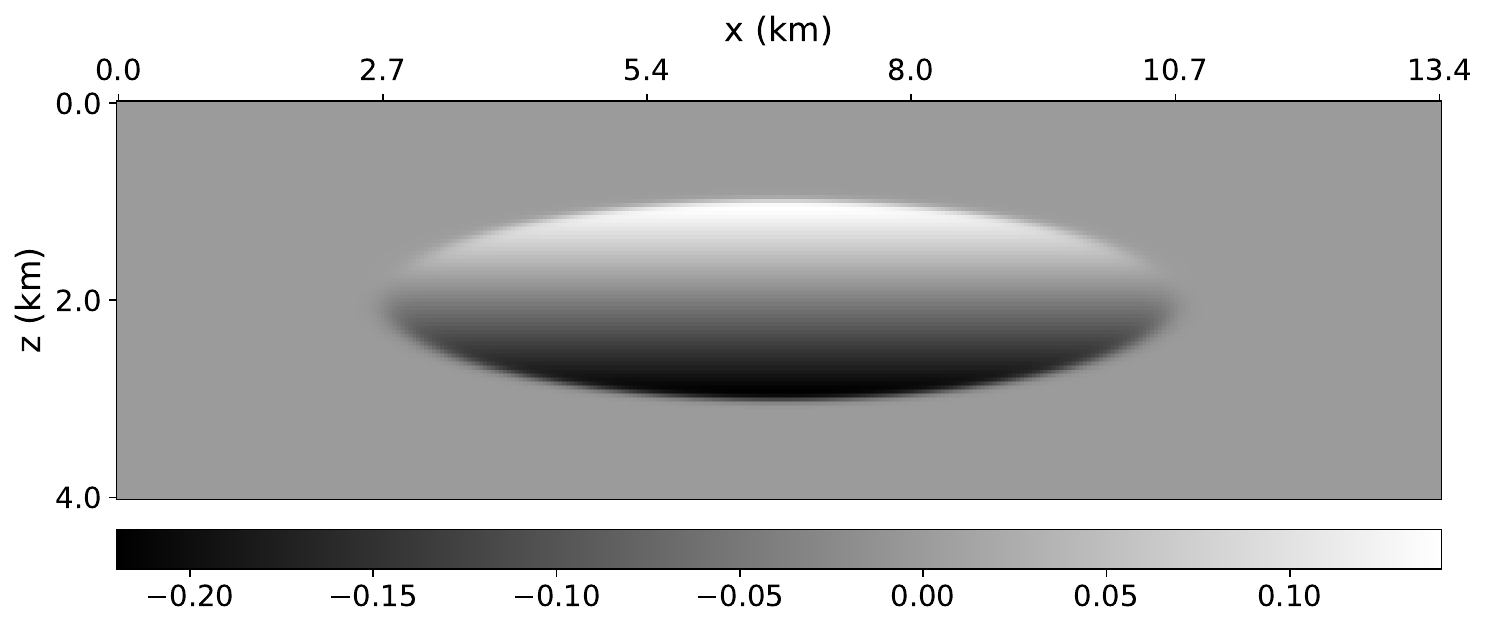}}}
(c){{\includegraphics[scale=0.24,angle=0]{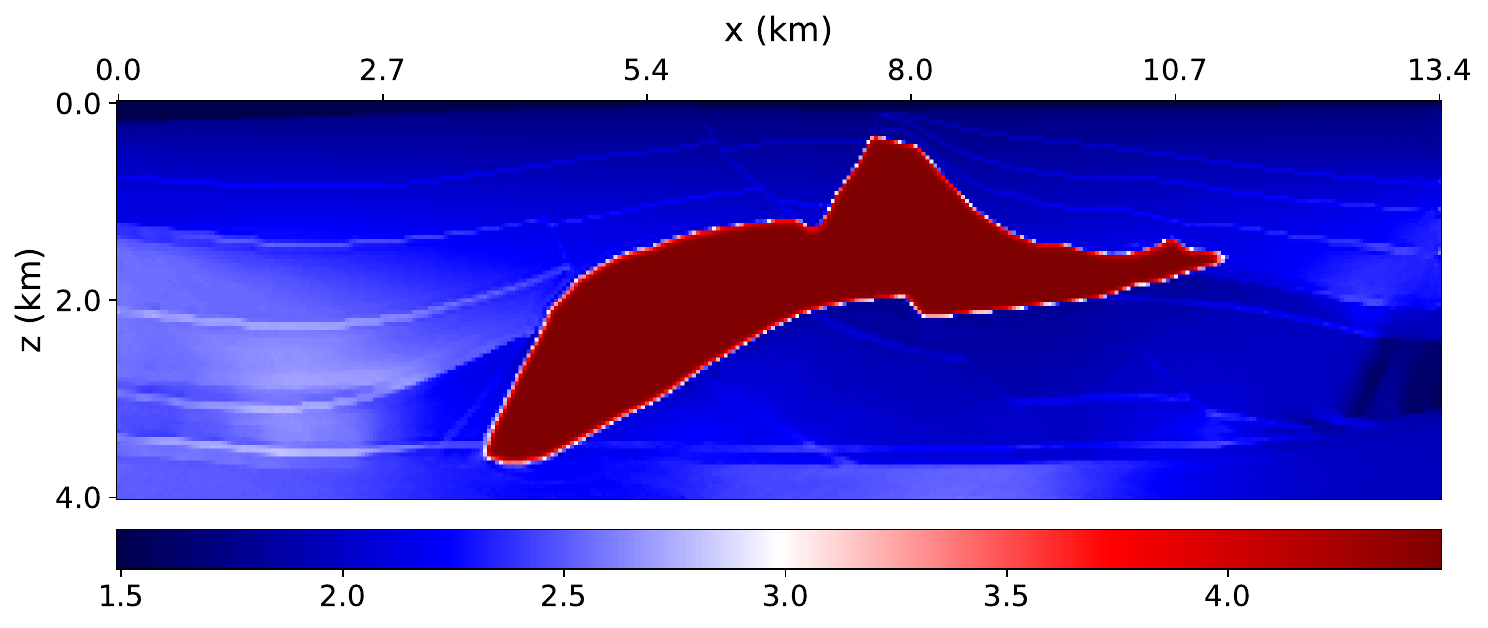}}} 
(d){{\includegraphics[scale=0.24,angle=0]{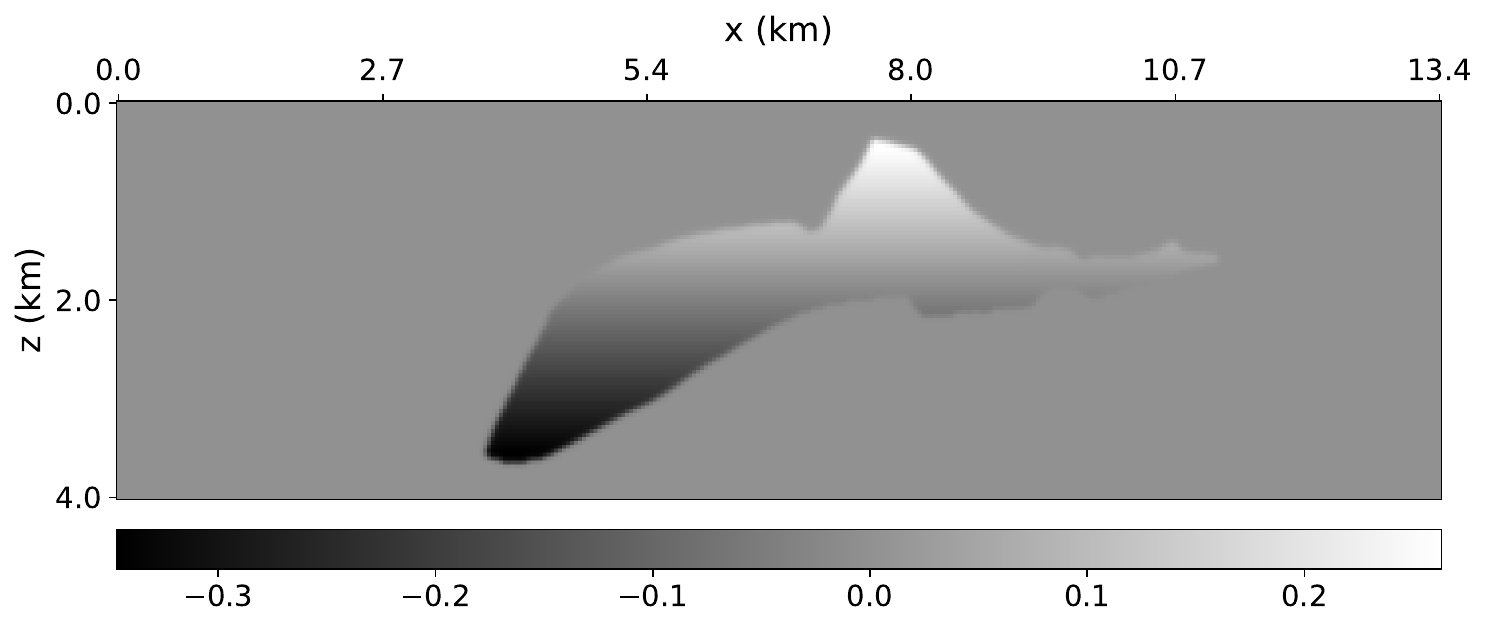}}}
(e){{\includegraphics[scale=0.24,angle=0]{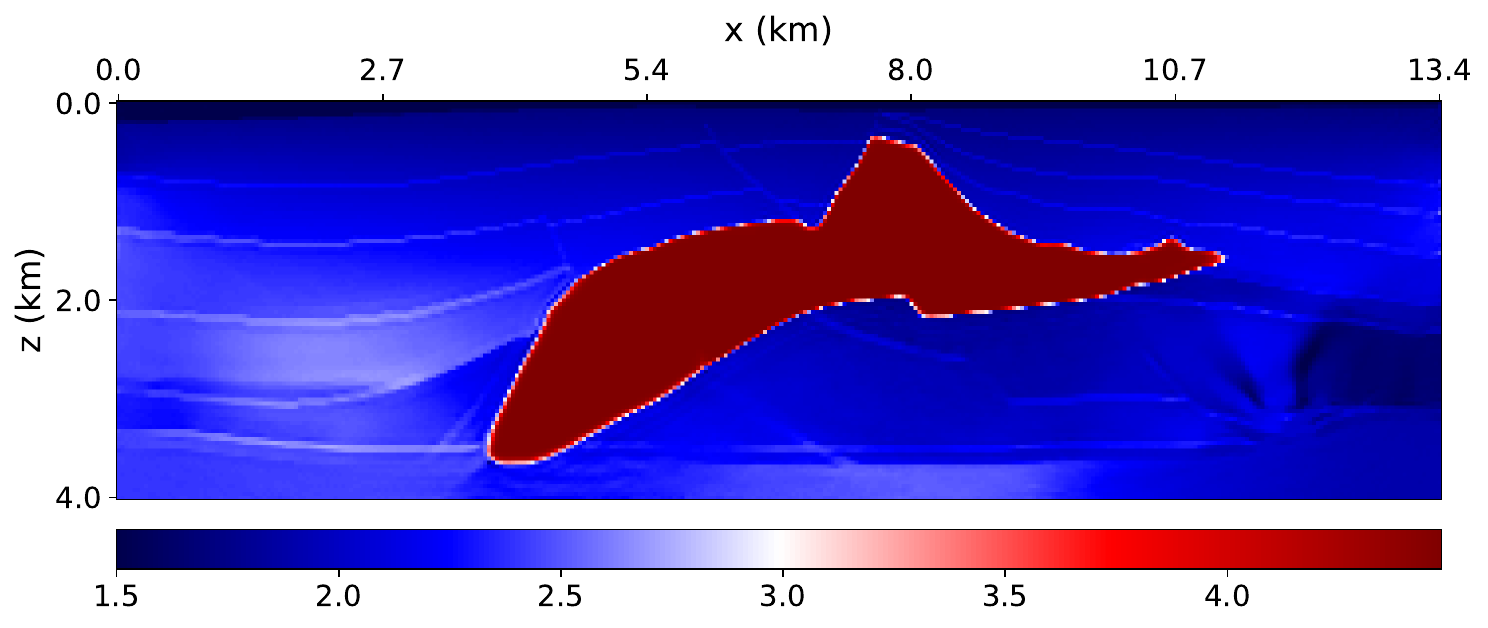}}}
(f){{\includegraphics[scale=0.24,angle=0]{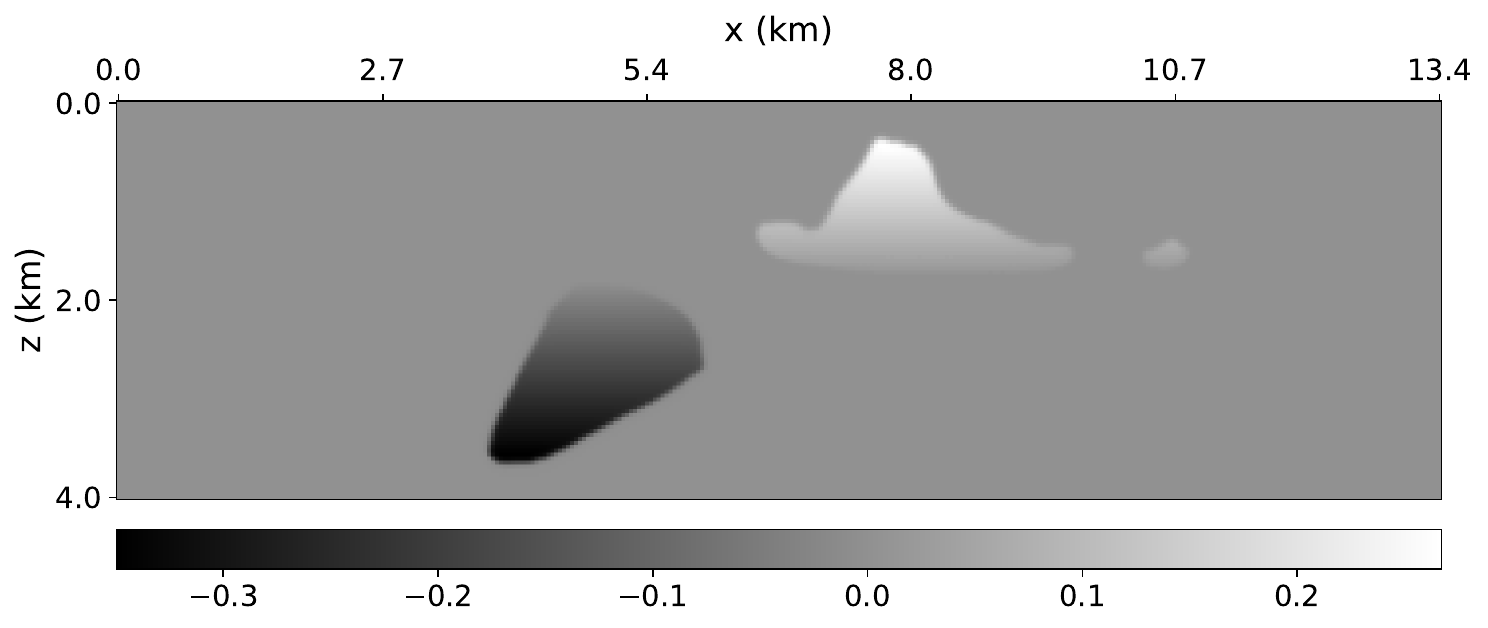}}}
\caption{Example 4: joint inversion results. (a) Initial model for velocity; (b) initial model for density; (c) recovered velocity of joint inversion from clean data; (d) recovered density of joint inversion from clean data; (e) recovered velocity from the data with $5\%$ Gaussian {\correct noise}; (f)  recovered density from the data with $5\%$ Gaussian {\correct noise}.}
\label{Fig10}
\end{figure}

\begin{figure}[htbp!]
\centering
(a){{\includegraphics[scale=0.24,angle=0]{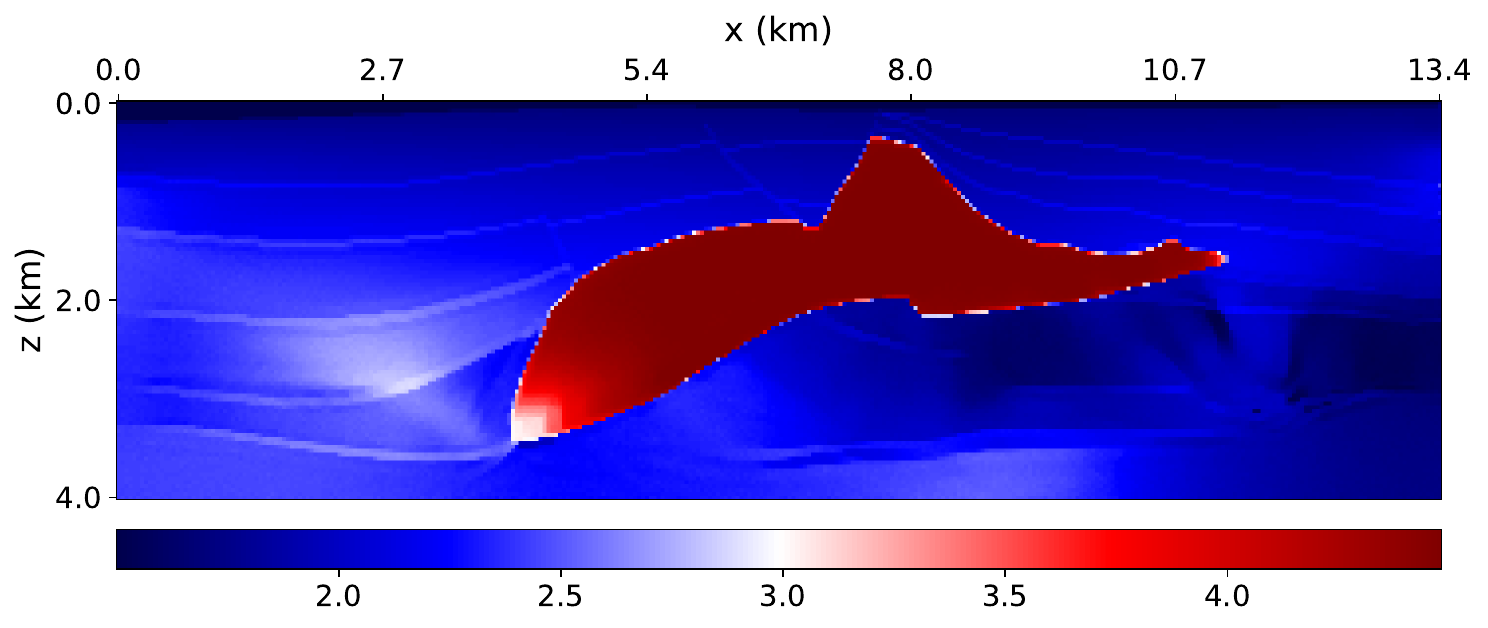}}} 
(b){{\includegraphics[scale=0.24,angle=0]{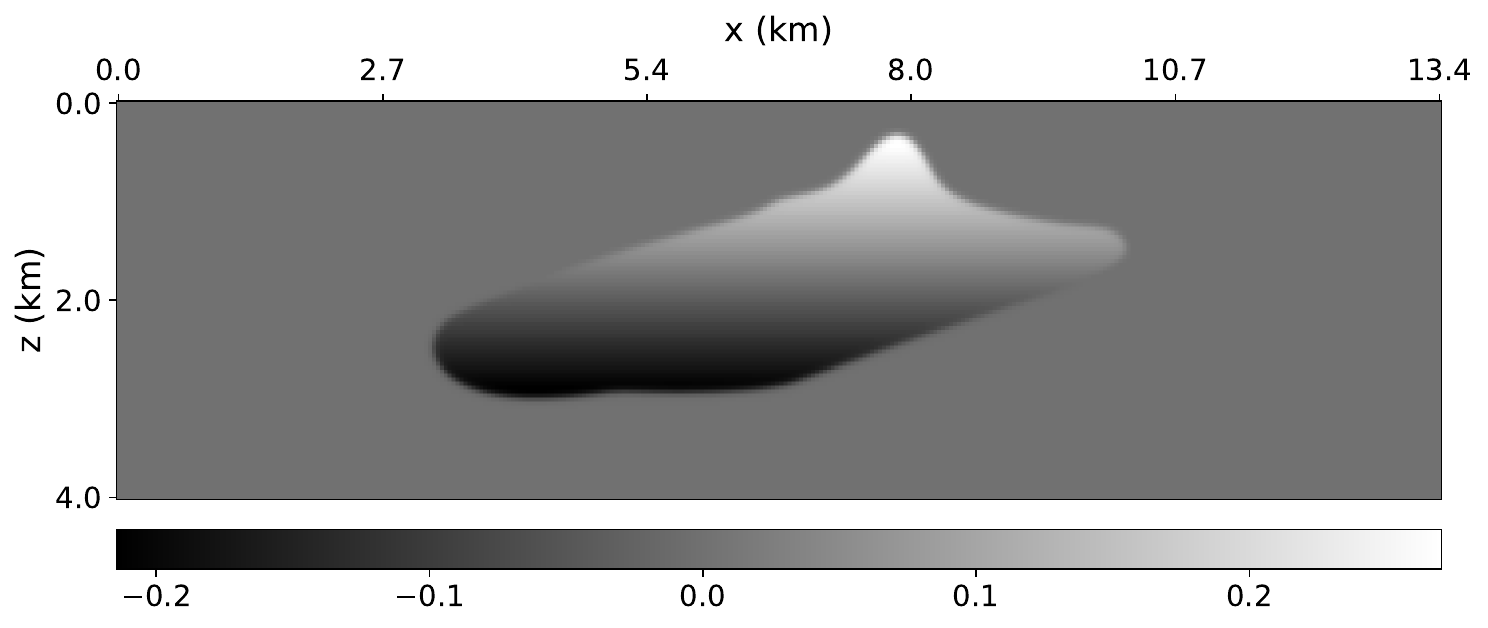}}}
(c){{\includegraphics[scale=0.24,angle=0]{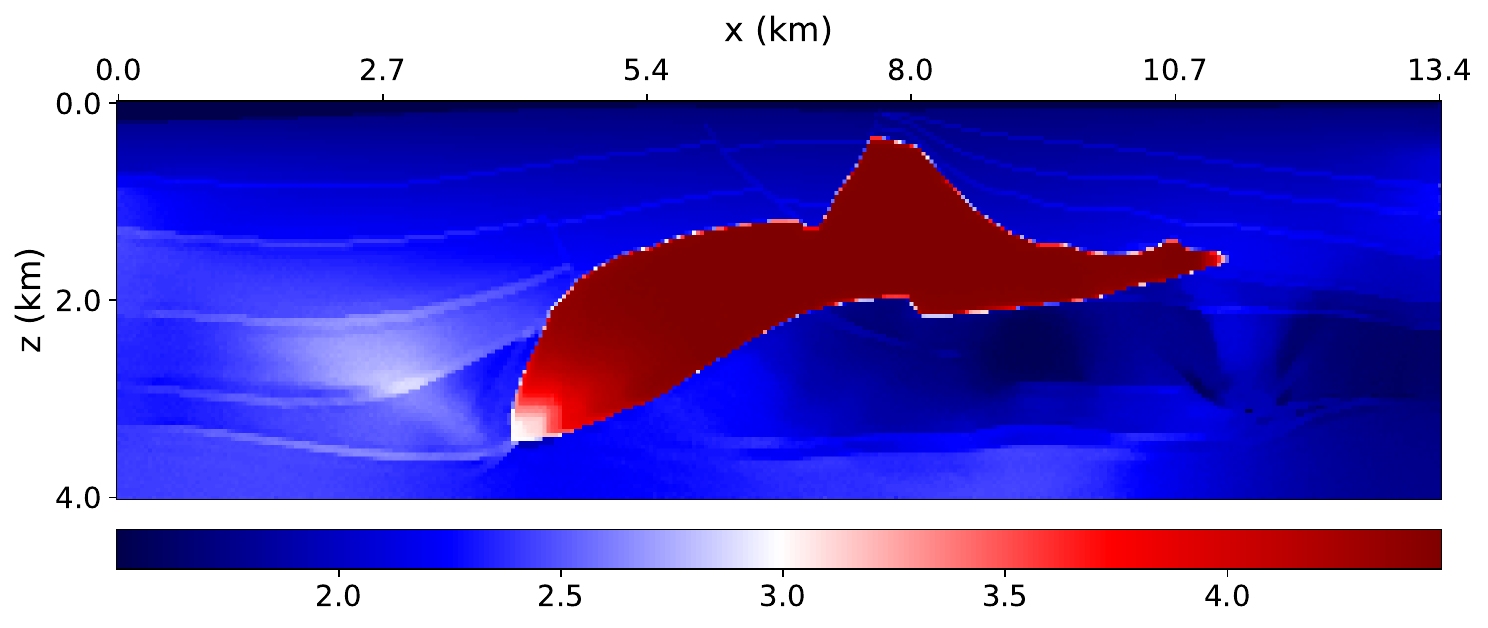}}}
(d){{\includegraphics[scale=0.24,angle=0]{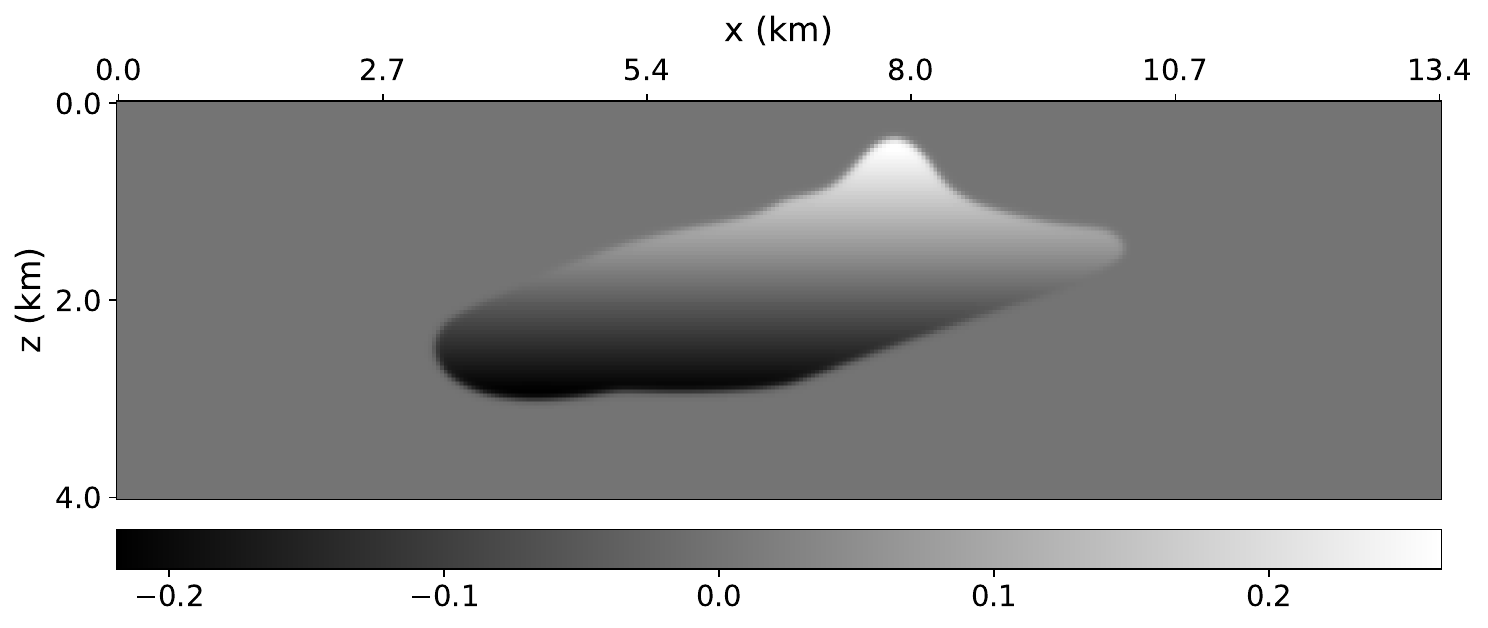}}}
\caption{Example 4: solutions of pure full-waveform inversion and pure gravity inversion, respectively. (a) Recovered velocity of FWI from clean data; (b) recovered density of gravity inversion from clean data; (c) recovered velocity of FWI from the data with $5\%$ Gaussian {\correct noise}; (d)  recovered density of gravity inversion from the data with $5\%$ Gaussian {\correct noise}.}
\label{Fig11}
\end{figure}

In the level-set joint inversion, we impose the density-contrast value $f(\mathbf{r})$ as a priori information; for simplicity, we further freeze the velocity value within the salt, $c_1(\mathbf{r})=4.482$\,km/s. Table \ref{Tab4} illustrates the values of algorithm parameters used in the joint inversion. Figure \ref{Fig10} provides the results, where \ref{Fig10}\,(a) and \ref{Fig10}\,(b) illustrate the initial guesses for velocity and density models, \ref{Fig10}\,(c) and \ref{Fig10}\,(d) plot the recovered solutions from clean data, and \ref{Fig10}\,(e) and \ref{Fig10}\,(f) plot the solutions from the data with $5\%$ Gaussian {\correct noise}. For comparison,  Figure \ref{Fig11} presents the solutions of pure full-waveform inversion and pure gravity inversion, respectively. To quantitatively assess the performance of joint inversion, we compute the structural similarity index (SSIM) for the recovered velocity models; Table \ref{Tab5} lists the SSIM values. 

We conclude that the joint inversion achieves good resolution for the salt model. This shallow-region scenario is actually favorable for full-waveform inversion; as shown in Figures \ref{Fig11}\,(a) and \ref{Fig11}\,(c), the solutions of single FWI are not bad. The level-set joint inversion further improves the quality of results.
When the data are contaminated by {\correct noise}, we employ a larger regularization parameter for the level-set function ($\lambda_\phi=5\times10^{-5}$), potentially leading to the distortion of recovered density in the joint inversion, as shown in Figure \ref{Fig10}\,(f); however, the velocity model is well recovered. It demonstrates that the velocity and density functions provide complementary information for understanding the subsurface structure in the joint inversion.

\begin{table}[!h]
		\centering
		\begin{tabular}{ccccccc} 
			\toprule
			$\epsilon$      &  $\alpha_\phi$  &  $\alpha_{v_2}$   &  $\omega_0$    &  $\lambda$  & $\lambda_\phi$ &  $\lambda_{c_2}$\\
			\midrule
			0.05       & 1 & 1  & 5  &  $\displaystyle \frac{\mathrm{ln}50}{n_{\mathrm{max}}}$ ($n_{\mathrm{max}}=2\times10^4$)  &  \makecell{for clean data: $2\times10^{-6}$ \\  for noisy data: $2\times10^{-5}$} &  $2\times10^{-5}$               \\
			\bottomrule
		\end{tabular}\caption{Example 4: values of algorithm parameters used in the joint inversion.}
		\label{Tab4}
\end{table}
\vspace{-8pt}

\begin{table}[!h]
		\centering
		\begin{tabular}{ccc} 
			\toprule
			      &  velocity from clean data  &  velocity from noisy data   \\
			\midrule
			SSIM of joint inversion       & 0.9146 & 0.9079    \\
			\midrule
			SSIM of FWI      &  0.8248      &  0.8232     \\
			\bottomrule
		\end{tabular}\caption{Example 4: SSIM values for the recovered velocity models.}
		\label{Tab5}
\end{table}
\vspace{-8pt}

\Needspace{2\baselineskip}

\section{Conclusions} \label{sec6}
We have proposed a level-set based structural approach for multi-physics joint inversion of full-waveform and gravity data. The inversion using full-waveform data can generate high-resolution subsurface structures, but it has difficulties in dealing with complex wave phenomena arising from sharp interfaces. Moreover, its imaging of large-scale regional structures, e.g., deep structures, is slow and inefficient. The gravity inversion, on the other hand, is sensitive to density contrasts associated with interfaces, and it is effective for imaging extensive regions. The joint inversion using the two datasets integrates the strengths of both. 

To enable gravity data to assist full-waveform data, we study the well-posedness theorem of gravity inversion, and we consider a volume mass distribution where the density-contrast value is imposed as a priori information. The level-set method is then proposed to express the volume mass in the joint inversion. By characterizing the shared interface of $\rho(\mathbf{r})$ and $c(\mathbf{r})$ through zero level set, the level-set function links the structural similarity of density and wave velocity functions. The joint inversion is achieved by recovering the level-set function and associated parameters. We propose an Adam algorithm with adjustment coefficients to solve the optimization problem of joint inversion, where useful regularizations are introduced into the inversion algorithm.

In addition, we develop a weighting term $\omega$ to manage the contributions of full-waveform and gravity data, which is crucial to the joint inversion involving multi-physics datasets. The parameter $\omega$ is designed to include a balanced part $\omega_1$ and a decaying part $\omega_2$. $\omega_1$ provides a balance between the data fitting terms $E_p$ and $E_g$, which are on different scales; $\omega_2$ enables gravity data to dominate in the initial inversions and full-waveform data to dominate in later stages, effectively utilizing the features and advantages of each dataset. The strategy of balanced and decaying weight is crucial to the success of joint inversion.

{\correct Plenty of numerical examples demonstrate the effectiveness of the level-set based joint inversion algorithm. Nevertheless, the proposed method still has limitations and challenges that require continuous improvement for practical applications. (i) The joint inversion algorithm converges slowly, often requiring thousands of iterations. This is primarily due to the nonlinear nature of level-set methods, which is a common challenge. Second-order optimization approaches, such as quasi-Newton methods like L-BFGS \cite{liunoc89}, have the potential to improve convergence speed. However, because of the inherent non-linearity, incorporating these approaches into the level-set joint inversion algorithm remains a difficult problem, which requires further study. (ii) The algorithm involves numerous hyper-parameters, such as $\omega_0$, $\lambda$, $\lambda_\phi$ and $\lambda_{c_i}$. While we provide empirical guidance for their selection, a more rigorous analysis should be investigated in future studies, which is helpful to improve the algorithm's robustness in practical applications. (iii) When programming the level-set joint inversion algorithm, we incorporate the code of Deepwave. One of the reviewers in peer-review suggests that Deepwave may have scalability limitations for large scale problems. More robust and scalable open-source solvers should be considered for practical applications.
Apart from the above, we will explore integrating deep learning methods \cite{yanma23,cheli24} into the joint inversion algorithm.}

\section*{Acknowledgments}
{\correct We thank the reviewers for their valuable suggestions, which have significantly improved the paper.}
Wenbin Li is supported by Natural Science Foundation of Shenzhen (JCYJ20240813104841055),  and the Fundamental Research Funds for the Central Universities (HIT.OCEF.2024017). Jianwei Ma is supported by National Natural Science Foundation of China (42230806, U23B6010), and China National Petroleum Corporation-Peking University Strategic Cooperation Project of Fundamental Research.

\newpage
\section*{References}
\bibliographystyle{plain}
%\bibliography{myref}

\begin{thebibliography}{10}

\bibitem{abugaohabliu12}
Aria Abubakar, G~Gao, Tarek~M Habashy, and J~Liu.
\newblock Joint inversion approaches for geophysical electromagnetic and
  elastic full-waveform data.
\newblock {\em Inverse problems}, 28(5):055016, 2012.

\bibitem{afnkoknak02}
Afnimar, Kazuki Koketsu, and Koichi Nakagawa.
\newblock Joint inversion of refraction and gravity data for the
  three-dimensional topography of a sediment--basement interface.
\newblock {\em Geophysical Journal International}, 151(1):243--254, 2002.

\bibitem{amibrakun97}
F.~Aminzadeh, J.~Brac, and T.~Kunz.
\newblock {\em 3-$D$ salt and overthrust models, SEG/EAGE 3-D Modeling Series
  No. 1}.
\newblock Society of Exploration Geophysicists, 1997.

\bibitem{bir61}
F.~Birch.
\newblock The velocity of compressional waves in rocks to 10 kilobars.
\newblock {\em Journal of Geophysical Research}, 66:2199--2224, 1961.

\bibitem{cheli24}
Yihang Chen and Wenbin Li.
\newblock Learning on the correctness class for domain inverse problems of
  gravimetry.
\newblock {\em Machine Learning: Science and Technology}, 5(3):035072, 2024.

\bibitem{crestagha18}
Benjamin Crestel, Georg Stadler, and Omar Ghattas.
\newblock A comparative study of structural similarity and regularization for
  joint inverse problems governed by pdes.
\newblock {\em Inverse Problems}, 35(2):024003, 2018.

\bibitem{delberchi16}
Paolo Dell'Aversana, Giancarlo Bernasconi, Fabio Chiappa, et~al.
\newblock A global integration platform for optimizing cooperative modeling and
  simultaneous joint inversion of multi-domain geophysical data.
\newblock {\em Aims Geosciences}, 2(1):1--31, 2016.

\bibitem{dorles06}
O.~Dorn and D.~Lesselier.
\newblock Level set methods for inverse scattering.
\newblock {\em Inverse Problems}, 22:{R}67--{R}131, 2006.

\bibitem{essguavanetc18}
Ernie Esser, Lluis Guasch, Tristan van Leeuwen, Aleksandr~Y Aravkin, and
  Felix~J Herrmann.
\newblock Total variation regularization strategies in full-waveform inversion.
\newblock {\em SIAM Journal on Imaging Sciences}, 11(1):376--406, 2018.

\bibitem{galmej04}
Luis~A Gallardo and Max~A Meju.
\newblock Joint two-dimensional dc resistivity and seismic travel time
  inversion with cross-gradients constraints.
\newblock {\em Journal of Geophysical Research: Solid Earth}, 109(B3), 2004.

\bibitem{galetc03}
L.~A. Gallardo-Delgado, M.~A. P{\'e}rez-Flores, and E.~G{\'o}mez-Trevi{\~n}o.
\newblock A versatile algorithm for joint {3D} inversion of gravity and
  magnetic data.
\newblock {\em Geophysics}, 68:949--959, 2003.

\bibitem{garetc74}
G.~H.~F. Gardner, L.~W. Gardner, and A.~R. Gregory.
\newblock Formation velocity and density-the diagnostic basics for
  stratigraphic traps.
\newblock {\em Geophysics}, 39:770--780, 1974.

\bibitem{golosh09}
Tom Goldstein and Stanley Osher.
\newblock The split {B}regman method for {L}1-regularized problems.
\newblock {\em SIAM journal on imaging sciences}, 2(2):323--343, 2009.

\bibitem{habold97}
E~Haber and D~Oldenburg.
\newblock Joint inversion: a structural approach.
\newblock {\em Inverse problems}, 13(1):63, 1997.

\bibitem{heretc95}
A~Hering, R~Misiek, A~Gyulai, T~Ormos, M~Dobr{\'o}ka, and L~Dresen.
\newblock A joint inversion algorithm to process geoelectric and surface wave
  seismic data. part i: basic ideas.
\newblock {\em Geophysical Prospecting}, 43:135--156, 1995.

\bibitem{isa90}
Victor Isakov.
\newblock {\em Inverse source problems}.
\newblock American Mathematical Society, Providence, Rhode Island, 1990.

\bibitem{isaleuqia11}
Victor Isakov, Shingyu Leung, and Jianliang Qian.
\newblock A fast local level set method for inverse gravimetry.
\newblock {\em {C}ommunications in {C}omputational {P}hysics}, 10:1044--1070,
  2011.

\bibitem{jia20}
Wenbin Jiang.
\newblock 3-d joint inversion of seismic waveform and airborne gravity
  gradiometry data.
\newblock {\em Geophysical Journal International}, 223(2):746--764, 2020.

\bibitem{joh21}
Steven~G Johnson.
\newblock Notes on perfectly matched layers ({PMLs}).
\newblock {\em arXiv preprint arXiv:2108.05348}, 2021.

\bibitem{kadvan19}
Ajinkya Kadu and Tristan van Leeuwen.
\newblock A convex formulation for binary tomography.
\newblock {\em IEEE Transactions on Computational Imaging}, 6:1--11, 2019.

\bibitem{kinba14}
Diederik~P Kingma and Jimmy Ba.
\newblock Adam: A method for stochastic optimization.
\newblock {\em arXiv preprint arXiv:1412.6980}, 2014.

\bibitem{konrovetc23}
Apostolos Kontakis, Diego Rovetta, Daniele Colombo, and Ernesto
  Sandoval-Curiel.
\newblock Efficient 1.5 d full waveform inversion in the laplace-fourier
  domain.
\newblock {\em Inverse Problems}, 39(7):075012, 2023.

\bibitem{leletc12}
P~G Leli{\`e}vre, C~G Farquharson, and C~A Hurich.
\newblock Joint inversion of seismic traveltimes and gravity data on
  unstructured grids with application to mineral exploration.
\newblock {\em Geophysics}, 77(1):K1--K15, 2012.

\bibitem{lileuqia14}
Wenbin Li, Shingyu Leung, and Jianliang Qian.
\newblock A level-set adjoint-state method for crosswell
  transmission-reflection traveltime tomography.
\newblock {\em Geophysical Journal International}, 199(1):348--367, 2014.

\bibitem{liluqia16}
Wenbin Li, Wangtao Lu, and Jianliang Qian.
\newblock A level set method for imaging salt structures using gravity data.
\newblock {\em Geophysics}, 81(2):G35--G51, 2016.

\bibitem{liqia21}
Wenbin Li and Jianliang Qian.
\newblock Simultaneously recovering both domain and varying density in inverse
  gravimetry by efficient level-set methods.
\newblock {\em Inverse Problems and Imaging}, 15(3):387--413, 2021.

\bibitem{liold98}
Y.~Li and D.~Oldenburg.
\newblock 3{D} inversion of gravity data.
\newblock {\em Geophysics}, 63:109--119, 1998.

\bibitem{litlessan98}
A.~Litman, D.~Lesselier, and F.~Santosa.
\newblock Reconstruction of a two-dimensional binary obstacle by controlled
  evolution of a level-set.
\newblock {\em Inverse Problems}, 14:685--706, 1998.

\bibitem{liunoc89}
Dong~C Liu and Jorge Nocedal.
\newblock On the limited memory {BFGS} method for large scale optimization.
\newblock {\em Mathematical programming}, 45(1):503--528, 1989.

\bibitem{metbroetc16}
Ludovic M{\'e}tivier, Romain Brossier, Quentin Merigot, Edouard Oudet, and Jean
  Virieux.
\newblock An optimal transport approach for seismic tomography: Application to
  3d full waveform inversion.
\newblock {\em Inverse Problems}, 32(11):115008, 2016.

\bibitem{mooetc11}
M.~Moorkamp, B.~Heincke, M.~Jegen, A.~W. Roberts, and R.~W. Hobbs.
\newblock A framework for {3-D} joint inversion of {MT}, gravity and seismic
  refraction data.
\newblock {\em Geophysical Journal International}, 184:477--493, 2011.

\bibitem{niejac00}
L.~Nielsen and B.~H. Jacobsen.
\newblock Integrated gravity and wide-angle seismic inversion for
  two-dimensional crustal modelling.
\newblock {\em Geophysical Journal International}, 140:222--232, 2000.

\bibitem{oshset88}
S.~J. Osher and J.~A. Sethian.
\newblock Fronts propagating with curvature dependent speed: algorithms based
  on {Hamilton-Jacobi} formulations.
\newblock {\em J. Comput. Phys.}, 79:12--49, 1988.

\bibitem{oshfed06}
Stanley Osher and Ronald Fedkiw.
\newblock {\em Level set methods and dynamic implicit surfaces}, volume 153.
\newblock Springer Science \& Business Media, 2006.

\bibitem{ple06}
R.~E. Plessix.
\newblock A review of the adjoint-state method for computing the gradient of a
  functional with geophysical applications.
\newblock {\em {G}eophysical {J}ournal {I}nternational}, 167:495--503, 2006.

\bibitem{richardson_alan_2023}
Alan Richardson.
\newblock Deepwave.
\newblock Software: https://doi.org/10.5281/zenodo.8381177, September 2023.

\bibitem{savrodmas82}
J~M Savino, W~L Rodi, and J~F Masso.
\newblock Inversion modeling of multiple geophysical data sets for geothermal
  exploration: application to roosevelt hot springs area. final report.
\newblock Technical report, S-Cubed, La Jolla, CA (USA), 1982.

\bibitem{siletc20}
Raul~U Silva, Jonas~D De~Basabe, Mrinal~K Sen, Mario Gonzalez-Escobar, Enrique
  Gomez-Trevino, and Selene Solorza-Calderon.
\newblock Cooperative full waveform and gravimetric inversion.
\newblock {\em J. Seismic Exploration}, 29(6):549--573, 2020.

\bibitem{sussmeosh94}
M.~Sussman, P.~Smereka, and S.~J. Osher.
\newblock A level set approach for computing solutions to incompressible
  two-phase flows.
\newblock {\em Journal of computational physics}, 114:146--159, 1994.

\bibitem{vanher13}
Tristan Van~Leeuwen and Felix~J Herrmann.
\newblock Mitigating local minima in full-waveform inversion by expanding the
  search space.
\newblock {\em Geophysical Journal International}, 195(1):661--667, 2013.

\bibitem{virope09}
Jean Virieux and St{\'e}phane Operto.
\newblock An overview of full-waveform inversion in exploration geophysics.
\newblock {\em Geophysics}, 74(6):WCC1--WCC26, 2009.

\bibitem{yanma23}
Fangshu Yang and Jianwei Ma.
\newblock Fwigan: Full-waveform inversion via a physics-informed generative
  adversarial network.
\newblock {\em Journal of Geophysical Research: Solid Earth},
  128(4):e2022JB025493, 2023.

\bibitem{yonliahuazhe18}
Peng Yong, Wenyuan Liao, Jianping Huang, and Zhenchun Li.
\newblock Total variation regularization for seismic waveform inversion using
  an adaptive primal dual hybrid gradient method.
\newblock {\em Inverse Problems}, 34(4):045006, 2018.

\end{thebibliography}

\end{document}